\documentclass[aps,prx,twocolumn,superscriptaddress,showpacs,floatfix,longbibliography]{revtex4-1}
\usepackage{amsmath,amssymb,wasysym,graphicx}

\usepackage[utf8]{inputenc}
\usepackage[T1]{fontenc}
\usepackage{xcolor}

\IfFileExists{newtxtext.sty}
   {\usepackage{newtxtext,newtxmath}}
   {\IfFileExists{stix.sty}
      {\usepackage{stix}}
      {\IfFileExists{mathptmx.sty}
      {\usepackage{mathptmx}}{} } }

\usepackage{textcomp}

\usepackage{bm}
\usepackage{lipsum} 
\usepackage{float}

\IfFileExists{siunitx.sty}{\usepackage{booktabs,siunitx}}{}

\pdfoutput=1
\usepackage{color}
\definecolor{LinkColor}{rgb}{0.256,0.439,0.588}
\usepackage{hyperref}
\hypersetup{
   pdfauthor={good guys},
   pdftitle={Correlation-induced insulating topological phases at charge neutrality in twisted bilayer graphene},
   colorlinks=true,
   citecolor=LinkColor,
   linkcolor=LinkColor,
   urlcolor=LinkColor
}

\renewcommand{\vec}[1]{\mathbf{#1}}

\definecolor{darkergreen}{rgb}{0.0, 0.6, 0.0}

\usepackage{pifont}
\usepackage[normalem]{ulem}

\usepackage{multirow}

\begin{document}

\title{Correlation-induced insulating topological phases at charge neutrality in twisted bilayer graphene} 

\author{Yuan Da Liao}
\affiliation{Beijing National Laboratory for Condensed Matter Physics and Institute of Physics, Chinese Academy of Sciences, Beijing 100190, China}
\affiliation{School of Physical Sciences, University of Chinese Academy of Sciences, Beijing 100190, China}

\author{Jian Kang}
\email{jkang@suda.edu.cn}
\affiliation{School of Physical Science and Technology \& Institute for Advanced Study, Soochow University, Suzhou, 215006, China}

\author{Clara N. Brei\o}
\affiliation{Niels Bohr Institute, University of Copenhagen, Lyngbyvej 2, 2100 Copenhagen, Denmark}

\author{Xiao Yan Xu}
\affiliation{Department of Physics, University of California at San Diego, La Jolla, California 92093, USA}

\author{Han-Qing Wu}
\affiliation{School of Physics, Sun Yat-Sen University, Guangzhou, 510275, China}

\author{\\Brian M. Andersen}
\email{bma@nbi.ku.dk}
\affiliation{Niels Bohr Institute, University of Copenhagen, Lyngbyvej 2, 2100 Copenhagen, Denmark}

\author{Rafael M. Fernandes}
\email{rfernand@umn.edu}
\affiliation{School of Physics and Astronomy, University of Minnesota, Minneapolis, MN 55455, USA}

\author{Zi Yang Meng}
\email{zymeng@hku.hk}
\affiliation{Department of Physics and HKU-UCAS Joint Institute of Theoretical and Computational Physics, The University of Hong Kong, Pokfulam Road, Hong Kong SAR, China}
\affiliation{Beijing National Laboratory for Condensed Matter Physics and Institute of Physics, Chinese Academy of Sciences, Beijing 100190, China}
\affiliation{Songshan Lake Materials Laboratory, Dongguan, Guangdong 523808, China}

\begin{abstract}
Twisted bilayer graphene (TBG) provides a unique framework to elucidate the interplay between strong correlations and topological phenomena in two-dimensional systems. The existence of multiple electronic degrees of freedom -- charge, spin, and valley -- gives rise to a plethora of possible ordered states and instabilities. Identifying which of them are realized in the regime of strong correlations is fundamental to shed light on the nature of the superconducting and correlated insulating states observed in the TBG experiments. Here, we use unbiased, sign-problem-free quantum Monte Carlo simulations to solve an effective interacting lattice model for TBG at charge neutrality. Besides the usual cluster Hubbard-like repulsion, this model also contains an assisted hopping interaction that emerges due to the non-trivial topological properties of TBG. Such a non-local interaction fundamentally alters the phase diagram at charge neutrality, gapping the Dirac cones even for infinitesimally small interaction. As the interaction strength increases, a sequence of different correlated insulating phases emerge, including a quantum valley Hall state with topological edge states, an intervalley-coherent insulator, and a valence bond solid. The charge-neutrality correlated insulating phases discovered here provide the sought-after reference states needed for a comprehensive understanding of the insulating states at integer fillings and the proximate superconducting states of TBG.
\end{abstract}

\date{\today}
\maketitle

\section{Introduction}
The recent discovery of correlated insulating and superconducting phases in twisted bilayer graphene (TBG)~\cite{bistritzer2011moire, cao2018correlated, cao2018unconventional} and other moir\'e systems~\cite{shen2019observation, liu2019spin, cao2019electric, chen2020tunable} sparked a flurry of activity to elucidate and predict the electronic quantum phases realized in their phase diagrams ~\cite{kerelsky2019maximized, tomarken2019electronic, lu2019superconductors, xie2019spectroscopic, jiang2019charge, wong2019cascade, zondiner2019cascade, saito2019decoupling, stepanov2019interplay, chen2019evidence, chen2019signatures, xu2018topological, kang2018symmetry, koshino2018maximally,yuan2018model,po2018origin, liu2018chiral,LadoPablo, ochi2018possible,dodaro2018phases,guo2018pairing,isobe2018unconventional,venderbos2018correlations,guinea2018electrostatic,liu2018pseudo,
LiuValley2019,Cea2019,tang2019spin,gonzalez2019kohn,kang2019strong,seo2019ferromagnetic,zhang2019nearly,lee2019theory,wucollective,
wu2019ferromagnetism,bultinck2019anomalous,liu2019nematic,alavirad2019ferromagnetism,chatterjee2019symmetry,chichinadze2019nematic,
bultinck2019ground,liu2019correlated,fernandes2019nematicity,zhang2020correlated,repellin2019ferromagnetism,LiuAnomalous2020,
kang2020nonabelian,huang2020slaverotor,lu2020chiral,li2019experimental,YuxuanWang2020,TianleWang2020,Christos2020,VKozii2020,WYHe2020,xu2018kekule,YDLiao2019,YDLiao2020review,soejima2020efficient,XieFang2020,VafekKang2020}. Because the low-energy bands of TBG have a very small bandwidth, of about $10$meV at the magic twist angle, the Coulomb interaction, which is of the order of $25$meV, is expected to play a fundamental role in shaping the phase diagram \cite{bistritzer2011moire,po2018origin,kang2018symmetry,wong2019cascade}. Indeed, insulating states have been reported at all commensurate fillings of the moir\'e superlattice \cite{lu2019superconductors}, signaling to the importance of strong correlations. Besides correlations, topological phenomena have also been reported, including a quantum anomalous Hall (QAH) phase~\cite{sharpe2019emergent,serlin2020intrinsic}. 

An important issue is the nature of the quantum ground state at charge neutrality, characterized in real space by 4 electrons per moir\'e unit cell, and in momentum space by Dirac points at the Fermi level.  Experimentally, a large charge gap characteristic of an insulating state was reported in transport measurements in Ref. \cite{lu2019superconductors} and in STM measurements in Ref. \cite{xie2019spectroscopic}, despite no obvious alignment with the underlying hBN layer. The fact that this gap is not observed in all devices has been attributed to inhomogeneity \cite{lu2019superconductors}. Theoretically, because the electronic states in TBG have several degrees of freedom -- spin, valley, and sublattice -- various possible ground states can emerge. Indeed, Hartree-Fock calculations of the continuum model at charge neutrality found various possible phases, such as orbital-magnetization density-waves, valley polarized states, and states that spontaneously break the three-fold rotational symmetry of the moir\'e lattice \cite{MacDonald2020,liu2019nematic,Guinea2020,bultinck2019ground,liu2019correlated,LiuAnomalous2020,TianleWang2020,VKozii2020}. To distinguish among these different possibilities, and to search for novel ordered states in TBG, it is desirable to employ a method that is not only unbiased, but that can also handle strong correlations. 

Large-scale quantum Monte Carlo (QMC) simulations provide an optimal tool, limited only by the finite lattice sizes. Although such a limitation makes it impossible to simulate a model with thousands of carbon atoms per moir\'e unit cell, it is very well suited to solve lattice models on the moir\'e length scale. At charge neutrality, the non-interacting part of the model has only Dirac points at the Fermi level. The crucial part of the model, however, is the interacting part, which governs the system's behavior in the strong-coupling regime. At first sight, based on the analogy with other strongly-correlated models, it would seem enough to consider a cluster Hubbard-like repulsion as the main interaction of the problem. Previously, some of us used QMC to simulate this model, which does not suffer from the infamous fermionic sign-problem \cite{xu2018kekule,YDLiao2019}. The result was a variety of valence-bond insulating states, which however only onset at relatively large values of the interaction $U$, of the order of several times the bandwidth $W$. Below these large values, the system remained in the Dirac semi-metal phase.

However, microscopically, the full interaction of the lattice model can be derived from projecting the screened Coulomb repulsion on the Wannier states (WSs) of TBG. The latter turn out to be quite different than in other correlated materials, as they have nodes on the sites of the moir\'e honeycomb superlattice and a three-peak structure that overlaps with Wannier functions centered at other sites \cite{kang2018symmetry, koshino2018maximally, po2018origin}. Recent work has shown that this leads to the emergence of an additional and sizable non-local interaction, of the form of an assisted-hopping term \cite{kang2019strong,kang2020nonabelian}. This new interaction ultimately arises from the fact that, in a lattice model, the symmetries of the continuum model cannot all be implemented locally, a phenomenon dubbed Wannier obstruction  \cite{po2018origin}. Therefore, the assisted-hopping interaction is not a simple perturbation, but a direct and unavoidable manifestation of the non-trivial topological properties of TBG. This important aspect of the TBG was not taken into consideration in the previous QMC simulations.

In this paper, we made this important step forward by studying the impact of the assisted-hopping interaction on the ground state of TBG at charge neutrality via sign-problem-free QMC simulations. We find that such a term qualitatively changes the phase diagram, as compared to the case where only the cluster Hubbard interaction is included. In particular, the Dirac semi-metal phase is no longer stable, but is gapped already at weak-coupling. We show that this gap is a manifestation of a quantum valley Hall (QVH) state, characterized by topological edge states. We confirm this weak-coupling result by unrestricted Hartree-Fock (HF) calculations of the same model simulated by QMC. The HF calculations, well suited for weak interactions, also show that the QVH state is a robust property of the weak-coupling regime, and is directly connected to the assisted-hopping term. As the interaction strength increases, a different type of insulating phase arises, displaying intervalley coherence (IVC) order. This onsite IVC order breaks the spin-valley SU(4) symmetry of the interacting part of the model, resembling recently proposed ferromagnetic-like SU(4) states proposed to emerge in TBG at charge neutrality and other integer fillings~\cite{kang2019strong,bultinck2019ground}. Upon further increasing the interaction, a columnar valence bond solid (cVBS) insulator state appears, favored by the Hubbard-like interaction~\cite{Lang2013,ZCZhou2016,xu2018kekule,YDLiao2019}. Importantly, the presence of the assisted-hopping term makes the QVH and IVC states accessible already for substantially smaller values of $U/W$, as compared to the case where there is only Hubbard repulsion. Therefore, the experimental observation of such quantum states in TBG at charge neutrality would provide strong evidence for the importance of non-local, topologically-driven interactions in this system.

\begin{figure*}[htp!]
	\includegraphics[width=0.7\textwidth]{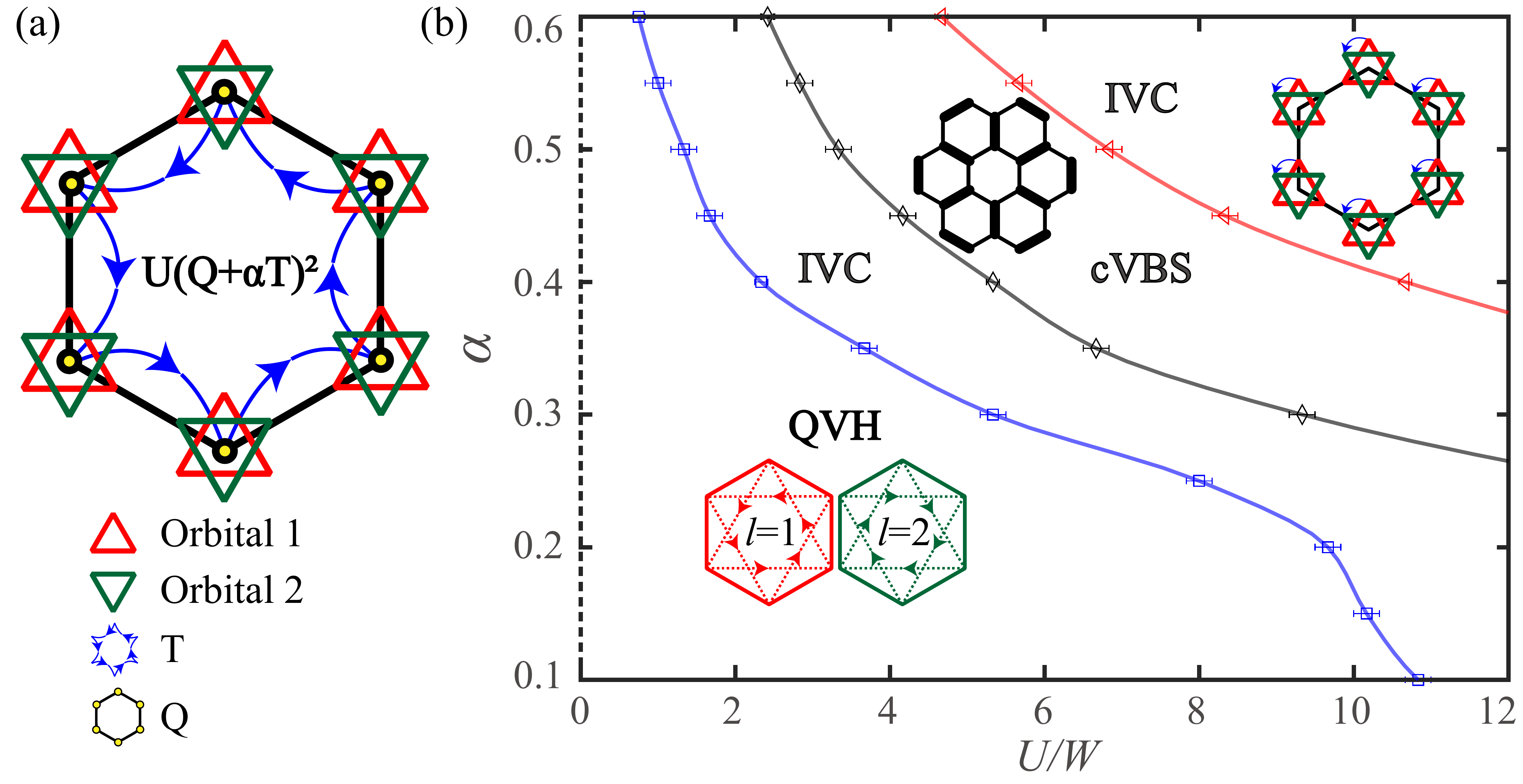}
	\caption{Ground state phase diagram at charge neutrality obtained via QMC simulations. (a) Schematics of the model: each lattice site  on the dual moir\'e honeycomb lattice contains two valleys $l=1,2$ (red and green triangles) and spins $\sigma=\uparrow,\downarrow$ (not shown), with spin-valley SU(4) symmetry.  The interactions act on every hexagon and consist of the cluster charge term $Q_{\varhexagon}$ (yellow dots) and the assisted-hopping interaction term $T_{\varhexagon}$ (blue arrows). (b) Ground state phase diagram, spanned by the $U/W$ and $\alpha$ axes, obtained from QMC simulations. 
		The y-axis at $U=0$ (dash line) stands for the Dirac semi-metal phase. 
		At very small $U$, the ground state is a quantum valley Hall (QVH) phase characterized by emergent imaginary next-nearest-neighbor hopping with complex conjugation at the valley index, as illustrated by the red and green dashed hoppings with opposite directions. The system has an insulating bulk but acquires topological edge states. Upon further increasing $U$, an intervalley-coherent (IVC) insulating state is found, which breaks the SU(4) symmetry at every lattice site by removing the valley symmetry. Because it preserves the lattice translational symmetry, it is ferromagnetic-like. The columnar valence bond solid (cVBS) insulator, which appears after the IVC phase, breaks the lattice translational symmetry and preserves the onsite SU(4) symmetry. Note that there is a re-emergence of the IVC phase for the largest interactions probed. The phase transitions between QVH and IVC (blue line), between the IVC and cVBS (black line), and between the cVBS and IVC (red line) are all first order.} \label{fig:fig1}
\end{figure*}

\section{Model, symmetry analysis and method}

Our lattice Hamiltonian $H$ for spinful fermions on the moir\'e superlattice consists of a non-interacting tight-binding term $H_{0}$ and an interaction term $H_{\varhexagon}$. An important property of the narrow bands of TBG is their fragile topology, resulting in the phenomenon known as Wannier obstruction, which prevents the construction of localized Wannier orbitals that locally implement the symmetries of the system of coupled Dirac fermions (i.e. the so-called continuum model) \cite{po2018origin}. There are essentially two ways to overcome the Wannier obstruction: (i) include additional remote bands (at the expense of adding more Wannier orbitals and the associated interactions) or (ii) implement one of the symmetries of the continuum model non-locally (at the expense of adding longer-range hopping parameters). In case (ii), the Wannier orbitals, denoted by the operators $c_{il\sigma}$, live on the sites $i$ of the dual honeycomb moir\'e superlattice and are labeled by spin $\sigma=\uparrow,\downarrow$ and two orbital degrees of freedom $l=1,2$ (roughly corresponding to the two valleys) \cite{kang2018symmetry, koshino2018maximally}. In case (i), the Wannier orbitals can live on the sites of the triangular moir\'e superlattice and additional orbital quantum numbers are required.

Since we are interested in the strong-coupling regime, it is crucial to project the screened Coulomb repulsion onto the non-obstructed low-energy WSs. This was done in Ref. \cite{kang2019strong} for case (ii), which implemented the $C_2 \mathcal{T}$ symmetry non-locally, where $C_2$ refers to two-fold rotations with respect to the $z$-axis and $\mathcal{T}$, to time-reversal. This was found to give rise to a non-local assisted hopping interaction, besides the more standard Hubbard-like repulsion. More specifically, the interacting Hamiltonian in this case is given by the sum of two contributions \cite{kang2019strong}
\begin{equation}
\label{eq:eq2}
H_{\varhexagon} = U\sum_{\varhexagon}(Q_{\varhexagon}+\alpha T_{\varhexagon}-4)^2,
\end{equation}
Here, $U$ sets the overall strength of the Coulomb interaction. The two terms in Eq.~(\ref{eq:eq2}), illustrated in Fig.~\ref{fig:fig1}(a), consist of the cluster charge $Q_{\varhexagon} \equiv \sum_{j\in \varhexagon}\frac{n_j}{3}$, with $n_j = \sum_{l\sigma}c^\dagger_{jl\sigma}c^{\phantom{\dagger}}_{jl\sigma}$, and the cluster assisted hopping 
$T_{\varhexagon} \equiv \sum_{j,\sigma} \left(i c_{j+1,1\sigma}^\dagger c^{\phantom{\dagger}}_{j,1\sigma} - i c_{j+1,2\sigma}^\dagger c^{\phantom{\dagger}}_{j,2\sigma} + h.c. \right)$. The index $j=1,\ldots,6$ sums over all six sites of the elemental hexagon in the honeycomb lattice. 

The cluster charge term $Q_{\varhexagon}$ is analogous to the Hubbard onsite repulsion in the standard Hubbard model; the reason why it extends over the entire hexagon is because of the screening length set by the separation between the gates in a TBG device and the overlap between WSs of neighboring sites. In particular, the Wannier wave-functions are not peaked at the honeycomb sites, but instead are extended and peaked at the centers of the three neighboring hexagons \cite{po2018origin,kang2018symmetry, koshino2018maximally}. 
Therefore, one single WS overlaps spatially with other WSs on neighboring sites, leading to the cluster charging term $Q_{\varhexagon}$. On the other hand, the origin of the assisted hopping term $T_{\varhexagon}$ is topological, i.e. it comes precisely from the fragile topology of TBG. This can be seen from the derivation of the coefficient $\alpha$, which controls the relative strength of the two interactions. It is the overlap of two neighboring WSs in a single hexagon, given as~\cite{kang2019strong}
\begin{equation}
   \alpha = -i \int_{\varhexagon} \mathrm{d} \mathbf{r}\ w_{1, 1}^*(\mathbf{r}) w_{2, 1}(\mathbf{r}) \ , 
\end{equation}
 where $w_{1, 1}(\mathbf{r})$ is the wave function of the WS at the site $1$ of the hexagon with the valley index $1$ and $w_{2,1}(\mathbf{r})$ is the WS at site $2$ of the hexagon. Note that the integral is taken only inside the single hexagon.  Although $\alpha$ is generally a complex number, its phase can be always removed by a gauge transformation. 
As argued in Ref.~\cite{kang2019strong}, the sizable value of $\alpha$ comes from the topological obstruction to the fully symmetric WSs. If the bands were topologically trivial, all the symmetries could be locally implemented for the WSs. As a consequence, the WSs are $C_2''$ symmetric and have the same parity. While two neighboring WSs overlap in two neighboring hexagons and sum to $0$ because of the orthogonality, the two verlaps are equal since $C_2''$ symmetry relates them together. Therefore, each one vanishes leading to $\alpha = 0$. The sizable value of $\alpha$ manifests the nontrivial topological properties of the narrow bands.  In Ref.~\cite{kang2019strong}, $\alpha$ was found to be $0.23$ based on Koshino's model without including the lattice relaxation~\cite{Koshino2012}. Since the relaxation will inevitably change $\alpha$, we will not fix its value here but study the phase diagram for a wider range of $\alpha$.


The non-local implementation of the $C_2 \mathcal{T}$ symmetry also results in many longer-range hopping parameters in $H_0$ \cite{kang2018symmetry}. While ideally one would like to solve the model containing all these tight-binding terms, but such a model would in general suffer from the sign-problem and cannot be efficiently simulated with QMC, which has the advantage of being unbiased and applicable even for large interaction values. In contrast, the interaction term $H_{\varhexagon}$ alone can be solved with QMC without the sign-problem, despite the presence of the non-local interaction $T_{\varhexagon}$ {(see discussions in Appendices~\ref{app:pqmc}). Therefore, because we are interested in the strong-coupling regime, we opt to keep the full non-trivial interaction term and simplify the tight-binding Hamiltonian in order to circumvent the sign problem.

\begin{equation}
\label{eq:eq1}
H_{0} =  -t\sum_{\langle ij \rangle l \sigma}\left(c^{\dagger}_{il\sigma}c^{\phantom{\dagger}}_{jl\sigma}+\rm{h.c}. \right),
\end{equation}

This simple nearest-neighbor band dispersion displays Dirac points at charge neutrality, and can be simulated with sign-problem-free QMC at charge neutrality (four electrons per hexagon once averaging over the lattice), which we assume hereafter. Moreover, we set the hopping parameter $t=1$ and use the bare bandwidth $W=6t$ as the energy unit in the remainder of the paper.

We emphasize that, in the strong-coupling regime, we expect that it is the non-trivial structure of the projected interactions that will determine the ground state, and not the bare tight-binding dispersion. Below, we provide evidence that this is indeed the case. Thus, the crucial point is that the topologically non-trivial properties of the TBG band structure are already incorporated in the interacting part of our model, which inherits them from the projection of the screened Coulomb interaction on the non-trivial WSs.

An interesting feature of $H_{\varhexagon}$ is its emergent SU(4) symmetry describing simultaneous rotations in spin and orbital spaces. To illustrate this, we introduce the spinor
$\psi_i = \left( c_{i1\uparrow} ,  c_{i1\downarrow} ,  c_{i2\uparrow} ,   c_{i2\downarrow}  \right)^T$ and rewrite the interactions as: 

\begin{align} 
Q_{\varhexagon} & = \frac13 \sum_{i \in \varhexagon} \psi_i^{\dagger} \psi^{\phantom{\dagger}}_i \\
T_{\varhexagon} & = i \sum_{i\in \varhexagon} \psi_{i+1}^{\dagger} T_0 \psi^{\phantom{\dagger}}_i + \rm{h.c.} 
\end{align}
with $T_0 = \mathrm{diag}(1, 1, -1, -1)$ denoting a diagonal matrix. Consider the unitary transformation 
\begin{align}
\psi_{i \in \mathcal{A}} \rightarrow  U \psi_i \quad \mbox{and} \quad \psi_{i \in \mathcal{B}} \rightarrow T_0 U T_0 \psi_i \ , \label{Eqn:SU4Int}
\end{align}
, where $U$ is an arbitrary $4\times 4$ unitary matrix and $\mathcal{A}$($\mathcal{B}$) are the two sublattices of the honeycomb lattice. 
It is clear that both  $Q_{\varhexagon}$ and $T_{\varhexagon}$ are invariant under this transformation. 
On the other hand, the kinetic term $H_0$ is not invariant under the transformation given by Eq.~\eqref{Eqn:SU4Int}, thus leaving the whole Hamiltonian only $\mathrm{U(1) \times SU(2) \times SU(2)}$ symmetric, 
i.e. the valley U(1) symmetry and the two independent spin SU(2) rotations for the two valleys~\cite{po2018origin,bultinck2019anomalous}. Thus, strictly speaking, the SU(4) symmetry is exact for $H_{\varhexagon}$ but only approximate for $H_{0}$.

To solve the model $H=H_{0}+H_{\varhexagon}$ non-perturbatively, we employ large-scale projection QMC simulations~\cite{xu2018kekule,YDLiao2019}. This QMC approach, employed in several previous studies ~\cite{meng2010quantum,Lang2013,xu2017topo,YuanYaoHe2018,xu2018kekule,YDLiao2019,YZLiu2020}, provides results about the $T=0$ ground state, the correlation functions (which are used to determine broken symmetries), and the electronic spectra (both single-particle and collective excitations). As explained above, despite the presence of the assisted hopping interaction, the model at charge-neutrality does not suffer from the sign-problem (see Appendix~\ref{app:pqmc} for details). Thus, it can be efficiently simulated by introducing an extended auxiliary bosonic field that dynamically couples to the electrons on a hexagon -- in contrast to the standard Hubbard model, where the auxiliary field is local. Details about the projection QMC implementation, as well as comparison with results from exact diagonalization, are discussed in the Appendix~\ref{app:pqmc} and Appendix~\ref{app:ed}. 

We also complemented the unbiased QMC simulations with self-consistent HF calculations, which are well-suited for the weak-coupling regime, and can be employed even when additional terms are included in $H$ that introduce a sign-problem for QMC. The HF approach is fully unrestricted in the sense that $H_0+H_{\varhexagon}$ is mean-field decoupled in all channels, and free to acquire any value in site-, spin-, and valley-space. 
Further technical details, including the resulting coupled set of (real space) self-consistency equations, can be found in the Appendix~\ref{app:hfm}. In the regime of weak interactions, we find excellent agreement between the results obtained from HF and QMC. Importantly, in the same appendix, we also extended the HF calculations to include longer-range hopping terms in $H_0$, and found that the results are similar. This supports our aforementioned expectation that the non-trivial structure of the projected interactions, arising from the fragile topology of TBG, dominates the ground state properties of the system, at least at charge neutrality.

\begin{figure}[htp!]
	\includegraphics[width=0.9\columnwidth]{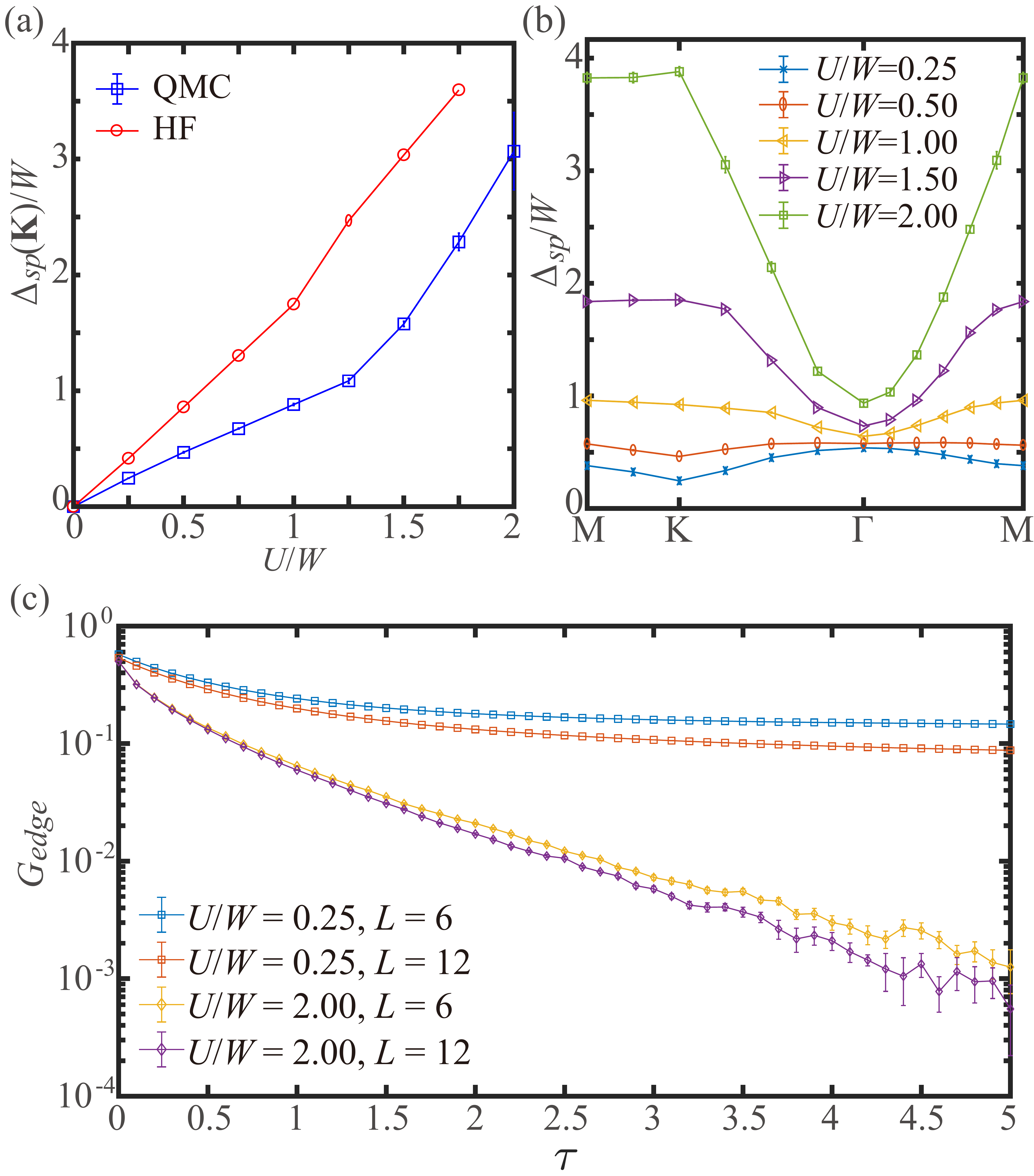}
\caption{Quantum valley Hall insulator (QVH) and gapless edge states. (a) The single-particle gap $\Delta_{\mathrm{sp}}(K)/W$ at the K point as a function of $U/W$ for $\alpha=0.45$, extracted from both QMC (blue points) and HF calculations (red points). For QMC, the spatial system size is $L=12$. The Dirac semi-metal is gapped out at the smallest $U$ values probed. (b) Single-particle gap extracted from QMC with $L=12$ along a high-symmetry path of the Brillouin zone. (c) The topological nature of the QVH phase is manifested by valley-polarized edge states. Here we compare the edge Green's function for valley $l=1$ and spin $\uparrow$ at $U/W=0.25$ (inside the QVH phase) and $U/W=2.0$ (inside the IVC phase). It is clear that gapless edge modes only appear in the former case, highlighting the topological nature of the QVH phase.}
\label{fig:fig2}
\end{figure}

\section{Quantum valley Hall phase, intervalley-coherent insulator, and valence-bond solid}  
The QMC-derived phase diagram for the ground states at charge neutrality is shown in Fig.~\ref{fig:fig1} (b) as a function of $U/W$ and $\alpha$. We emphasize that while $U$ gives the overall magnitude of the total interaction term, $\alpha$ is proportional to the relative strength between the assisted-hopping and cluster-charge terms. We find that three types of correlated insulating phases emerge in the phase diagram: the quantum valley Hall (QVH) phase, the intervalley-coherent (IVC) phase, and the columnar valence bond solid (cVBS).

The QVH phase is the ground state for small $U$ values and is characterized by a gap in the single-particle spectrum. This gap can be extracted from the imaginary-time decay of the Green's function along a high-symmetry path of the Brillouin zone (BZ), $G(\mathbf{k},\tau)\sim e^{-\Delta_{\mathrm{sp}}(\mathbf{k})\tau}$. Fig.~\ref{fig:fig2}(a) shows the enhancement of the single-particle gap at the $K$ point of the BZ as a function of $U$ for a fixed $\alpha = 0.45$ (blue points).  Together with Fig.~\ref{fig:fig2}(b), one sees the gap opens at the entire BZ at infinitesimally small $U$. 
In many honeycomb lattice models, the Dirac cone at the $K$ point is protected by a symmetry, and the semi-metal phase is robust against weak interactions~\cite{meng2010quantum,Lang2013,xu2018kekule,YDLiao2019}. In TBG, however, the relevant symmetry, $C_2 \mathcal{T}$, cannot be implemented locally due to the topological Wannier obstruction. This opens up the possibility of very weak interactions gapping out the Dirac cone.

In our QMC simulations, for any non-zero $\alpha$ that we investigated, a gap appeared even for the smallest values of $U$ probed. This suggests a weak-coupling origin of this phase. To verify it, we performed HF calculations on the same lattice model. The results, shown by the red points in Fig.~\ref{fig:fig2}(a), are in very good agreement with the QMC results. We also used HF to investigate the stability of the gap against changing the phase that appears in the assisted-hopping term $T_{\varhexagon}$ \cite{kang2019strong}. This phase can be gauged away, at the expense of introducing complex hopping terms in $H_0$, which introduce a sign-problem to the QMC simulations. However, they do not affect the efficiency of the HF algorithm. As discussed in the Appendix~\ref{app:hfm}, our analysis confirm that the onset of the QVH phase is robust and appears regardless of the phase of $T_{\varhexagon}$.

Importantly, we find that the gap completely disappears when $\alpha=0$, in agreement with Ref. \cite{YDLiao2019}. Combined with the fact that the gap onsets for small interaction values when $\alpha \neq 0$, this suggests that the origin of the gap can be understood from a mean-field decoupling of the cross-term $\sum_{\varhexagon} Q_{\varhexagon} T_{\varhexagon}$ of the interaction in Eq.~\ref{eq:eq2}. This cross-term can be rewritten as:

\begin{equation}
\sum_{\varhexagon} Q_{\varhexagon} T_{\varhexagon} = i \sum_{\varhexagon} \sum_{i, j = 1}^6 \sum_{l, m = 1}^2 (-1)^{m} \left( c^{\dagger}_{i,l} c^{\dagger}_{j+1, m} c^{\phantom{\dagger}}_{j, m} c^{\phantom{\dagger}}_{i,l} - h.c. \right)  \label{Eq:ExpandInt}
\end{equation}
where $l$ and $m$ are valley indices and the spin index is omitted for simplicity. The terms with $j = i - 1$ and $j = i$ vanish after summing over different hexagons. In the weak-coupling limit, we can do a mean-field decoupling and use $\langle c^{\dagger}_{i, l} c^{\phantom{\dagger}}_{i + 1, m} \rangle \propto \delta_{lm}$, due to the nearest-neighbor hopping term present in $H_0$. The cross-term then becomes: 

\begin{equation}
 \sum_{\varhexagon} Q_{\varhexagon} T_{\varhexagon} \propto -i \sum_{\varhexagon} \sum_{i = 1}^6 \sum_{l = 1}^2 (-1)^{l} \left( c^{\dagger}_{i, l} c^{\phantom{\dagger}}_{i + 2, l} + c^{\dagger}_{i -2, l} c^{\phantom{\dagger}}_{i, l} - h.c.  \right) \label{Eq:Cross}
 \end{equation}
Thus, the cross-term of the interaction naturally induces an imaginary hopping between next-nearest-neighbors in the weak-coupling limit. As a consequence, the mean-field Hamiltonian becomes two copies (four, if we consider the spin degeneracy) of the Haldane model \cite{Haldane88,Hohenadler2012}, leading to a Chern number of $\pm 1$ for the two different valleys. For this reason, we call this state a QVH phase; it is illustrated in the corresponding inset in Fig. \ref{fig:fig1} (b). We verified that our self-consistent HF calculation generates the same pattern of imaginary NNN hopping. 

One of the hallmarks of the Haldane model is the existence of gapless edge modes, despite the bulk being gapped. In the QVH phase, these edge states should be valley-polarized. To probe them, we performed QMC simulations with open boundary conditions and extracted the imaginary-time Green's functions on the edge, $G_{\mathrm{edge}}(\tau)\sim e^{-\Delta_{\mathrm{sp}}\tau}$. As shown in Fig.~\ref{fig:fig2} (c), in the regime of small $U$ ($U/W=0.25$), the Green's function on the edge decays to a constant in the long imaginary-time limit, demonstrating the existence of a gapless edge mode in the QVH phase. 
To verify the existence of edge states, we also use HF to capture the topological nature of the QVH phase. In practice, we open the boundaries in the system and compute a self-consistent result with parameters as in Table \ref{tab:RSres1} from Appendix~\ref{app:hfm} ($t=1$ ,$\alpha = 0.45$, $U/W = 0.5$, $T = 2.5\cdot10^{-5}$ and $N = 4 \times 600$). We find clear evidence of edge states as seen in Fig. \ref{fig:figS4}.
Note that a Chern number can be defined separately for each valley $l=1$ and $l=2$ (with spin degeneracy). Because the valley $U(1)$ symmetry guarantees that these two Chern numbers must be equal, the whole system is characterized by one Chern number that takes integer values, i.e. it belongs to a $\mathcal{Z}$ classification~\cite{YYHe2016}.

\begin{figure}[htp!]
\includegraphics[width=0.9\columnwidth]{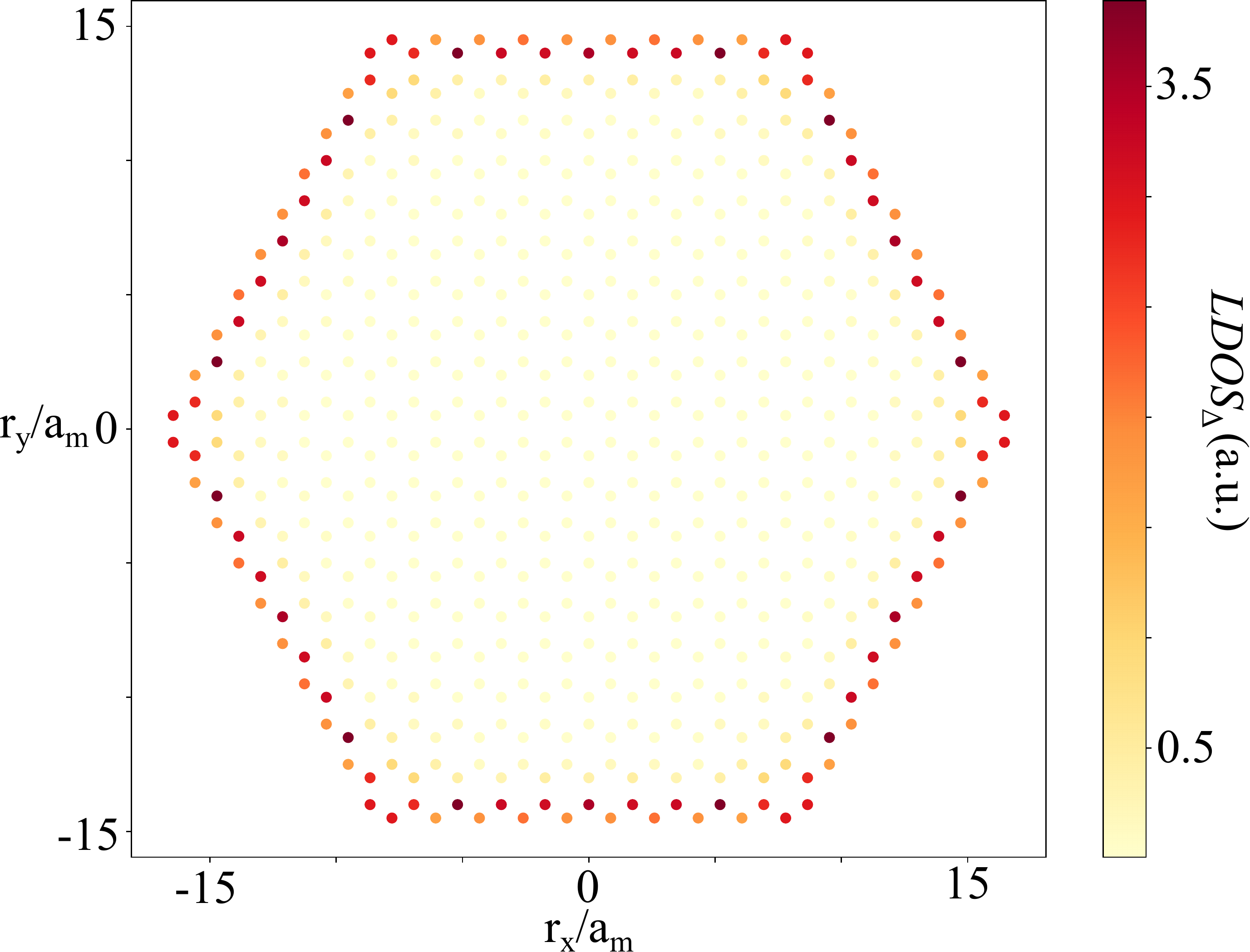}
\caption{In-gap local density of states in the QVH phase. Real space plot of the local density of states integrated over $1.66 < E/W < 2.00$ with $U/W = 0.5$. Here $r_{x,y}$ is the position of the lattice sites in units of the moiré lattice spacing, $a_m$. The result is computed with open boundary conditions and parameters as in Table \ref{tab:RSres1} from Appendix~\ref{app:hfm}.}
\label{fig:figS4}
\end{figure}

Fig.~\ref{fig:fig2}(c) also shows that, as $U$ increases ($U/W = 2$), the gapless edge mode disappears, signaling a departure from the topological QVH phase. Clearly, the bulk remains gapped, as shown in Fig.~\ref{fig:fig2}(a). The new insulating phase is an intervalley coherent (IVC) state, which spontaneously breaks the onsite spin-valley SU(4) symmetry. In the QMC simulations, IVC order is signalled by an enhancement of the correlation function $C_I(\mathbf{k})=\frac{1}{L^4}\sum_{i,j \in \mathcal{A}(\mathcal{B})}e^{i\vec{k}\cdot(\mathbf{r}_i-\mathbf{r}_j)}\left\langle I_{i}I_{j} \right\rangle$, here, the operator $I_{i}=\sum_{\sigma}( c^{\dagger}_{i,l,\sigma}c^{}_{i,l',\sigma}+h.c.)$, $ l \neq l' $, represents an ``onsite hopping" between the two different valleys. Thus, the correlation function is a $2\times 2$ matrix in sublattice space, i.e. $
\begin{pmatrix}
	C^{\mathcal{A}\mathcal{A}}_I & C^{\mathcal{A}\mathcal{B}}_I \\
	C^{\mathcal{B}\mathcal{A}}_I & C^{\mathcal{B}\mathcal{B}}_I \\
\end{pmatrix}
$, which has the relation $C^{\mathcal{A}\mathcal{A}}_I=C^{\mathcal{B}\mathcal{B}}_I=-C^{\mathcal{A}\mathcal{B}}_I=-C^{\mathcal{B}\mathcal{A}}_I$. In the upper panels of Figs.~\ref{fig:fig3}(a) and \ref{fig:fig3}(b), we show the diagonal component $C^{\mathcal{A}\mathcal{A}}_I(\mathbf{k})$. The fact that the correlation function is peaked at $\mathbf{k} = \boldsymbol{\Gamma}$ implies that the IVC order is ferromagnetic-like, i.e. it does not break translational symmetry. Such an onsite coupling between opposite valleys (see the corresponding inset in the phase diagram in Fig. \ref{fig:fig1} (b)) breaks the valley $U(1)$ symmetry, and hence the SU(4) symmetry of the model. The fact that the SU(4) symmetry-breaking pattern is ferromagnetic-like is similar to recent analytical results \cite{kang2019strong,seo2019ferromagnetic}, which focused, however, at integer fillings away from charge neutrality. We also note that our IVC state is different from that of Ref.~\cite{bultinck2019ground}, since our IVC phase does not have the edge modes protected by a modified Kramers time-reversal symmetry, as is the case of the IVC state proposed in Ref.~\cite{bultinck2019ground}.
 
For larger values of $U/W$, as shown in Fig.~\ref{fig:fig3}, the IVC order fades away, but the system remains insulating. The new state that emerges is the columnar valence-bond solid (cVBS) insulator, characterized by the appearance of strong nearest-neighbor bonds forming the pattern illustrated in the corresponding inset of Fig. \ref{fig:fig1} (b). The onset of cVBS order is signalled by an enhancement of the bond-bond correlation function~\cite{Lang2013,ZCZhou2016,xu2018kekule,YDLiao2019}, $C_B(\mathbf{k}) = \frac{1}{L^4}\sum_{i,j}e^{i\vec{k}\cdot(\mathbf{r}_i - \mathbf{r}_j)}\left\langle B_{i,\delta} B_{j,\delta} \right\rangle,$ 
with bond operator $B_{i,\delta}=\sum_{l,\sigma} (c_{i,l,\sigma}^\dagger c_{i+\delta,l,\sigma}+h.c.)$ and $\delta$ denoting one of the three nearest-neighbor bond directions of the honeycomb lattice ($\hat{e_1}$, $\hat{e}_2$ and $\hat{e}_3$). For this particular calculation, $\hat{e}_1$ was chosen. 

As shown in the lower panels of Figs.~\ref{fig:fig3}(a) and \ref{fig:fig3}(b), we find an enhanced $C_B(\mathbf{k})$ at momenta $\mathbf{K}$ and $\mathbf{K'}$, demonstrating that the bond-order pattern breaks translational symmetry. However, a peak of $C_B(\mathbf{k})$ at these momenta does not allow us to unambiguously identify the cVBS state, as the plaquette valence-bond solid (pVBS) also displays peaks at the same momenta~\cite{Lang2013,xu2018kekule,YDLiao2019}. To further distinguish the two types of VBS phases, we construct the complex order parameter $D_{\mathbf{K}} = \frac{1}{L^2}\sum_{i}\left( B_{i,\hat{e}_1}+\omega B_{i,\hat{e}_2} +\omega^2 B_{i,\hat{e}_3}\right) e^{i\mathbf{K}\cdot\mathbf{r}_i}$ with $\omega = e^{i\frac{2\pi}{3}}$. The Monte Carlo histogram of $D_{\mathbf{K}}$ is different for the two VBS phases~\cite{Lang2013,ZCZhou2016}: for the pVBS state, the angular distribution of $D_{\mathbf{K}}$ is peaked at ${\rm arg} (D_{\mathbf{K}}) = \frac{\pi}{3}, \pi, \frac{5\pi}{3}$, whereas for the cVBS state, it is peaked at ${\rm arg} (D_{\mathbf{K}}) = 0,\frac{2\pi}{3},\frac{4\pi}{3}$. Our results, shown in the inset of Fig.~\ref{fig:fig3}(a), clearly demonstrate that the cVBS order is realized in our phase diagram.

\begin{figure}[htp!]
\includegraphics[width=0.85\columnwidth]{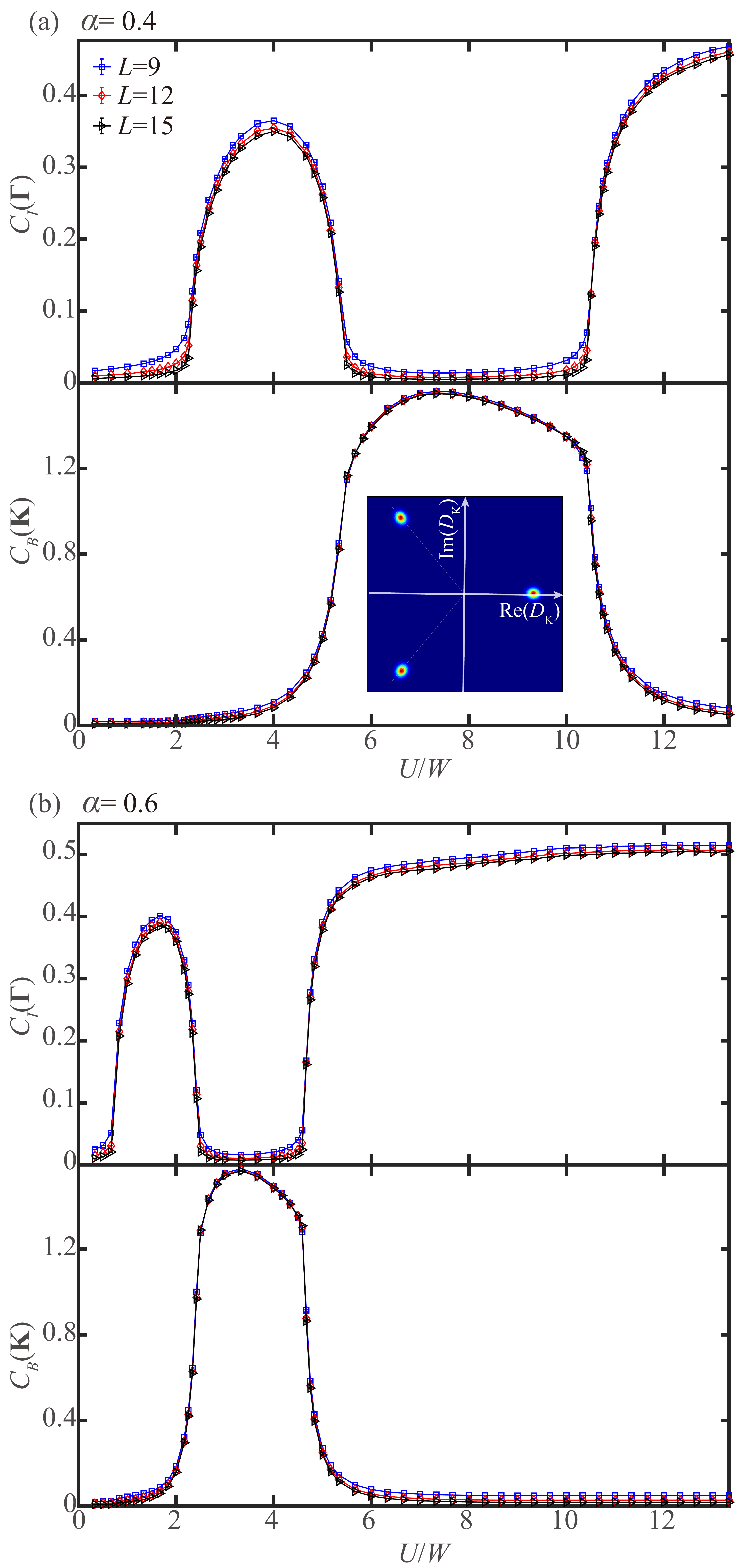}
\caption{Intervalley coherent (IVC) and columnar valence bond solid (cVBS) insulating states. Correlation functions $C_I(\boldsymbol{\Gamma})$ and $C_B(\mathbf{K})$, indicative of IVC and cVBS orders, respectively, as a function of $U/W$ for (a) $\alpha=0.4$ and (b) $\alpha=0.6$. Linear system sizes are indicated in the legend. In both panels, the QVH-IVC transition, the IVC-cVBS transition, and the cVBS-IVC transition are all first-order. The inset in panel (a) presents the histogram of the complex bond order parameter $D_{\mathbf{K}}$ at $U/W\sim5.3$. The positions of the three peaks are those expected for a cVBS phase, instead of a pVBS state.}
\label{fig:fig3}
\end{figure}

The phase boundaries in Fig.~\ref{fig:fig1}(b) are obtained by scanning the correlation functions $C_I(\boldsymbol{\Gamma})$ and $C_B(\mathbf{K})$ as a function of $U/W$ for fixed values of $\alpha$. Two of these scans are shown in Fig.~\ref{fig:fig3}, for $\alpha=0.4$ (panel (a)) and $\alpha=0.6$ (panel (b)). It is clear that, as $U/W$ increases, in both cases the ground state evolves from QVH to IVC to cVBS and then back to IVC. Furthermore, in the strong-coupling limit $U/W \rightarrow \infty$, the IVC order $C_I(\mathbf{k} = 0)$ is independent of $\alpha$ and saturates at $0.5$, consistent with our analytical calculation at the charge neutrality point, see Appendix~\ref{app:scl}. The transitions between IVC to cVBS are first order, as signaled by the fact that as the system size $L$ increases, the suppression of the IVC order becomes sharper (see for instance the region around $U/W\sim5$ and $U/W\sim11$ in panel (a)). A similar sharp drop is also featured at the QVH-IVC transition (region around $U/W\sim2.5$ in panel (a)), indicating that the QVH-IVC and IVC-cVBS transitions are all first-order. It is interesting to note that, as $\alpha$ increases, the values of $U/W$ for which the IVC and cVBS phases emerge are strongly reduced.

We note that in the limit of vanishing bandwidth our analyses in Appendices~\ref{app:pqmc} and ~\ref{app:scl} reveal that in this very strong-coupling limit the ground state of the system is in the IVC phase for the range of $\alpha$ considered here.

\section{Discussion}
In this paper, we employed QMC simulations, which are exact and unbiased, to obtain the phase diagram of a lattice model of TBG at charge neutrality. Our main result is that even very small interaction values trigger a transition from the non-interacting Dirac semi-metal phase to an insulating state. Upon increasing $U$, the nature of the insulator changes from a non-symmetry-breaking topological QVH phase, to an onsite SU(4) symmetry-breaking IVC state, to a translational symmetry-breaking cVBS phase, and then finally back to a reentrant IVC state. This rich phase diagram is a consequence of the interplay between two different types of interaction terms: a cluster-charge repulsion $Q_{\varhexagon}$ and a non-local assisted-hopping interaction $T_{\varhexagon}$. The former is analogous to the standard Hubbard repulsion and, as such, is expected to promote either SU(4) antiferromagnetic order or valence-bond order in the strong-coupling regime. The latter, on the other hand, arises from the topological properties of the flat bands in TBG. When combined with $Q_{\varhexagon}$, it gives rise not only to SU(4) ferromagnetic-like order, but also to correlated insulating phases with topological properties, such as the QVH phase. 

While the precise value of $U/W$ in TBG is not known, a widely used estimate is that this ratio is of order $1$ \cite{kang2019strong}. Referring to our phase diagram in Fig. \ref{fig:fig1}(b), this means that certainly the QVH phase and possibly the IVC phase can be realized at charge neutrality, provided that $\alpha$ is not too small. While some experimental probes do report a gap at charge neutrality Ref. \cite{lu2019superconductors,xie2019spectroscopic}, additional experiments are needed to establish its ubiquity among different devices and the nature of the insulating state. The main manifestation of the QVH phase would be the appearance of gapless edge states, whereas in the case of the IVC state, it would be the emergence of a $\mathbf{k}=0$ order with onsite coupling between the two different valleys.

A number of recent insightful Hartree-Fock studies have also reported several unusual ordered states at charge neutrality\cite{MacDonald2020,liu2019nematic,Guinea2020,bultinck2019ground,liu2019correlated}. In their approach, starting from the Bistritzer-MacDonald (BM) continuum wave functions, the Coulomb interactions are projected by use of the continuum model, and typically includes several remote bands. As usual, HF studies can depend crucially on the restrictions imposed in the search for ordered states. This may explain the rich variety of proposed spontaneously symmetry-broken phases identified from the continuum-approach, including a semi-metal phase, quantum Hall insulator, valley-Hall and spin- and valley-polarized phases.  Recently, Ref.~\cite{bultinck2019ground}, allowing for coherence between the two valleys, argued that the resulting insulating IVC phase is the ground state at charge neutrality for a broad parameter range. As discussed in the Introduction of this paper, we have presented a complementary approach; starting from the strongly-interacting limit we have applied the topologically nontrivial projected Coulomb interaction and utilized fully unrestricted and unbiased numerical methods able to handle cases where the scale of interactions exceeds the kinetic bandwidth, to identify the ordered states at charge neutrality. In qualitative agreement with some earlier studies, we locate an IVC phase from this strong-coupling approach, but additionally identify both the QVH and a translationally symmetry-breaking cVBS phase. While HF calculations with the Bloch states usually produce homogeneous phases without breaking the translation symmetry, more recent DMRG calculation with hybrid-WSs has identified the stripe phase as a strong candidate for a ground state in a toy BM model without spin and valley degrees of freedom~\cite{kang2020nonabelian,soejima2020efficient}. In agreement with the DMRG calculations, our QMC study found that the increasing kinetic terms drive the system from the IVC phase in the strong-coupling limit into the cVBS phase in a more intermediate coupling regime.
	


In a more general context beyond TBG, our work offers a promising route to realize correlation-driven topological phases. As explained above, the topological QVH insulating state appears due to the cross-term in the interaction Hamiltonian that contains both $Q_{\varhexagon}$ and $T_{\varhexagon}$. While repulsive interactions similar to the charge-cluster term are generally expected to appear in any correlated electronic system, an interesting question is about the necessary conditions for the emergence of an interaction similar to the assisted-hopping. In our case, it arises from the projection of the standard Coulomb repulsion on WSs that suffer from topological obstruction. The latter, in turn, is a manifestation of the phenomenon of fragile topology \cite{fragile_topology}. Thus, interacting systems with fragile topology may offer an appealing route to search for interaction-driven topological states. While here the Wannier obstruction arising from the fragile topology is circumvented by implementing the $C_2 \mathcal{T}$ symmetry of the continuous model non-locally, another route is to include the remote bands, separated from the narrow bands of TBG by a sizable gap. While we expect the ground state to be the same regardless of how the Wannier obstruction is avoided, it is an interesting open question to establish the strong-coupling phase diagram of TBG starting from a model containing both the narrow and remote bands.

\section*{Acknowledgements}
We thank Eslam Khalaf, Ashvin Vishwanath, and Yi Zhang for insightful conversations on the subject, especially on the nature of the IVC phase. We also thank Oskar Vafek for valuable suggestions and pointing out a missing factor in the IVC correlation function. YDL and ZYM acknowledge support from the National Key Research and Development Program of China (Grant No.~2016YFA0300502) and Research Grants Council of Hong Kong SAR China (Grant No.~17303019). HQW is supported by NSFC through Grant No. 11804401 and the Fundamental Research Funds for the Central Universities. JK acknowledges the support from the NSFC Grant No.~12074276, and Priority Academic Program Development (PAPD) of Jiangsu Higher Education Institutions. RMF is supported by the U. S. Department of Energy, Office of Science, Basic Energy Sciences, Materials Sciences and Engineering Division, under Award No. DE-SC0020045. YDL and ZYM thank the Center for Quantum Simulation Sciences in the Institute of Physics, Chinese Academy of Sciences, the Computational Initiative at the Faculty of Science and Information Technology Service at the University of Hong Kong, the Platform for Data-Driven Computational Materials Discovery at the Songshan Lake Materials Laboratory and the National Supercomputer Centers in Tianjin and Guangzhou for their technical support and generous allocation of CPU time. JK thanks the Kavli Institute for Theoretical Sciences for hospitality during the completion of this work. ZYM, JK, and RMF thank the hospitality of the Aspen Center for Physics, where part of this work was developed. The Aspen Center for Physics is supported by National Science Foundation grant PHY-1607611.


\appendix

\section{Projection QMC method}
\label{app:pqmc}
\subsection{Construction}
Since we are interested in the ground state properties of the system, the projection QMC (PQMC) is the method of choice~\cite{Assaad2008,meng2010quantum,YZLiu2020}.  In PQMC, one can obtain a ground state wave function $\vert \Psi_0 \rangle$ from projecting a trial wave function $\vert \Psi_T \rangle$  along the imaginary axis $\vert \Psi_0 \rangle = \lim\limits_{\Theta \to \infty} e^{-\frac{\Theta}{2} \mathbf{H}} \vert \Psi_T \rangle$, then observable can be calculated as
\begin{equation}
\label{eq:observablepqmc}
\langle \hat{O} \rangle = \frac{\langle \Psi_0 \vert \hat{O} \vert \Psi_0 \rangle}{\langle \Psi_0 \vert \Psi_0 \rangle} 
						= \lim\limits_{\Theta \to \infty} \frac{\langle \Psi_T \vert  e^{-\frac{\Theta}{2} \mathbf{H}} \hat{O}  e^{-\frac{\Theta}{2} \mathbf{H}} \vert \Psi_T \rangle}{\langle \Psi_T \vert  e^{-\Theta \mathbf{H}} \vert \Psi_T \rangle} .
\end{equation}
To evaluate overlaps in the above equation, we performed Trotter decomposition to discretize $\Theta$ into $L_\tau$ slices ($\Theta=L_\tau \Delta\tau$). Each slices $\Delta\tau$ is small and the systematic error is $\mathcal{O}(\Delta\tau^2)$. After the Trotter decomposition, we have
\begin{equation}
\langle\Psi_{T}|e^{-\Theta H}|\Psi_{T}\rangle=\langle\Psi_{T}|\left(e^{-\Delta\tau H_{U}}e^{-\Delta\tau H_{0}}\right)^{L_\tau}|\Psi_{T}\rangle+\mathcal{O}(\Delta{\tau}^{2})
\end{equation}
where the non-interacting and interacting parts of the Hamiltonian is separated. To treat the interacting part, one usually employ a Hubbard Stratonovich (HS) transformation to decouple the interacting quartic fermion term to fermion bilinears coupled to auxiliary fields. 

For the cluster interaction in Eq.~\eqref{eq:eq2} of the main text, we make use of a fourth order $SU(2)$ symmetric decoupling
\begin{equation}
e^{-\Delta\tau U(Q_{\varhexagon}+\alpha T_{\varhexagon}-4)^{2}}=\frac{1}{4}\sum_{\{s_{\varhexagon}\}}\gamma(s_{\varhexagon})e^{\lambda\eta(s_{\varhexagon})\left(Q_{\varhexagon}+\alpha T_{\varhexagon}-4\right)}
\label{eq:decompo}
\end{equation}
with $\lambda=\sqrt{-\Delta\tau U}$, $\gamma(\pm1)=1+\sqrt{6}/3$,
$\gamma(\pm2)=1-\sqrt{6}/3$, $\eta(\pm1)=\pm\sqrt{2(3-\sqrt{6})}$,
$\eta(\pm2)=\pm\sqrt{2(3+\sqrt{6})}$ and the sum is taken over the auxiliary fields $s_{\varhexagon}$ on each hexagon which can take four values $\pm2$ and $\pm1$. After tracing out the free fermionic degrees of freedom, we obtain the following formula with a constant factor omitted
\begin{widetext}
   \begin{eqnarray}
\langle\Psi_{T}|e^{-\Theta H}|\Psi_{T}\rangle=\sum_{\{s_{\varhexagon,\tau}\}}\left[\left(\prod_{\tau}\prod_{\varhexagon}\gamma(s_{\varhexagon,\tau})e^{-4\lambda\eta(s_{\varhexagon,\tau})}\right)\det\left[P^{\dagger}B(\Theta,0)P\right]\right]
\label{eq:mcweight}
   \end{eqnarray}
\end{widetext}
where $P$ is the coefficient matrix of trial wave function $|\Psi_T\rangle$. In the simulation, we make use of the real space ground state wavefunction of the tight-binding Hamiltonian $H_0$ as the trial wave function $|\Psi_T\rangle$. In the above formula, the $B$ matrix is defined as
\begin{equation}
B(\tau+1,\tau)=\sum_{\{s_{\varhexagon,\tau}\}}e^{\lambda\eta(s_{\varhexagon,\tau})V}\cdot e^{-\Delta_\tau K}
\end{equation}
and has properties $B(\tau_3,\tau_1)=B(\tau_3,\tau_2)B(\tau_2,\tau_1)$, i.e. the $B$ matrix is an imaginary time propagator, where we have written the coefficient matrix of interaction part as $V$ and $K$ is the hopping matrix from the $H_0$.

Every hexagon contains six sites, as shown in the figure below, so our $V$ matrix is a block matrix, every block contributes a $6\times 6$ matrix,

\begin{figure}[htb] 
  \begin{minipage}[h]{0.35\columnwidth} 
    \centering 
    \includegraphics[width=0.7\columnwidth]{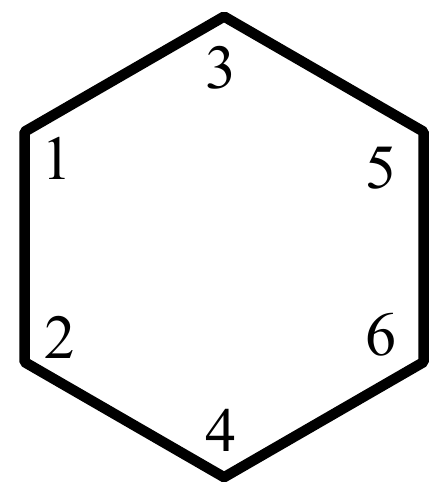} 
    \label{fig:66} 
  \end{minipage}%
  \hfill
  \begin{minipage}[h]{0.65\columnwidth} 
    \centering
\[\label{eq:66change}
  \begin{pmatrix}
   \frac{1}{3} & i\alpha    & -i\alpha   &  0           & 0           & 0            \\
   -i\alpha    & \frac{1}{3} & 0           &  i\alpha   & 0           & 0            \\
   i\alpha   & 0           & \frac{1}{3} &  0           & -i\alpha    & 0            \\
   0           & -i\alpha   & 0           &  \frac{1}{3} & 0           & i\alpha     \\
   0           & 0           & i\alpha    &  0           & \frac{1}{3} & -i\alpha    \\
   0           & 0           & 0           &  -i\alpha    & i\alpha   &  \frac{1}{3} 
  \end{pmatrix} \]
  \end{minipage} 
\end{figure}
The configurational space $\{s_{\varhexagon}(i,\tau)\}$ with size $L\times L\times \Theta$ is the space in which the physical observables in Eq.~\eqref{eq:observablepqmc} are computed with ensemble average. We choose the projection length $\Theta=2L/t$ and discretize it with a step $\Delta\tau=0.1/t$. The spatial system sizes are $L=6,9,12,15$. 

The Monte Carlo sampling of auxiliary fields are further performed based on the weight defined in the sum of  Eq.~\eqref{eq:mcweight}. The measurements are performed near $\tau=\Theta/2$. Single particle observables are measured by Green's function directly and many body correlation functions are measured from the products of single-particle Green's function based on their corresponding form after Wick-decomposition. The equal time Green's function are calculated as
\begin{equation}
G(\tau,\tau)=1-R(\tau)\left(L(\tau)R(\tau)\right)^{-1}L(\tau)
\end{equation}
with $R(\tau)=B(\tau,0)P$, $L(\tau)=P^{\dagger}B(\Theta,\tau)$. \\

\subsection{Absence of sign-problem}
At the charge neutrality point, the model is sign-problem-free, as can be seen from the following analysis. Define $W_{\sigma,l,S_i}$ as the update weight of one fixed auxiliary field at the $i$-th hexagon, where $l=1,2$ is a valley/orbital index and $\sigma=\uparrow ,\downarrow $ is a spin index. From the symmetry of the Hamiltonian, $W_{\uparrow , l,S_i}=W_{\downarrow, l, S_i}$. Since the model is particle-hole symmetric at charge neutrality, one can perform a particle-hole transformation (PHS) only for the valley $l=2$. Then one can focus on a fixed auxiliary field, and focus only on one spin flavor, such as spin up. Eq.~\eqref{eq:decompo} in the main text can then be abbreviated as Eq.~\eqref{eq:1} and Eq.~\eqref{eq:2}. Applying PHS for valley 2 and using the relation Eq.~\eqref{eq:relation}, we find that Eq. (\ref{eq:2}) becomes Eq.~\eqref{eq:3}.
\begin{widetext}
   \begin{eqnarray}
\label{eq:1}
For\  l=1,\  \exp\left( i\alpha\eta(S_i)\left[ \sum_{p=1}^6 \left( ic_{p+1,1,\uparrow}^\dagger c_{p,1,\uparrow} - ic_{p,1,\uparrow}^\dagger c_{p+1,1,\uparrow} \right)+\frac{1}{3}\sum_{p=1}^6 \left( c_{p,1,\uparrow}^\dagger c_{p,1,\uparrow} - \frac{1}{2}\right) \right] \right)
   \end{eqnarray}
   \begin{eqnarray}
\label{eq:2}
For\  l=2,\  \exp\left( i\alpha\eta(S_i)\left[ \sum_{p=1}^6 \left( -ic_{p+1,2,\uparrow}^\dagger c_{p,2,\uparrow} + ic_{p,2,\uparrow}^\dagger c_{p+1,2,\uparrow} \right)+\frac{1}{3}\sum_{p=1}^6 \left( c_{p,2,\uparrow}^\dagger c_{p,2,\uparrow} - \frac{1}{2}\right) \right] \right)
   \end{eqnarray}
   \begin{eqnarray}
\label{eq:relation}
\begin{aligned}
-ic_{p+1,2,\uparrow}^\dagger c_{p,2,\uparrow} + ic_{p,2,\uparrow}^\dagger c_{p+1,2,\uparrow} &\stackrel{PHS}{\longrightarrow} ic_{p+1,2,\uparrow}^\dagger c_{p,2,\uparrow} - ic_{p,2,\uparrow}^\dagger c_{p+1,2,\uparrow} \\
c_{p,2,\uparrow}^\dagger c_{p,2,\uparrow} - \frac{1}{2} &\stackrel{PHS}{\longrightarrow} \frac{1}{2} - c_{p,2,\uparrow}^\dagger c_{p,2,\uparrow}
\end{aligned}
   \end{eqnarray}
   \begin{eqnarray}
\label{eq:3}
For\  l=2\ PHS,\  \exp\left( i\alpha\eta(S_i)\left[ \sum_{p=1}^6 \left( ic_{p+1,2,\uparrow}^\dagger c_{p,2,\uparrow} - ic_{p,2,\uparrow}^\dagger c_{p+1,2,\uparrow} \right)+\frac{1}{3}\sum_{p=1}^6 \left( \frac{1}{2} - c_{p,2,\uparrow}^\dagger c_{p,2,\uparrow} \right) \right] \right)
   \end{eqnarray}
      \begin{eqnarray}
\label{eq:ab}
\begin{aligned}
For\ l=1,\  e^{-\alpha\eta(S_i)\mathbf{B}+i\alpha\eta(s_{\varhexagon})\mathbf{A}} &\to e^{-\alpha\eta(s_{\varhexagon})\mathbf{B}}e^{i\alpha\eta(s_{\varhexagon})\mathbf{A}} \\
For\  l=2\ PHS,\  e^{-\alpha\eta(S_i)\mathbf{B}+i\alpha\eta(s_{\varhexagon})\mathbf{A}} &\to e^{-\alpha\eta(s_{\varhexagon})\mathbf{B}}e^{-i\alpha\eta(s_{\varhexagon})\mathbf{A}} 
\end{aligned}
   \end{eqnarray}
\end{widetext}

Let us define the matrices $\mathbf{A} = \frac{1}{3}\sum_{p=1}^6 \left( c_{p,2,\uparrow}^\dagger c_{p,2,\uparrow} - \frac{1}{2}\right) $ and $i\mathbf{B} = \sum_{p=1}^6 \left( ic_{p+1,1,\uparrow}^\dagger c_{p,1,\uparrow} - ic_{p,1,\uparrow}^\dagger c_{p+1,1,\uparrow} \right)$ -- or, equivalenty,  $\mathbf{B} = \sum_{p=1}^6 \left( c_{p+1,1,\uparrow}^\dagger c_{p,1,\uparrow} - c_{p,1,\uparrow}^\dagger c_{p+1,1,\uparrow} \right)$. The matrices $\mathbf{A}$ and $\mathbf{B}$ are real matrices. Then, due to the fact that the matrix $\mathbf{A}$ is a diagonal matrix and $\mathbf{A}_{ii}=\mathbf{A}_{jj}$, Eqs. (\ref{eq:1}) and (\ref{eq:3}) can be writen as Eq.~\eqref{eq:ab}.

Because of the relations above, the total weight of the model is $\sum_{S_i} W_{\uparrow,1,S_i}*W_{\uparrow,2,S_i}*W_{\downarrow,1,S_i}*W_{\downarrow,1,S_i} = \sum_{S_i} \left( W_{\uparrow,1,S_i}W_{\uparrow,1,S_i}^* \right)^2 $, which is a real positive number. This implies that the QMC simulations are sign-problem-free.

\subsection{Strong-coupling limit}
The PQMC simulations can also be applied at the strong-coupling limit, where the Hamiltonian only
contains the interaction part $H_{\varhexagon}$. We have performed the corresponding simulations and found as a function of $\alpha$ that the system is always in the IVC phase, since the corresponding correlation function $C_{I}(\Gamma)$ is close to the saturation value of 0.5. At the same time, the correlation function of the cVBS phase, $C_{B}(\mathbf{K})$ approaches zero as the system size increases. The results are shown in Fig.~\ref{fig:H0}. This is consistent with the theoretical analysis in this limit discussed in Appendix.~\ref{app:scl}.

\begin{figure}[h]
\centering
\includegraphics[width=\columnwidth]{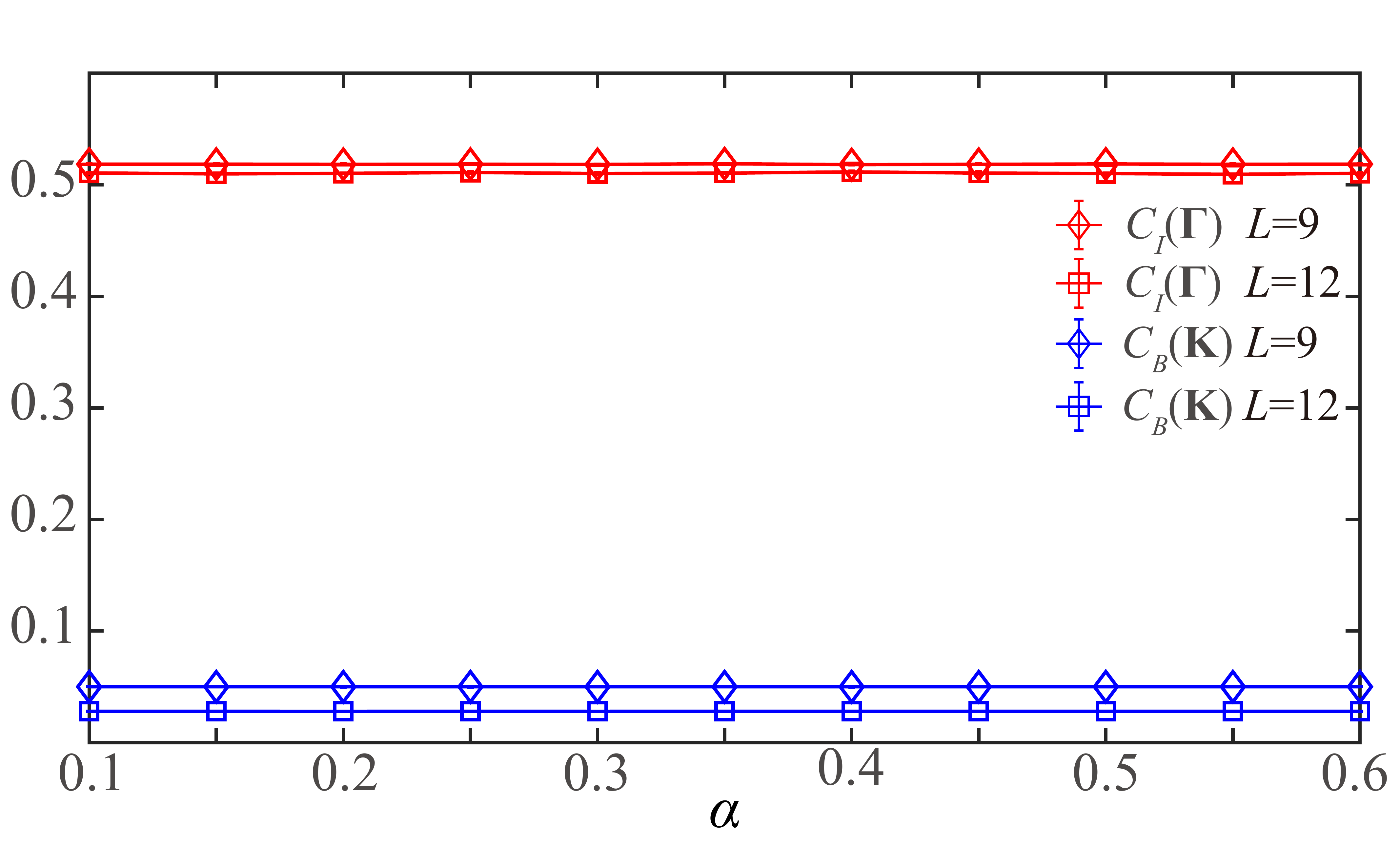}
\caption{$C_{I}(\Gamma)$ and $C_B(\mathbf{K})$ as a function of $\alpha$ at the strong-coupling limit. We perform the QMC simulation with projection length $\Theta=200L$, interval of time-slice $\Delta\tau=0.1$, spatial system sizes $L=9,12$, and setting $U=1$ as a dimensionless constant. The corresponding correlation function $C_{I}(\Gamma)$ of IVC is close to the saturation value of 0.5. And the correlation function $C_B(\mathbf{K})$ of cVBS is close to 0, which means cVBS disappear at the strong-coupling limit.}
\label{fig:H0}
\end{figure}

\section{Benchmark with exact diagonalization}
\label{app:ed}
We employ Lanczos exact diagonalization (ED) to benchmark the PQMC results, shown in Fig.~\ref{fig:ed}. The system contains $2\times 2$ unit cells of the honeycomb lattice with periodic boundary condition (16 electrons in total). We make use of symmetries, such as the valley $U(1)$ symmetry and the total $S_z$ conservation for each valley, to reduce the computational cost of the ED. The ground state lays in the subspace with $N_{\uparrow}=4$, $N_{\downarrow}=4$ in both valleys, where $N_{\uparrow}(N_{\downarrow})$ is the number of electrons with spin up (down) in each valley. The dimension of the ground-state subspace is about 24 million. In the PQMC simulations, we choose the linear system size $L=2$ and the projection length $\Theta=100/t$ with Trotter slice $\Delta \tau=0.0005/t$. We compared the ground-state expectation values of $\langle H_0 \rangle$ and of the double occupation as a function of $U/t$ at $\alpha=0.3$, which are shown below. The results of both methods agree very well.

\begin{figure}[H]
    \centering
    \includegraphics[width=\columnwidth]{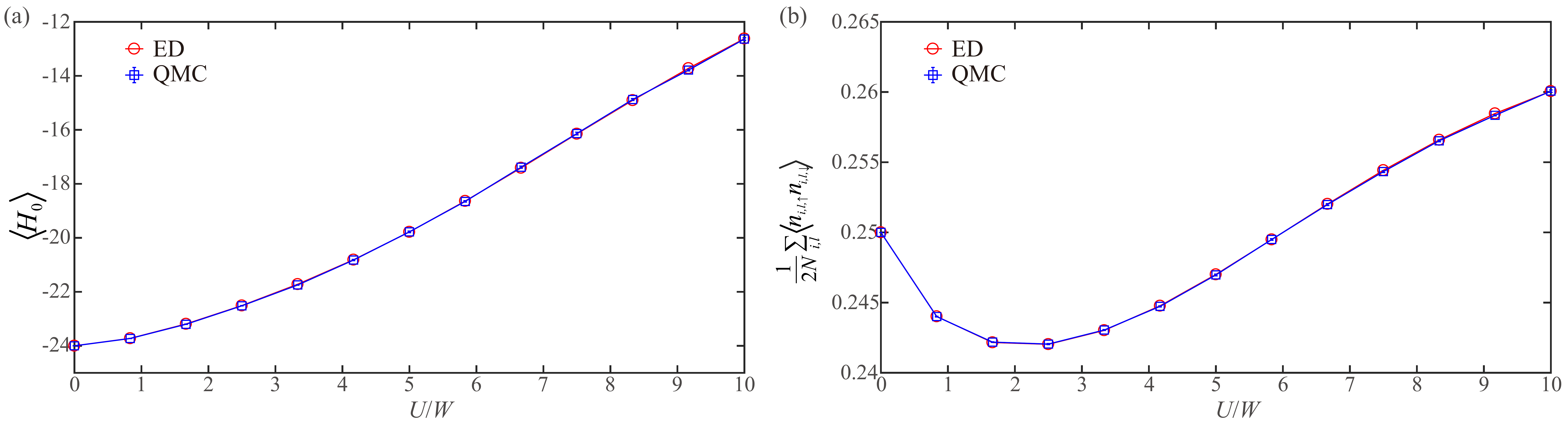}
    \caption{(a) Kinetic energy and (b) Double occupancy as a function of $U/W$ for $\alpha=0.3$. Red circles and blue squares with error bars are obtained from ED and QMC, respectively.}
    \label{fig:ed}
\end{figure}

\section{Hartree-Fock method}
\label{app:hfm}
To solve the TBG model within the Hartree-Fock approach, we write the Hamiltonian in Eq. (\ref{eq:eq2}) of the main text as
\begin{align}
H_{\varhexagon} = U\sum_{\varhexagon}(Q'_{\varhexagon}+\alpha T'_{\varhexagon})(Q_{\varhexagon}+\alpha T_{\varhexagon}),
\end{align}
where primes indicate independent index summations. The direct terms immediately give
\begin{align}
H^{H}_{\varhexagon} = 2 U \sum_{\varhexagon} \big\langle Q'_{\varhexagon}+\alpha T'_{\varhexagon} \big\rangle (Q_{\varhexagon}+\alpha T_{\varhexagon}). \label{eq:Hartree}
\end{align}
The exchange terms are Eq.~\eqref{eq:Fock}, where $\sum_{\mathrm{all}} = \Big(\sum_{l, l' = 1,2}\sum_{\sigma\sigma'}\sum^{6}_{i, i'=1}\Big)$. Manipulating and collecting terms in (\ref{eq:Hartree}) and (\ref{eq:Fock}) yields the Hartree-Fock Hamiltonian Eq.~\eqref{eq:hfhamiltonian}.
\begin{widetext}
   \begin{eqnarray}
\begin{aligned}
H^{F}_{\varhexagon} = &- U \sum_{\varhexagon}\sum_{\mathrm{all}} \Big[ \frac{1}{9}[\langle c^{\dagger}_{ i' l'\sigma'} c_{i l \sigma} \rangle c^{\dagger}_{ i l \sigma} c_{i' l' \sigma'} + h.c. ]\\
&+ \frac{\alpha i }{3}\big\{ [ (-)^{l'+1}\langle c^{\dagger}_{ i' + 1 l'\sigma'} c_{i l \sigma} \rangle c^{\dagger}_{ i l \sigma} c_{i' l' \sigma'} + h.c.] +  [ (-)^{l'}\langle c^{\dagger}_{ i' l'\sigma'} c_{i l \sigma} \rangle c^{\dagger}_{ i l \sigma} c_{i'+1 l' \sigma'} + h.c.] \big\} \nonumber \\
&+  \alpha^2 i^2 \big\{ [(-)^{l'+l}\langle c^{\dagger}_{ i' + 1 l'\sigma'} c_{i l \sigma} \rangle c^{\dagger}_{ i + 1 l \sigma} c_{i' l' \sigma'}+h.c.] - [(-1)^{l'+l} \langle c^{\dagger}_{ i' + 1 l'\sigma'} c_{i+1 l \sigma} \rangle c^{\dagger}_{ i l \sigma} c_{i' l' \sigma'} +h.c.] \nonumber \\
&- [(-1)^{l'+l} \langle c^{\dagger}_{ i' l'\sigma'} c_{i l \sigma} \rangle c^{\dagger}_{ i +1 l \sigma} c_{i' +1  l' \sigma'} +h.c.] + [(-)^{l'+l}\langle c^{\dagger}_{ i' l'\sigma'} c_{i +1 l \sigma} \rangle c^{\dagger}_{ i l \sigma} c_{i'+1 l' \sigma'}+h.c.]   \big\} \Big], \label{eq:Fock}
\end{aligned}
   \end{eqnarray}
   \begin{eqnarray}
   \label{eq:hfhamiltonian}
\begin{aligned}
H^{HF}_{\varhexagon} &= 2 U  \sum_{\varhexagon} \Big\{\bar{n}_{\varhexagon} (Q_{\varhexagon} + \alpha T_{\varhexagon}) -  \sum_{\mathrm{all}} \Big[\sum_{n,m} \alpha_n(l')\alpha_m(l+1) \langle c^{\dagger}_{ i'+n l'\sigma'} c_{i+m l \sigma} \rangle \Big]c^{\dagger}_{i l \sigma} c_{i' l' \sigma'}\Big\}.
\end{aligned}
   \end{eqnarray}
\end{widetext}

Here $n,m = \{-1,0,1\}$ and we have defined
\begin{align*}
\bar{n}_{\varhexagon} &=  \langle Q_{\varhexagon} + \alpha T_{\varhexagon} \rangle, \\
\vec{\alpha}(l) &= \begin{pmatrix} \alpha_{-1}\\ \alpha_0 \\ \alpha_1 \end{pmatrix} =  \begin{pmatrix} (-1)^l \, i \, \alpha \\ 1/3 \\ (-1)^{l+1} \, i \, \alpha\end{pmatrix}.
\end{align*}
We solve the full Hartree-Fock Hamiltonian ($H = H_0 + H^{HF}_{\varhexagon}$) self-consistently using that
\begin{align}
 \langle c^{\dagger}_{\kappa} c_{\lambda} \rangle = \sum^N_{\epsilon, \eta = 1} U^{\dagger}_{\epsilon \kappa} U_{\lambda \eta} \langle \gamma^{\dagger}_{\epsilon} \gamma_{\eta} \rangle =  \sum_{\epsilon} U^{\dagger}_{\epsilon \kappa} U_{\lambda \epsilon} f(E_{\epsilon}, \mu),
\end{align}
where $\kappa,\lambda = \{il\sigma\}$, $U$ is the unitary tranformation diagonalizing $H$, $\gamma$'s are the eigenvectors and $f(E_{\epsilon},\mu)$ is the Fermi-Dirac distribution of the excitation energies, $E_{\epsilon}$. We explicitly write the dependence on the chemical potential, $\mu$, as we iterate this value to fulfil $N^{-1}\sum_{\epsilon} f(E_{\epsilon},\mu) = \nu$, where $\nu$ is the filling.

We compute results at charge neutrality ($\nu = 0.5$) with a total of 600 lattice sites and periodic boundary conditions. The calculations are fully unrestricted; thus we iterate all $(4\times600)^2$ mean-fields, and define convergence by the condition that $\sum |\Delta E_{\epsilon}| < N\times10^{-10}$, where $\Delta E_{\epsilon}$ is the change of the excitation energies from one iteration to the next, and $N$ is the total number of states ($4\times600$).
In Table \ref{tab:RSres1} we present an example of the HF calculations, displaying results for $t=1$ ,$\alpha = 0.45$ and $U/W = 0.5$. We set the temperature $T = 2.5\cdot10^{-5}$ in all computations. The values in the table are the renormalized mean-fields. It is evident that all hoppings within each hexagon are renormalized due to the interactions. The simple hopping renormalizations, however, do not open a gap in the Dirac cones. The gap is generated directly by the mean-fields $\langle c^{\dagger}_{i, 1, \sigma} c_{i\pm2, 1, \sigma} \rangle = -\langle c^{\dagger}_{i, 2, \sigma} c_{i\pm2, 2, \sigma} \rangle =  \pm 0.0914 i$, which explicitly display a spin degenerate quantum valley Hall (QVH) phase, as illustrated in Fig. \ref{fig:fig1}(b) of the main text.

In Fig. \ref{fig:figS3}(a) we show the single-particle gap in the QVH phase with $\alpha = 0.45$ for several interaction strengths. Fig. \ref{fig:figS3}(b) displays the corresponding band structures. The Dirac cone at $K$ in the bare bands is immediately gapped out when including interactions. The renormalization initially flattens the bands with a significant gap at all high-symmetry points. As $U$ increases the valence band gradually develops a peak at $\Gamma$ while it is pushed down correspondingly at $K$. This behavior results in a gradual shift of the maximal gap value from $\Gamma$ to $K$.

\begin{figure}[htp!]
\includegraphics[width=\columnwidth]{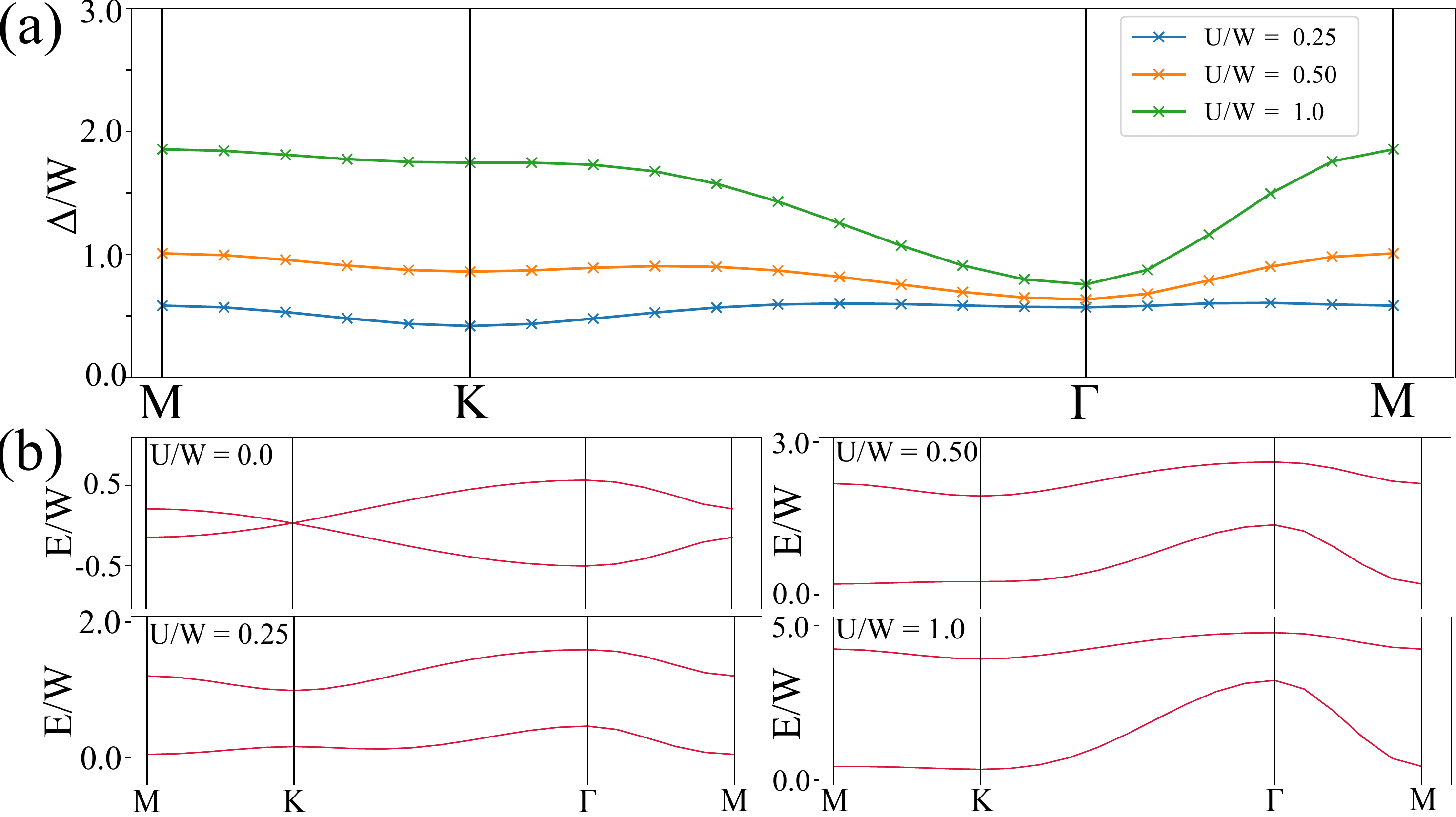}
\caption{QVH insulator and band strutures (a) The single-particle gap $\Delta/W$ along the high-symmetry path of the BZ with $\alpha = 0.45$ and various interaction strengths. The Dirac cones are gapped at infinitely small $U$ and the system enters the QVH state. The maximal gap value gradually shifts from $\Gamma$ to the $K$. (b) Band structures at various interaction strengths with $\alpha = 0.45$. Top left plot displays the bare kinetic bands in the absence of any interactions. The Dirac cone at $K$ is evident and confirms the semi-metallic phase of the bare bands. The remaining three plots of (b) present the renormalized band structures with increasing $U$. The two bands flattens for small $U$ and gradually develops a peak at $\Gamma$. All bands are fourfold degenerate, as the QVH phase does not break the approximate $SU(4)$ symmetry.}
\label{fig:figS3}
\end{figure}

Furthermore, while QMC cannot handle a longer-range tight-binding model due to the sign-problem, the same is not true for our unrestricted Hartree-Fock calculation. We thus have computed a check with the tight-binding model suggested in Refs.~\cite{koshino2018maximally,yuan2018model} including complex fifth nearest-neighbor hopping ($t_2/t = 0.025 \pm 0.1 i$), which breaks particle-hole symmetry and introduces a splitting along the $\Gamma M$-line. We find a complete, quantitative agreement with the renormalized mean-field results presented in Table.~\ref{tab:RSres1}. Thus, in the weak-coupling regime, the QVH phase is very robust to the addition of long-range hoppings in $H_0$. The resulting bands with and without long-range hopping can be seen in Fig. \ref{fig:figS6}.

\begin{figure}[htp!]
\includegraphics[width=\columnwidth]{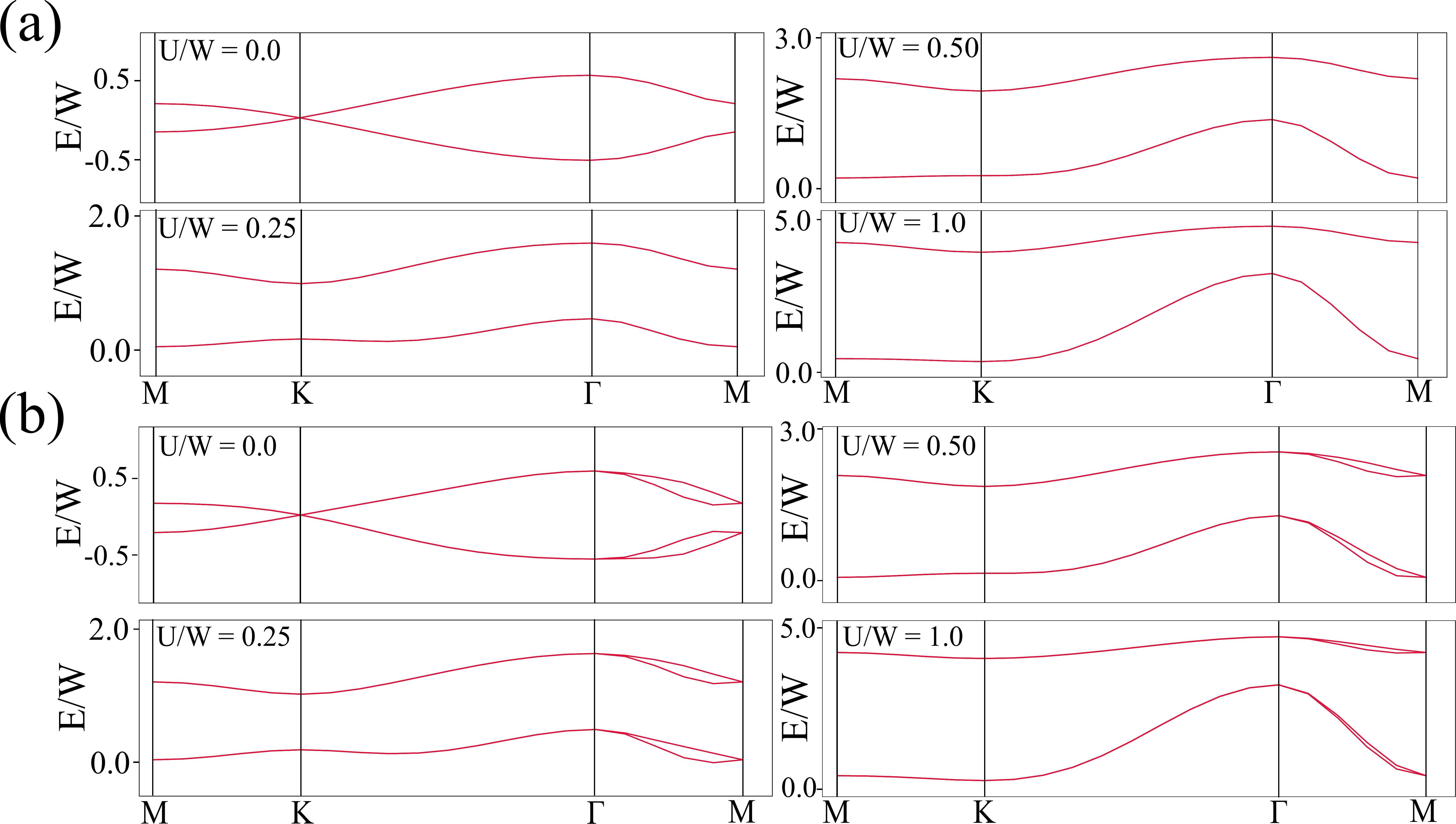}
\caption{\textbf{QVH insulator and band strutures with longer hoppings.} Band structures at various interaction strengths with $\alpha = 0.45$ and fifth nearest-neighbor hopping \textbf{(a)}$ t_2/t = 0$ and  \textbf{(b)} $t_2/t = 0.025 \pm 0.1 i$. Top left plot in \textbf{(a,b)} display the bare kinetic bands in the absence of any interactions. The Dirac cones at $K$ are evident and confirms the semi-metallic phase of the bare bands in both cases. The remaining six plots present the renormalized band structures with increasing $U$. It is evident that the only effect of including the fifth nearest-neighbor hopping is an emergent splitting along $\Gamma-M$. This splitting decreases with increasing $U/W$.}
\label{fig:figS6}
\end{figure}

Finally we present results obtained by implementing the interaction terms found in Ref. ~\cite{kang2019strong}. As mentioned in the discussion section of the main text, we are able to solve this model within the HF approach as it does not suffer from sign problems. The assisted hopping reads
\begin{align}
T_{\varhexagon} = \sum^6_{i=1} \sum_{l,\sigma} (-)^{i-1} \big( c^{\dagger}_{i l \sigma} c_{i+1 l \sigma} + \rm{h.c.}\big).
\end{align}
The other terms, $Q_{\varhexagon}$ and $H_0$, remain unchanged. To reach this expression for $T_{\varhexagon}$, we have performed the following gauge transformation,
\begin{align}
&c_{i l \sigma} \longrightarrow e^{i\theta_l/2}c_{i l \sigma}, \hspace{0.5cm} i\, \mathrm{odd} \nonumber\\
&c_{i l \sigma} \longrightarrow e^{-i\theta_l/2}c_{i l \sigma}. \hspace{0.3cm} i\, \mathrm{even} \nonumber\\
\end{align}
The transformation introduces phases in $H_0$ effectively causing $t$ to become complex. We set the phases according to Ref. ~\cite{kang2019strong}, that is $\theta_1 = - \theta_2 =  0.743 \pi$. The renormalized mean-fields are presented in Table \ref{tab:RSres2}, where we have performed the inverse gauge transformation for direct comparison with Table \ref{tab:RSres1}. Input parameters are the same as those used to generate Table \ref{tab:RSres1}.
The result is consistent with the values presented in Table \ref{tab:RSres1} and clearly also features a QVH phase.

\onecolumngrid
\ 
\begin{table}[H]
\centering
\small
\begin{tabular}{|c|c|c|c|c|c|c|c|c|}
\hline
 &\multicolumn{8}{c|}{U/W = 0.50} \\  \hline
                        & $ c_{i,1,\sigma}$ & $ c_{i,2,\sigma}$ & $ c_{i \pm1,1,\sigma} $ & $ c_{i\pm1,2,\sigma}  $ & $ c_{i \pm2,1,\sigma} $ & $ c_{i\pm2,2,\sigma}  $ & $ c_{i +3,1,\sigma} $ & $ c_{i+3,2,\sigma}  $ \\ \hline
$c^{\dagger}_{i,1,\sigma}$ &  - 	                          &   -                    &  $- 0.0209$                    &  	-                  & $\pm  0.0914 i $ &           -               &  $0.0024 $ &       -    \\
$c^{\dagger}_{i,2,\sigma}$ & -                                 &   -                &  -                                    &$- 0.0209 $           &            -             & $\mp  0.0914  i$  &       -           & $0.0024$  \\  \hline
\end{tabular}
\caption{Mean-field renormalization with $U/W = 0.50$ and $\alpha = 0.45$. The input on $[c^{\dagger}_{il\sigma}, c_{jl'\sigma}]$ represents the renormalized mean-field parameter $\langle c^{\dagger}_{il\sigma} c_{jl'\sigma} \rangle$. The result is homogeneous and spin degenerate, hence the listed values contain information about all sites and flavours. Note that interactions have generated neither spin- nor valley-mixing. We have subtracted the bare band contributions evaluated at $\nu = 0.5$ and ignored all mean-fields with $\frac{|MF|_{max}}{|MF|} > 100$.}
\label{tab:RSres1}
\end{table}

\begin{table}[H]
\centering
\small
\begin{tabular}{|c|c|c|c|c|c|c|c|c|}
\hline
 &\multicolumn{8}{c|}{U/W = 0.50} \\  \hline
                        & $ c_{i,1,\sigma}$ & $ c_{i,2,\sigma}$ & $ c_{i \pm1,1,\sigma} $ & $ c_{i\pm1,2,\sigma}  $ & $ c_{i \pm2,1,\sigma} $ & $ c_{i\pm2,2,\sigma}  $ & $ c_{i +3,1,\sigma} $ & $ c_{i+3,2,\sigma}  $ \\ \hline
$c^{\dagger}_{i,1,\sigma}$ &  - 	                          &   -                    &  $- 0.1015 - (-)^i 0.0555 i$                    &  	-                  & $\mp  0.0945 i $ &           -               &  $- 0.0806 + (-)^i 0.0300 i$  &       -    \\
$c^{\dagger}_{i,2,\sigma}$ & -                                 &   -                &  -                                    &$- 0.1015 + (-)^i 0.0555 i$           &            -             & $\pm  0.0945  i$  &       -           & $- 0.0806 - (-)^i 0.0300 i$   \\  \hline
\end{tabular}
\caption{Mean-field renormalization with $U/W = 0.50$ and $\alpha = 0.45$ using model from Ref. ~\cite{kang2019strong}. The input on $[c^{\dagger}_{il\sigma}, c_{jl'\sigma}]$ represents the renormalized mean-field parameter $\langle c^{\dagger}_{il\sigma} c_{jl'\sigma} \rangle$. The result is homogeneous and spin degenerate, hence the listed values contain information about all sites and flavours. We have subtracted the bare band contributions evaluated at $\nu = 0.5$ and ignored all mean-fields with $\frac{|MF|_{max}}{|MF|} > 100$.}
\label{tab:RSres2}
\end{table}
\twocolumngrid

\section{Strong-coupling limit at the charge neutrality point}
\label{app:scl}
For the system at the charge neutrality point, each unit cell contains four fermions in average. Following the method applied in Ref.~\cite{kang2019strong}, the ground state $| \Psi_{gr} \rangle$ of the interaction $H_{\varhexagon}$ should be annihilated by the assisted hopping operator $T_{\varhexagon}$ for any hexagon. The most general form of the wavefunction is 
\begin{align}
  | \Psi_{gr} \rangle = \prod_i \sum_{\alpha \beta} U_{1 \alpha} U_{2\beta} \psi_{i, \alpha}'^{\dagger}  \psi_{i, \beta}'^{\dagger} | \emptyset \rangle \label{Eqn:FerroWaveFunc}
\end{align}
where $\psi_i' = \left( c_{i, 1, \uparrow},\ c_{i, 1, \downarrow},\ (-)^{s(i)} c_{i, 2, \uparrow},\ (-)^{s(i)} c_{i, 2, \downarrow}\right)^T$, and $U_{a \gamma}$ is an arbitrary $4 \times 4$ matrix.

\subsection{$H_0 = 0$}
In the case of zero kinetic energy, the manifold of the ground states is described by Eq.~\eqref{Eqn:FerroWaveFunc}, and thus, can be compared with the numerical results produced by QMC. For this purpose, we consider the correlation function:
\begin{align} 
I_{i, \uparrow} & =  \langle c^{\dagger}_{i, 1, \uparrow} c_{i, 2, \uparrow} + h.c. \rangle =  (-)^{s(i)} \left(U_{11}^* U_{13} + U_{21}^* U_{23} + c.c \right) \\
I_{i, \downarrow} & =  \langle c^{\dagger}_{i, 1, \downarrow} c_{i, 2, \downarrow} + h.c. \rangle  =  (-)^{s(i)} \left( U_{12}^* U_{14} + U_{22}^* U_{24} + c.c  \right)
\end{align}
This leads to
\begin{widetext}
   \begin{eqnarray}
	C_I^{\mathcal{A}\mathcal{A}} & =  \frac1{L^4}\sum_{i, j \in \mathcal{A}}  \langle\!\langle \left( I_{i, \uparrow} + I_{i, \downarrow} \right)  \left( I_{j, \uparrow} + I_{j, \downarrow} \right) \rangle\!\rangle  \nonumber  =  \frac1{L^4}\sum_{i, j \in \mathcal{A}}  \left( \langle\!\langle I_{i, \uparrow}  I_{j, \uparrow}  \rangle\!\rangle +   \langle\!\langle I_{i, \downarrow} I_{j, \downarrow} \rangle\!\rangle  \right)  \nonumber \\
& =  2\times \left( |U_{11}|^2 |U_{13}|^2 + |U_{21}|^2 |U_{23}|^2 +|U_{12}|^2 |U_{14}|^2 +|U_{22}|^2 |U_{24}|^2  \right) \nonumber  = \frac12
   \end{eqnarray}
\end{widetext}
Note here $\langle\!\langle \cdots  \rangle\!\rangle$ is the average over all possible $4 \times 4$ unitary matrices and therefor $U_{ij}^* U_{k l} = \frac14 \delta_{ik} \delta_{lj}$. Similarly, we can obtain $C_I^{\mathcal{B}\mathcal{B}} = -C_I^{\mathcal{A}\mathcal{B}} = -C_I^{\mathcal{B}\mathcal{A}} = \frac12$. 

\subsection{Strong-coupling limit}
In this subsection, we assume that $H_0$ is finite but small compared with $H_{\varhexagon}$. Since the kinetic terms break $SU(4)$ symmetry, the ground state manifold shrinks and becomes a subset of the manifold described in Eq.~\eqref{Eqn:FerroWaveFunc}. Our purpose here is to identify the new manifold of the ground states and show that it is independent of the exact form of kinetic terms as along as it breaks the $SU(4)$ symmetry described in the main text.

For the convenience of calculation, we write the Eq.~\eqref{Eqn:FerroWaveFunc} as the following form, 
\begin{widetext}
   \begin{eqnarray}
 | \Psi_{gr} \rangle = \prod_i \left( \alpha_1 c^{\dagger}_{i,1,\hat n} + (-)^{s(i)} \alpha_2 c^{\dagger}_{i, 2,\hat m}  \right)  \left( \gamma \left( \alpha_2^* c^{\dagger}_{i,1,\hat n} - (-)^{s(i)} \alpha_1^* c^{\dagger}_{i, 2,\hat m} \right) + \beta_1 c^{\dagger}_{i,1, -\hat n} + (-)^{s(i)} \beta_2 c^{\dagger}_{i,2, -\hat m}  \right) | \emptyset \rangle   \label{Eqn:WaveFuncForm}
   \end{eqnarray}
\end{widetext}
where $s(i) = 0$ and $1$ if the site $i$ is on sublattice $\mathcal{A}$ and $\mathcal{B}$ respectively. $\hat n$ and $\hat m$ are two arbitrary spin quantization directions. $\alpha_1$, $\alpha_2$, $\beta_1$, and $\beta_2$ are four complex variables that satisfy $|\alpha_1|^2 + |\alpha_2|^2 = |\gamma|^2 + |\beta_1|^2 + |\beta_2|^2 = 1$. 

Furthermore, consider the hopping between two sites. We will show next that the energy is minimized when $|\alpha_1| = |\alpha_2| = 1/\sqrt{2}$, $\gamma = 0$, and $|\beta_1| = |\beta_2| = 1/\sqrt{2}$. Applying the second order perturbation theory,  the correction to the energy of the ground state is found to be
\[  \delta E = \sum_n \frac{|\langle n | H_0 | \Psi_{GS} \rangle |^2}{E_0 - E_n}  \]
where $n$ sums over all the excited states. Since it is almost impossible to obtain the exact spectrum of the excited states, we will maximize the term $\sum_n |\langle n | H_0 | \Psi_{GS} \rangle |^2$ instead of $\delta E$. Furthermore, note that $\langle GS | H_0 | \Psi_{GS} \rangle = 0$, and write  $H_0 = \sum_{i j} K(i, j)$ where $i$ and $j$ refer to the honeycomb lattice site, we obtain
\[ \sum_n |\langle n | H_0 | \Psi_{GS} \rangle |^2 = \big\lVert H_0 | \Psi_{GS} \rangle \big\rVert^2 = \sum_{i j}  \big\lVert K(i, j) | \Psi_{GS} \rangle \big\rVert^2  \ . \]
For notation convenience, it is worth to introduce $f_{i, a}^{\dagger} = \sum_{\alpha} \psi_{i, \alpha}'^{\dagger} U_{\alpha a}$, where $U$ is an unitary $4 \times 4$ matrix, and the ground state is given by
\[  | \Psi_{GS} \rangle = \prod_i \prod_{a = 1}^2 f_{i, 1}^{\dagger} f_{i, 2}^{\dagger} | \emptyset \rangle \ . \]
From Eq.~\eqref{Eqn:WaveFuncForm}, it is obvious that
\[ U_{1\alpha} = \left( \alpha_1,\ 0,\ \alpha_2,\ 0 \right) \quad U_{2\alpha} = \left( \gamma \alpha_2^*,\ \beta_1,\ -\gamma\alpha_1^*,\ \beta_2 \right) \ . \]
Introducing the diagonal matrix 
\[ T = \mathrm{diag}\left( t,\ t,\ (-)^{s(i)+s(j)}t^*,\ (-)^{s(i)+s(j)}t^*  \right) \ , \] 
where $t$ is the hopping constant from site $j$ to site $i$ for the valley $1$. The kinetic terms can be written as
\[ K(i, j) = \psi_{i, \alpha}'^{\dagger} T_{\alpha \beta} \psi_{j, \beta} = f_{i, a}^{\dagger} \left( U^{\dagger} T U \right)_{a b} f_{j, b} \ . \]
Now, it is easy to derive that 
\begin{align}
 & \big\lVert K(i, j) | \Psi_{GS} \rangle \big\rVert^2 = \sum_{i, j}\sum_{a = 3}^4 \sum_{b = 1}^2 T_{ii} T_{jj}^* U_{i a}^* U_{i b} U_{j a} U_{j b}^* \nonumber \\
= & \sum_{i j} T_{ii} T_{jj}^* \left( \delta_{i j} - \sum_{a = 1}^2 U_{i a}^* U_{j a} \right) U_{i b} U_{j b}^* \nonumber \\
= & \sum_i \left| T_{i i} \right|^2 \sum_{b = 1}^2 \left| U_{i b} \right|^2 - \sum_{i j} \sum_{a b = 1}^2  T_{ii} T_{jj}^* U_{i a}^* U_{i b} U_{j a} U_{j b}^* \nonumber \\
= & 8 |t|^2 -  \sum_{i j} \sum_{a b = 1}^2  T_{ii} T_{jj}^* U_{i a}^* U_{i b} U_{j a} U_{j b}^*
\end{align}
Since the first term is independent of the form of the ground state, we need to minimize the last term. Notice the $T$ matrix can be written as
\[ T = t_0 I_{4 \times 4} + t_1 \mathrm{diag}(1,\ 1, \ -1,\ -1)  \]
and $t_{0, 1} = \frac12 \left( t \pm (-)^{s(i) + s(j)} t^* \right)$ that leads to $t_0 t_1^*$ is a pure imaginary number. With this, we found
\begin{align}
& \big\lVert K(i, j) | \Psi_{GS} \rangle \big\rVert^2 = 8 |t|^2- 8 |t_0|^2 - |t_1|^2 \sum_{a b} \left| M_{a b} \right|^2 \ ,
\end{align}
where $M$ is a $2 \times 2$ matrix gives by
\[ M = \begin{pmatrix}
|\alpha_1|^2 - |\alpha_2|^2 & 2 \gamma \alpha_1^* \alpha_2^* \\ 2 \gamma^* \alpha_1 \alpha_2 & - |\gamma|^2 \left( |\alpha_1|^2 - |\alpha_2|^2 \right) + |\beta_1|^2 - |\beta_2|
\end{pmatrix}  \]
Clearly, the energy due to the second order perturbation is minimized when  $|\alpha_1| = |\alpha_2| = 1/\sqrt{2}$, $\gamma = 0$, and $|\beta_1| = |\beta_2| = 1/\sqrt{2}$

It is worth to emphasize that this result is independent of the exact form of the kinetic terms. As long as the hoppings break the $SU(4)$ symmetry, the second order perturbation always leads to the same manifold of the ground states.

As a consequence, the ground state is an equal mixture of two valleys. It is easy to obtain that
\begin{eqnarray}
	\begin{aligned}
	\langle c^{\dagger}_{i, 1, \hat n} c_{i, 2, \hat m} \rangle &= (-)^{s(i)}  \alpha_1^* \alpha_2 \\
   \langle c^{\dagger}_{i, 1, \hat n} c_{i, 2, -\hat m} \rangle &=  \langle c^{\dagger}_{i, 1, -\hat n} c_{i, 2, \hat m} \rangle =  0 \\
   \langle c^{\dagger}_{i, 1, -\hat n} c_{i, 2, -\hat m} \rangle &= (-)^{s(i)} \beta_1^* \beta_2 
	\end{aligned}
\end{eqnarray}
\begin{widetext}
Suppose that $\hat n = (\sin\theta \cos\phi, \sin\theta \sin\phi, \cos\theta)$ and $\hat m = (\sin\theta' \cos\phi', \sin\theta' \sin\phi', \cos\theta')$. We obtain that the operator
\begin{eqnarray} 
I_{i, \uparrow} & = & \langle c^{\dagger}_{i, 1, \uparrow} c_{i, 2, \uparrow} + h.c. \rangle =  (-)^{s(i)} \left( \cos\frac{\theta}2 \cos\frac{\theta'}2 \alpha_1^* \alpha_2 +  \sin\frac{\theta}2 \sin\frac{\theta'}2 e^{i (\phi -\phi')} \beta_1^* \beta_2  + c.c \right) \\
I_{i, \downarrow} & = & \langle c^{\dagger}_{i, 1, \downarrow} c_{i, 2, \downarrow} + h.c. \rangle  =  (-)^{s(i)} \left( \cos\frac{\theta}2 \cos\frac{\theta'}2 \beta_1^* \beta_2 -  \sin\frac{\theta}2 \sin\frac{\theta'}2 e^{i (\phi -\phi')} \alpha_1^* \alpha_2  + c.c  \right)
\end{eqnarray}
As a consequence, when average over all the possible configurations of the ground states, we obtain
\begin{eqnarray}
C_I^{\mathcal{A}\mathcal{A}} & = & \frac1{L^4}\sum_{i, j \in \mathcal{A}}  \langle\!\langle \left( I_{i, \uparrow} + I_{i, \downarrow} \right)  \left( I_{j, \uparrow} + I_{j, \downarrow} \right) \rangle\!\rangle  =  \frac1{L^4}\sum_{i, j \in \mathcal{A}}  \left( \langle\!\langle I_{i, \uparrow}  I_{j, \uparrow}  \rangle\!\rangle +   \langle\!\langle I_{i, \downarrow} I_{j, \downarrow} \rangle\!\rangle  \right)  \nonumber \\
& = &  2\times \left( \langle\!\langle \cos^2\frac{\theta}2 \cos^2\frac{\theta'}2 \rangle\!\rangle \left( |\alpha_1|^2 |\alpha_2|^2 + |\beta_1|^2 |\beta_2|^2 \right) + \langle\!\langle \sin^2\frac{\theta}2 \sin^2\frac{\theta'}2 \rangle\!\rangle \left( |\alpha_1|^2 |\alpha_2|^2 + |\beta_1|^2 |\beta_2|^2 \right) \right)
\end{eqnarray}
\end{widetext}
where $\langle\!\langle \cdots \rangle \!\rangle$ refers to the average over the direction $\hat n$ and $\hat m$, as well as the phases of $\alpha_1$, $\alpha_2$, $\beta_1$, and $\beta_2$. Averaging over $\hat n$ and $\hat m$ on the sphere, we obtain $\langle\!\langle \cos^2\frac{\theta}2  \rangle\!\rangle = \langle\!\langle \cos^2\frac{\theta'}2  \rangle\!\rangle =  \langle\!\langle \sin^2\frac{\theta}2  \rangle\!\rangle = \langle\!\langle \sin^2\frac{\theta'}2  \rangle\!\rangle = \frac12$. Thus, 
$ C_I^{\mathcal{A}\mathcal{A}} = \frac12 \ . $
Similarly, we can obtain $C_I^{\mathcal{B}\mathcal{B}} = -C_I^{\mathcal{A}\mathcal{B}} = -C_I^{\mathcal{B}\mathcal{A}} = \frac12$.  This is the same as the case of zero kinetic energy and consistent with the QMC result in the limit $U/W \rightarrow \infty$ in Appendix~\ref{app:pqmc}.

\bibliographystyle{apsrev4-1}
\bibliography{main}

\begin{thebibliography}{84}%
\makeatletter
\providecommand \@ifxundefined [1]{%
 \@ifx{#1\undefined}
}%
\providecommand \@ifnum [1]{%
 \ifnum #1\expandafter \@firstoftwo
 \else \expandafter \@secondoftwo
 \fi
}%
\providecommand \@ifx [1]{%
 \ifx #1\expandafter \@firstoftwo
 \else \expandafter \@secondoftwo
 \fi
}%
\providecommand \natexlab [1]{#1}%
\providecommand \enquote  [1]{``#1''}%
\providecommand \bibnamefont  [1]{#1}%
\providecommand \bibfnamefont [1]{#1}%
\providecommand \citenamefont [1]{#1}%
\providecommand \href@noop [0]{\@secondoftwo}%
\providecommand \href [0]{\begingroup \@sanitize@url \@href}%
\providecommand \@href[1]{\@@startlink{#1}\@@href}%
\providecommand \@@href[1]{\endgroup#1\@@endlink}%
\providecommand \@sanitize@url [0]{\catcode `\\12\catcode `\$12\catcode
  `\&12\catcode `\#12\catcode `\^12\catcode `\_12\catcode `\%12\relax}%
\providecommand \@@startlink[1]{}%
\providecommand \@@endlink[0]{}%
\providecommand \url  [0]{\begingroup\@sanitize@url \@url }%
\providecommand \@url [1]{\endgroup\@href {#1}{\urlprefix }}%
\providecommand \urlprefix  [0]{URL }%
\providecommand \Eprint [0]{\href }%
\providecommand \doibase [0]{http://dx.doi.org/}%
\providecommand \selectlanguage [0]{\@gobble}%
\providecommand \bibinfo  [0]{\@secondoftwo}%
\providecommand \bibfield  [0]{\@secondoftwo}%
\providecommand \translation [1]{[#1]}%
\providecommand \BibitemOpen [0]{}%
\providecommand \bibitemStop [0]{}%
\providecommand \bibitemNoStop [0]{.\EOS\space}%
\providecommand \EOS [0]{\spacefactor3000\relax}%
\providecommand \BibitemShut  [1]{\csname bibitem#1\endcsname}%
\let\auto@bib@innerbib\@empty
\bibitem [{\citenamefont {Bistritzer}\ and\ \citenamefont
  {MacDonald}(2011)}]{bistritzer2011moire}%
  \BibitemOpen
  \bibfield  {author} {\bibinfo {author} {\bibfnamefont {R.}~\bibnamefont
  {Bistritzer}}\ and\ \bibinfo {author} {\bibfnamefont {A.~H.}\ \bibnamefont
  {MacDonald}},\ }\href {\doibase 10.1073/pnas.1108174108} {\bibfield
  {journal} {\bibinfo  {journal} {Proceedings of the National Academy of
  Sciences}\ }\textbf {\bibinfo {volume} {108}},\ \bibinfo {pages} {12233}
  (\bibinfo {year} {2011})}\BibitemShut {NoStop}%
\bibitem [{\citenamefont {Cao}\ \emph {et~al.}(2018{\natexlab{a}})\citenamefont
  {Cao}, \citenamefont {Fatemi}, \citenamefont {Demir}, \citenamefont {Fang},
  \citenamefont {Tomarken}, \citenamefont {Luo}, \citenamefont
  {Sanchez-Yamagishi}, \citenamefont {Watanabe}, \citenamefont {Taniguchi},
  \citenamefont {Kaxiras}, \citenamefont {Ashoori},\ and\ \citenamefont
  {Jarillo-Herrero}}]{cao2018correlated}%
  \BibitemOpen
  \bibfield  {author} {\bibinfo {author} {\bibfnamefont {Y.}~\bibnamefont
  {Cao}}, \bibinfo {author} {\bibfnamefont {V.}~\bibnamefont {Fatemi}},
  \bibinfo {author} {\bibfnamefont {A.}~\bibnamefont {Demir}}, \bibinfo
  {author} {\bibfnamefont {S.}~\bibnamefont {Fang}}, \bibinfo {author}
  {\bibfnamefont {S.~L.}\ \bibnamefont {Tomarken}}, \bibinfo {author}
  {\bibfnamefont {J.~Y.}\ \bibnamefont {Luo}}, \bibinfo {author} {\bibfnamefont
  {J.~D.}\ \bibnamefont {Sanchez-Yamagishi}}, \bibinfo {author} {\bibfnamefont
  {K.}~\bibnamefont {Watanabe}}, \bibinfo {author} {\bibfnamefont
  {T.}~\bibnamefont {Taniguchi}}, \bibinfo {author} {\bibfnamefont
  {E.}~\bibnamefont {Kaxiras}}, \bibinfo {author} {\bibfnamefont {R.~C.}\
  \bibnamefont {Ashoori}}, \ and\ \bibinfo {author} {\bibfnamefont
  {P.}~\bibnamefont {Jarillo-Herrero}},\ }\href {\doibase 10.1038/nature26154}
  {\bibfield  {journal} {\bibinfo  {journal} {Nature}\ }\textbf {\bibinfo
  {volume} {556}},\ \bibinfo {pages} {80} (\bibinfo {year}
  {2018}{\natexlab{a}})}\BibitemShut {NoStop}%
\bibitem [{\citenamefont {Cao}\ \emph {et~al.}(2018{\natexlab{b}})\citenamefont
  {Cao}, \citenamefont {Fatemi}, \citenamefont {Fang}, \citenamefont
  {Watanabe}, \citenamefont {Taniguchi}, \citenamefont {Kaxiras},\ and\
  \citenamefont {Jarillo-Herrero}}]{cao2018unconventional}%
  \BibitemOpen
  \bibfield  {author} {\bibinfo {author} {\bibfnamefont {Y.}~\bibnamefont
  {Cao}}, \bibinfo {author} {\bibfnamefont {V.}~\bibnamefont {Fatemi}},
  \bibinfo {author} {\bibfnamefont {S.}~\bibnamefont {Fang}}, \bibinfo {author}
  {\bibfnamefont {K.}~\bibnamefont {Watanabe}}, \bibinfo {author}
  {\bibfnamefont {T.}~\bibnamefont {Taniguchi}}, \bibinfo {author}
  {\bibfnamefont {E.}~\bibnamefont {Kaxiras}}, \ and\ \bibinfo {author}
  {\bibfnamefont {P.}~\bibnamefont {Jarillo-Herrero}},\ }\href {\doibase
  10.1038/nature26160} {\bibfield  {journal} {\bibinfo  {journal} {Nature}\
  }\textbf {\bibinfo {volume} {556}},\ \bibinfo {pages} {43} (\bibinfo {year}
  {2018}{\natexlab{b}})}\BibitemShut {NoStop}%
\bibitem [{\citenamefont {Shen}\ \emph {et~al.}(2020)\citenamefont {Shen},
  \citenamefont {Chu}, \citenamefont {Wu}, \citenamefont {Li}, \citenamefont
  {Wang}, \citenamefont {Zhao}, \citenamefont {Tang}, \citenamefont {Liu},
  \citenamefont {Tian}, \citenamefont {Watanabe}, \citenamefont {Taniguchi},
  \citenamefont {Yang}, \citenamefont {Meng}, \citenamefont {Shi},
  \citenamefont {Yazyev},\ and\ \citenamefont {Zhang}}]{shen2019observation}%
  \BibitemOpen
  \bibfield  {author} {\bibinfo {author} {\bibfnamefont {C.}~\bibnamefont
  {Shen}}, \bibinfo {author} {\bibfnamefont {Y.}~\bibnamefont {Chu}}, \bibinfo
  {author} {\bibfnamefont {Q.}~\bibnamefont {Wu}}, \bibinfo {author}
  {\bibfnamefont {N.}~\bibnamefont {Li}}, \bibinfo {author} {\bibfnamefont
  {S.}~\bibnamefont {Wang}}, \bibinfo {author} {\bibfnamefont {Y.}~\bibnamefont
  {Zhao}}, \bibinfo {author} {\bibfnamefont {J.}~\bibnamefont {Tang}}, \bibinfo
  {author} {\bibfnamefont {J.}~\bibnamefont {Liu}}, \bibinfo {author}
  {\bibfnamefont {J.}~\bibnamefont {Tian}}, \bibinfo {author} {\bibfnamefont
  {K.}~\bibnamefont {Watanabe}}, \bibinfo {author} {\bibfnamefont
  {T.}~\bibnamefont {Taniguchi}}, \bibinfo {author} {\bibfnamefont
  {R.}~\bibnamefont {Yang}}, \bibinfo {author} {\bibfnamefont {Z.~Y.}\
  \bibnamefont {Meng}}, \bibinfo {author} {\bibfnamefont {D.}~\bibnamefont
  {Shi}}, \bibinfo {author} {\bibfnamefont {O.~V.}\ \bibnamefont {Yazyev}}, \
  and\ \bibinfo {author} {\bibfnamefont {G.}~\bibnamefont {Zhang}},\ }\href
  {\doibase 10.1038/s41567-020-0825-9} {\bibfield  {journal} {\bibinfo
  {journal} {Nature Physics}\ }\textbf {\bibinfo {volume} {16}},\ \bibinfo
  {pages} {520} (\bibinfo {year} {2020})}\BibitemShut {NoStop}%
\bibitem [{\citenamefont {Liu}\ \emph {et~al.}(2020{\natexlab{a}})\citenamefont
  {Liu}, \citenamefont {Hao}, \citenamefont {Khalaf}, \citenamefont {Lee},
  \citenamefont {Watanabe}, \citenamefont {Taniguchi}, \citenamefont
  {Vishwanath},\ and\ \citenamefont {Kim}}]{liu2019spin}%
  \BibitemOpen
  \bibfield  {author} {\bibinfo {author} {\bibfnamefont {X.}~\bibnamefont
  {Liu}}, \bibinfo {author} {\bibfnamefont {Z.}~\bibnamefont {Hao}}, \bibinfo
  {author} {\bibfnamefont {E.}~\bibnamefont {Khalaf}}, \bibinfo {author}
  {\bibfnamefont {J.~Y.}\ \bibnamefont {Lee}}, \bibinfo {author} {\bibfnamefont
  {K.}~\bibnamefont {Watanabe}}, \bibinfo {author} {\bibfnamefont
  {T.}~\bibnamefont {Taniguchi}}, \bibinfo {author} {\bibfnamefont
  {A.}~\bibnamefont {Vishwanath}}, \ and\ \bibinfo {author} {\bibfnamefont
  {P.}~\bibnamefont {Kim}},\ }\href {\doibase 10.1038/s41586-020-2458-7}
  {\bibfield  {journal} {\bibinfo  {journal} {Nature}\ }\textbf {\bibinfo
  {volume} {583}},\ \bibinfo {pages} {221} (\bibinfo {year}
  {2020}{\natexlab{a}})}\BibitemShut {NoStop}%
\bibitem [{\citenamefont {Cao}\ \emph {et~al.}(2020)\citenamefont {Cao},
  \citenamefont {Rodan-Legrain}, \citenamefont {Rubies-Bigorda}, \citenamefont
  {Park}, \citenamefont {Watanabe}, \citenamefont {Taniguchi},\ and\
  \citenamefont {Jarillo-Herrero}}]{cao2019electric}%
  \BibitemOpen
  \bibfield  {author} {\bibinfo {author} {\bibfnamefont {Y.}~\bibnamefont
  {Cao}}, \bibinfo {author} {\bibfnamefont {D.}~\bibnamefont {Rodan-Legrain}},
  \bibinfo {author} {\bibfnamefont {O.}~\bibnamefont {Rubies-Bigorda}},
  \bibinfo {author} {\bibfnamefont {J.~M.}\ \bibnamefont {Park}}, \bibinfo
  {author} {\bibfnamefont {K.}~\bibnamefont {Watanabe}}, \bibinfo {author}
  {\bibfnamefont {T.}~\bibnamefont {Taniguchi}}, \ and\ \bibinfo {author}
  {\bibfnamefont {P.}~\bibnamefont {Jarillo-Herrero}},\ }\href {\doibase
  10.1038/s41586-020-2260-6} {\bibfield  {journal} {\bibinfo  {journal}
  {Nature}\ }\textbf {\bibinfo {volume} {583}},\ \bibinfo {pages} {215}
  (\bibinfo {year} {2020})}\BibitemShut {NoStop}%
\bibitem [{\citenamefont {Chen}\ \emph {et~al.}(2020)\citenamefont {Chen},
  \citenamefont {Sharpe}, \citenamefont {Fox}, \citenamefont {Zhang},
  \citenamefont {Wang}, \citenamefont {Jiang}, \citenamefont {Lyu},
  \citenamefont {Li}, \citenamefont {Watanabe}, \citenamefont {Taniguchi} \emph
  {et~al.}}]{chen2020tunable}%
  \BibitemOpen
  \bibfield  {author} {\bibinfo {author} {\bibfnamefont {G.}~\bibnamefont
  {Chen}}, \bibinfo {author} {\bibfnamefont {A.~L.}\ \bibnamefont {Sharpe}},
  \bibinfo {author} {\bibfnamefont {E.~J.}\ \bibnamefont {Fox}}, \bibinfo
  {author} {\bibfnamefont {Y.-H.}\ \bibnamefont {Zhang}}, \bibinfo {author}
  {\bibfnamefont {S.}~\bibnamefont {Wang}}, \bibinfo {author} {\bibfnamefont
  {L.}~\bibnamefont {Jiang}}, \bibinfo {author} {\bibfnamefont
  {B.}~\bibnamefont {Lyu}}, \bibinfo {author} {\bibfnamefont {H.}~\bibnamefont
  {Li}}, \bibinfo {author} {\bibfnamefont {K.}~\bibnamefont {Watanabe}},
  \bibinfo {author} {\bibfnamefont {T.}~\bibnamefont {Taniguchi}},  \emph
  {et~al.},\ }\href {https://www.nature.com/articles/s41586-020-2049-7}
  {\bibfield  {journal} {\bibinfo  {journal} {Nature}\ }\textbf {\bibinfo
  {volume} {579}},\ \bibinfo {pages} {56} (\bibinfo {year} {2020})}\BibitemShut
  {NoStop}%
\bibitem [{\citenamefont {Kerelsky}\ \emph {et~al.}(2019)\citenamefont
  {Kerelsky}, \citenamefont {McGilly}, \citenamefont {Kennes}, \citenamefont
  {Xian}, \citenamefont {Yankowitz}, \citenamefont {Chen}, \citenamefont
  {Watanabe}, \citenamefont {Taniguchi}, \citenamefont {Hone}, \citenamefont
  {Dean} \emph {et~al.}}]{kerelsky2019maximized}%
  \BibitemOpen
  \bibfield  {author} {\bibinfo {author} {\bibfnamefont {A.}~\bibnamefont
  {Kerelsky}}, \bibinfo {author} {\bibfnamefont {L.~J.}\ \bibnamefont
  {McGilly}}, \bibinfo {author} {\bibfnamefont {D.~M.}\ \bibnamefont {Kennes}},
  \bibinfo {author} {\bibfnamefont {L.}~\bibnamefont {Xian}}, \bibinfo {author}
  {\bibfnamefont {M.}~\bibnamefont {Yankowitz}}, \bibinfo {author}
  {\bibfnamefont {S.}~\bibnamefont {Chen}}, \bibinfo {author} {\bibfnamefont
  {K.}~\bibnamefont {Watanabe}}, \bibinfo {author} {\bibfnamefont
  {T.}~\bibnamefont {Taniguchi}}, \bibinfo {author} {\bibfnamefont
  {J.}~\bibnamefont {Hone}}, \bibinfo {author} {\bibfnamefont {C.}~\bibnamefont
  {Dean}},  \emph {et~al.},\ }\href
  {https://www.nature.com/articles/s41586-019-1431-9} {\bibfield  {journal}
  {\bibinfo  {journal} {Nature}\ }\textbf {\bibinfo {volume} {572}},\ \bibinfo
  {pages} {95} (\bibinfo {year} {2019})}\BibitemShut {NoStop}%
\bibitem [{\citenamefont {Tomarken}\ \emph {et~al.}(2019)\citenamefont
  {Tomarken}, \citenamefont {Cao}, \citenamefont {Demir}, \citenamefont
  {Watanabe}, \citenamefont {Taniguchi}, \citenamefont {Jarillo-Herrero},\ and\
  \citenamefont {Ashoori}}]{tomarken2019electronic}%
  \BibitemOpen
  \bibfield  {author} {\bibinfo {author} {\bibfnamefont {S.~L.}\ \bibnamefont
  {Tomarken}}, \bibinfo {author} {\bibfnamefont {Y.}~\bibnamefont {Cao}},
  \bibinfo {author} {\bibfnamefont {A.}~\bibnamefont {Demir}}, \bibinfo
  {author} {\bibfnamefont {K.}~\bibnamefont {Watanabe}}, \bibinfo {author}
  {\bibfnamefont {T.}~\bibnamefont {Taniguchi}}, \bibinfo {author}
  {\bibfnamefont {P.}~\bibnamefont {Jarillo-Herrero}}, \ and\ \bibinfo {author}
  {\bibfnamefont {R.~C.}\ \bibnamefont {Ashoori}},\ }\href {\doibase
  10.1103/PhysRevLett.123.046601} {\bibfield  {journal} {\bibinfo  {journal}
  {Phys. Rev. Lett.}\ }\textbf {\bibinfo {volume} {123}},\ \bibinfo {pages}
  {046601} (\bibinfo {year} {2019})}\BibitemShut {NoStop}%
\bibitem [{\citenamefont {Lu}\ \emph {et~al.}(2019)\citenamefont {Lu},
  \citenamefont {Stepanov}, \citenamefont {Yang}, \citenamefont {Xie},
  \citenamefont {Aamir}, \citenamefont {Das}, \citenamefont {Urgell},
  \citenamefont {Watanabe}, \citenamefont {Taniguchi}, \citenamefont {Zhang}
  \emph {et~al.}}]{lu2019superconductors}%
  \BibitemOpen
  \bibfield  {author} {\bibinfo {author} {\bibfnamefont {X.}~\bibnamefont
  {Lu}}, \bibinfo {author} {\bibfnamefont {P.}~\bibnamefont {Stepanov}},
  \bibinfo {author} {\bibfnamefont {W.}~\bibnamefont {Yang}}, \bibinfo {author}
  {\bibfnamefont {M.}~\bibnamefont {Xie}}, \bibinfo {author} {\bibfnamefont
  {M.~A.}\ \bibnamefont {Aamir}}, \bibinfo {author} {\bibfnamefont
  {I.}~\bibnamefont {Das}}, \bibinfo {author} {\bibfnamefont {C.}~\bibnamefont
  {Urgell}}, \bibinfo {author} {\bibfnamefont {K.}~\bibnamefont {Watanabe}},
  \bibinfo {author} {\bibfnamefont {T.}~\bibnamefont {Taniguchi}}, \bibinfo
  {author} {\bibfnamefont {G.}~\bibnamefont {Zhang}},  \emph {et~al.},\ }\href
  {https://www.nature.com/articles/s41586-019-1695-0} {\bibfield  {journal}
  {\bibinfo  {journal} {Nature}\ }\textbf {\bibinfo {volume} {574}},\ \bibinfo
  {pages} {653} (\bibinfo {year} {2019})}\BibitemShut {NoStop}%
\bibitem [{\citenamefont {Xie}\ \emph {et~al.}(2019)\citenamefont {Xie},
  \citenamefont {Lian}, \citenamefont {J{\"a}ck}, \citenamefont {Liu},
  \citenamefont {Chiu}, \citenamefont {Watanabe}, \citenamefont {Taniguchi},
  \citenamefont {Bernevig},\ and\ \citenamefont
  {Yazdani}}]{xie2019spectroscopic}%
  \BibitemOpen
  \bibfield  {author} {\bibinfo {author} {\bibfnamefont {Y.}~\bibnamefont
  {Xie}}, \bibinfo {author} {\bibfnamefont {B.}~\bibnamefont {Lian}}, \bibinfo
  {author} {\bibfnamefont {B.}~\bibnamefont {J{\"a}ck}}, \bibinfo {author}
  {\bibfnamefont {X.}~\bibnamefont {Liu}}, \bibinfo {author} {\bibfnamefont
  {C.-L.}\ \bibnamefont {Chiu}}, \bibinfo {author} {\bibfnamefont
  {K.}~\bibnamefont {Watanabe}}, \bibinfo {author} {\bibfnamefont
  {T.}~\bibnamefont {Taniguchi}}, \bibinfo {author} {\bibfnamefont {B.~A.}\
  \bibnamefont {Bernevig}}, \ and\ \bibinfo {author} {\bibfnamefont
  {A.}~\bibnamefont {Yazdani}},\ }\href
  {https://www.nature.com/articles/s41586-019-1422-x} {\bibfield  {journal}
  {\bibinfo  {journal} {Nature}\ }\textbf {\bibinfo {volume} {572}},\ \bibinfo
  {pages} {101} (\bibinfo {year} {2019})}\BibitemShut {NoStop}%
\bibitem [{\citenamefont {Jiang}\ \emph {et~al.}(2019)\citenamefont {Jiang},
  \citenamefont {Lai}, \citenamefont {Watanabe}, \citenamefont {Taniguchi},
  \citenamefont {Haule}, \citenamefont {Mao},\ and\ \citenamefont
  {Andrei}}]{jiang2019charge}%
  \BibitemOpen
  \bibfield  {author} {\bibinfo {author} {\bibfnamefont {Y.}~\bibnamefont
  {Jiang}}, \bibinfo {author} {\bibfnamefont {X.}~\bibnamefont {Lai}}, \bibinfo
  {author} {\bibfnamefont {K.}~\bibnamefont {Watanabe}}, \bibinfo {author}
  {\bibfnamefont {T.}~\bibnamefont {Taniguchi}}, \bibinfo {author}
  {\bibfnamefont {K.}~\bibnamefont {Haule}}, \bibinfo {author} {\bibfnamefont
  {J.}~\bibnamefont {Mao}}, \ and\ \bibinfo {author} {\bibfnamefont {E.~Y.}\
  \bibnamefont {Andrei}},\ }\href
  {https://www.nature.com/articles/s41586-019-1460-4} {\bibfield  {journal}
  {\bibinfo  {journal} {Nature}\ }\textbf {\bibinfo {volume} {573}},\ \bibinfo
  {pages} {91} (\bibinfo {year} {2019})}\BibitemShut {NoStop}%
\bibitem [{\citenamefont {Wong}\ \emph {et~al.}(2020)\citenamefont {Wong},
  \citenamefont {Nuckolls}, \citenamefont {Oh}, \citenamefont {Lian},
  \citenamefont {Xie}, \citenamefont {Jeon}, \citenamefont {Watanabe},
  \citenamefont {Taniguchi}, \citenamefont {Bernevig},\ and\ \citenamefont
  {Yazdani}}]{wong2019cascade}%
  \BibitemOpen
  \bibfield  {author} {\bibinfo {author} {\bibfnamefont {D.}~\bibnamefont
  {Wong}}, \bibinfo {author} {\bibfnamefont {K.~P.}\ \bibnamefont {Nuckolls}},
  \bibinfo {author} {\bibfnamefont {M.}~\bibnamefont {Oh}}, \bibinfo {author}
  {\bibfnamefont {B.}~\bibnamefont {Lian}}, \bibinfo {author} {\bibfnamefont
  {Y.}~\bibnamefont {Xie}}, \bibinfo {author} {\bibfnamefont {S.}~\bibnamefont
  {Jeon}}, \bibinfo {author} {\bibfnamefont {K.}~\bibnamefont {Watanabe}},
  \bibinfo {author} {\bibfnamefont {T.}~\bibnamefont {Taniguchi}}, \bibinfo
  {author} {\bibfnamefont {B.~A.}\ \bibnamefont {Bernevig}}, \ and\ \bibinfo
  {author} {\bibfnamefont {A.}~\bibnamefont {Yazdani}},\ }\href
  {https://www.nature.com/articles/s41586-020-2339-0} {\bibfield  {journal}
  {\bibinfo  {journal} {Nature}\ }\textbf {\bibinfo {volume} {582}},\ \bibinfo
  {pages} {198} (\bibinfo {year} {2020})}\BibitemShut {NoStop}%
\bibitem [{\citenamefont {Zondiner}\ \emph {et~al.}(2020)\citenamefont
  {Zondiner}, \citenamefont {Rozen}, \citenamefont {Rodan-Legrain},
  \citenamefont {Cao}, \citenamefont {Queiroz}, \citenamefont {Taniguchi},
  \citenamefont {Watanabe}, \citenamefont {Oreg}, \citenamefont {von Oppen},
  \citenamefont {Stern} \emph {et~al.}}]{zondiner2019cascade}%
  \BibitemOpen
  \bibfield  {author} {\bibinfo {author} {\bibfnamefont {U.}~\bibnamefont
  {Zondiner}}, \bibinfo {author} {\bibfnamefont {A.}~\bibnamefont {Rozen}},
  \bibinfo {author} {\bibfnamefont {D.}~\bibnamefont {Rodan-Legrain}}, \bibinfo
  {author} {\bibfnamefont {Y.}~\bibnamefont {Cao}}, \bibinfo {author}
  {\bibfnamefont {R.}~\bibnamefont {Queiroz}}, \bibinfo {author} {\bibfnamefont
  {T.}~\bibnamefont {Taniguchi}}, \bibinfo {author} {\bibfnamefont
  {K.}~\bibnamefont {Watanabe}}, \bibinfo {author} {\bibfnamefont
  {Y.}~\bibnamefont {Oreg}}, \bibinfo {author} {\bibfnamefont {F.}~\bibnamefont
  {von Oppen}}, \bibinfo {author} {\bibfnamefont {A.}~\bibnamefont {Stern}},
  \emph {et~al.},\ }\href {https://www.nature.com/articles/s41586-020-2373-y}
  {\bibfield  {journal} {\bibinfo  {journal} {Nature}\ }\textbf {\bibinfo
  {volume} {582}},\ \bibinfo {pages} {203} (\bibinfo {year}
  {2020})}\BibitemShut {NoStop}%
\bibitem [{\citenamefont {Saito}\ \emph {et~al.}(2020)\citenamefont {Saito},
  \citenamefont {Ge}, \citenamefont {Watanabe}, \citenamefont {Taniguchi},\
  and\ \citenamefont {Young}}]{saito2019decoupling}%
  \BibitemOpen
  \bibfield  {author} {\bibinfo {author} {\bibfnamefont {Y.}~\bibnamefont
  {Saito}}, \bibinfo {author} {\bibfnamefont {J.}~\bibnamefont {Ge}}, \bibinfo
  {author} {\bibfnamefont {K.}~\bibnamefont {Watanabe}}, \bibinfo {author}
  {\bibfnamefont {T.}~\bibnamefont {Taniguchi}}, \ and\ \bibinfo {author}
  {\bibfnamefont {A.~F.}\ \bibnamefont {Young}},\ }\href
  {https://www.nature.com/articles/s41567-020-0928-3} {\bibfield  {journal}
  {\bibinfo  {journal} {Nature Physics}\ }\textbf {\bibinfo {volume} {16}},\
  \bibinfo {pages} {1} (\bibinfo {year} {2020})}\BibitemShut {NoStop}%
\bibitem [{\citenamefont {Stepanov}\ \emph {et~al.}(2020)\citenamefont
  {Stepanov}, \citenamefont {Das}, \citenamefont {Lu}, \citenamefont
  {Fahimniya}, \citenamefont {Watanabe}, \citenamefont {Taniguchi},
  \citenamefont {Koppens}, \citenamefont {Lischner}, \citenamefont {Levitov},\
  and\ \citenamefont {Efetov}}]{stepanov2019interplay}%
  \BibitemOpen
  \bibfield  {author} {\bibinfo {author} {\bibfnamefont {P.}~\bibnamefont
  {Stepanov}}, \bibinfo {author} {\bibfnamefont {I.}~\bibnamefont {Das}},
  \bibinfo {author} {\bibfnamefont {X.}~\bibnamefont {Lu}}, \bibinfo {author}
  {\bibfnamefont {A.}~\bibnamefont {Fahimniya}}, \bibinfo {author}
  {\bibfnamefont {K.}~\bibnamefont {Watanabe}}, \bibinfo {author}
  {\bibfnamefont {T.}~\bibnamefont {Taniguchi}}, \bibinfo {author}
  {\bibfnamefont {F.~H.}\ \bibnamefont {Koppens}}, \bibinfo {author}
  {\bibfnamefont {J.}~\bibnamefont {Lischner}}, \bibinfo {author}
  {\bibfnamefont {L.}~\bibnamefont {Levitov}}, \ and\ \bibinfo {author}
  {\bibfnamefont {D.~K.}\ \bibnamefont {Efetov}},\ }\href
  {https://www.nature.com/articles/s41586-020-2459-6} {\bibfield  {journal}
  {\bibinfo  {journal} {Nature}\ }\textbf {\bibinfo {volume} {583}},\ \bibinfo
  {pages} {375} (\bibinfo {year} {2020})}\BibitemShut {NoStop}%
\bibitem [{\citenamefont {Chen}\ \emph
  {et~al.}(2019{\natexlab{a}})\citenamefont {Chen}, \citenamefont {Jiang},
  \citenamefont {Wu}, \citenamefont {Lyu}, \citenamefont {Li}, \citenamefont
  {Chittari}, \citenamefont {Watanabe}, \citenamefont {Taniguchi},
  \citenamefont {Shi}, \citenamefont {Jung} \emph {et~al.}}]{chen2019evidence}%
  \BibitemOpen
  \bibfield  {author} {\bibinfo {author} {\bibfnamefont {G.}~\bibnamefont
  {Chen}}, \bibinfo {author} {\bibfnamefont {L.}~\bibnamefont {Jiang}},
  \bibinfo {author} {\bibfnamefont {S.}~\bibnamefont {Wu}}, \bibinfo {author}
  {\bibfnamefont {B.}~\bibnamefont {Lyu}}, \bibinfo {author} {\bibfnamefont
  {H.}~\bibnamefont {Li}}, \bibinfo {author} {\bibfnamefont {B.~L.}\
  \bibnamefont {Chittari}}, \bibinfo {author} {\bibfnamefont {K.}~\bibnamefont
  {Watanabe}}, \bibinfo {author} {\bibfnamefont {T.}~\bibnamefont {Taniguchi}},
  \bibinfo {author} {\bibfnamefont {Z.}~\bibnamefont {Shi}}, \bibinfo {author}
  {\bibfnamefont {J.}~\bibnamefont {Jung}},  \emph {et~al.},\ }\href
  {https://www.nature.com/articles/s41567-018-0387-2} {\bibfield  {journal}
  {\bibinfo  {journal} {Nature Physics}\ }\textbf {\bibinfo {volume} {15}},\
  \bibinfo {pages} {237} (\bibinfo {year} {2019}{\natexlab{a}})}\BibitemShut
  {NoStop}%
\bibitem [{\citenamefont {Chen}\ \emph
  {et~al.}(2019{\natexlab{b}})\citenamefont {Chen}, \citenamefont {Sharpe},
  \citenamefont {Gallagher}, \citenamefont {Rosen}, \citenamefont {Fox},
  \citenamefont {Jiang}, \citenamefont {Lyu}, \citenamefont {Li}, \citenamefont
  {Watanabe}, \citenamefont {Taniguchi} \emph {et~al.}}]{chen2019signatures}%
  \BibitemOpen
  \bibfield  {author} {\bibinfo {author} {\bibfnamefont {G.}~\bibnamefont
  {Chen}}, \bibinfo {author} {\bibfnamefont {A.~L.}\ \bibnamefont {Sharpe}},
  \bibinfo {author} {\bibfnamefont {P.}~\bibnamefont {Gallagher}}, \bibinfo
  {author} {\bibfnamefont {I.~T.}\ \bibnamefont {Rosen}}, \bibinfo {author}
  {\bibfnamefont {E.~J.}\ \bibnamefont {Fox}}, \bibinfo {author} {\bibfnamefont
  {L.}~\bibnamefont {Jiang}}, \bibinfo {author} {\bibfnamefont
  {B.}~\bibnamefont {Lyu}}, \bibinfo {author} {\bibfnamefont {H.}~\bibnamefont
  {Li}}, \bibinfo {author} {\bibfnamefont {K.}~\bibnamefont {Watanabe}},
  \bibinfo {author} {\bibfnamefont {T.}~\bibnamefont {Taniguchi}},  \emph
  {et~al.},\ }\href {https://www.nature.com/articles/s41586-019-1393-y}
  {\bibfield  {journal} {\bibinfo  {journal} {Nature}\ }\textbf {\bibinfo
  {volume} {572}},\ \bibinfo {pages} {215} (\bibinfo {year}
  {2019}{\natexlab{b}})}\BibitemShut {NoStop}%
\bibitem [{\citenamefont {Xu}\ and\ \citenamefont
  {Balents}(2018)}]{xu2018topological}%
  \BibitemOpen
  \bibfield  {author} {\bibinfo {author} {\bibfnamefont {C.}~\bibnamefont
  {Xu}}\ and\ \bibinfo {author} {\bibfnamefont {L.}~\bibnamefont {Balents}},\
  }\href {\doibase 10.1103/PhysRevLett.121.087001} {\bibfield  {journal}
  {\bibinfo  {journal} {Phys. Rev. Lett.}\ }\textbf {\bibinfo {volume} {121}},\
  \bibinfo {pages} {087001} (\bibinfo {year} {2018})}\BibitemShut {NoStop}%
\bibitem [{\citenamefont {Kang}\ and\ \citenamefont
  {Vafek}(2018)}]{kang2018symmetry}%
  \BibitemOpen
  \bibfield  {author} {\bibinfo {author} {\bibfnamefont {J.}~\bibnamefont
  {Kang}}\ and\ \bibinfo {author} {\bibfnamefont {O.}~\bibnamefont {Vafek}},\
  }\href {\doibase 10.1103/PhysRevX.8.031088} {\bibfield  {journal} {\bibinfo
  {journal} {Phys. Rev. X}\ }\textbf {\bibinfo {volume} {8}},\ \bibinfo {pages}
  {031088} (\bibinfo {year} {2018})}\BibitemShut {NoStop}%
\bibitem [{\citenamefont {Koshino}\ \emph {et~al.}(2018)\citenamefont
  {Koshino}, \citenamefont {Yuan}, \citenamefont {Koretsune}, \citenamefont
  {Ochi}, \citenamefont {Kuroki},\ and\ \citenamefont
  {Fu}}]{koshino2018maximally}%
  \BibitemOpen
  \bibfield  {author} {\bibinfo {author} {\bibfnamefont {M.}~\bibnamefont
  {Koshino}}, \bibinfo {author} {\bibfnamefont {N.~F.~Q.}\ \bibnamefont
  {Yuan}}, \bibinfo {author} {\bibfnamefont {T.}~\bibnamefont {Koretsune}},
  \bibinfo {author} {\bibfnamefont {M.}~\bibnamefont {Ochi}}, \bibinfo {author}
  {\bibfnamefont {K.}~\bibnamefont {Kuroki}}, \ and\ \bibinfo {author}
  {\bibfnamefont {L.}~\bibnamefont {Fu}},\ }\href {\doibase
  10.1103/PhysRevX.8.031087} {\bibfield  {journal} {\bibinfo  {journal} {Phys.
  Rev. X}\ }\textbf {\bibinfo {volume} {8}},\ \bibinfo {pages} {031087}
  (\bibinfo {year} {2018})}\BibitemShut {NoStop}%
\bibitem [{\citenamefont {Yuan}\ and\ \citenamefont
  {Fu}(2018)}]{yuan2018model}%
  \BibitemOpen
  \bibfield  {author} {\bibinfo {author} {\bibfnamefont {N.~F.~Q.}\
  \bibnamefont {Yuan}}\ and\ \bibinfo {author} {\bibfnamefont {L.}~\bibnamefont
  {Fu}},\ }\href {\doibase 10.1103/PhysRevB.98.045103} {\bibfield  {journal}
  {\bibinfo  {journal} {Phys. Rev. B}\ }\textbf {\bibinfo {volume} {98}},\
  \bibinfo {pages} {045103} (\bibinfo {year} {2018})}\BibitemShut {NoStop}%
\bibitem [{\citenamefont {Po}\ \emph {et~al.}(2018{\natexlab{a}})\citenamefont
  {Po}, \citenamefont {Zou}, \citenamefont {Vishwanath},\ and\ \citenamefont
  {Senthil}}]{po2018origin}%
  \BibitemOpen
  \bibfield  {author} {\bibinfo {author} {\bibfnamefont {H.~C.}\ \bibnamefont
  {Po}}, \bibinfo {author} {\bibfnamefont {L.}~\bibnamefont {Zou}}, \bibinfo
  {author} {\bibfnamefont {A.}~\bibnamefont {Vishwanath}}, \ and\ \bibinfo
  {author} {\bibfnamefont {T.}~\bibnamefont {Senthil}},\ }\href {\doibase
  10.1103/PhysRevX.8.031089} {\bibfield  {journal} {\bibinfo  {journal} {Phys.
  Rev. X}\ }\textbf {\bibinfo {volume} {8}},\ \bibinfo {pages} {031089}
  (\bibinfo {year} {2018}{\natexlab{a}})}\BibitemShut {NoStop}%
\bibitem [{\citenamefont {Liu}\ \emph {et~al.}(2018)\citenamefont {Liu},
  \citenamefont {Zhang}, \citenamefont {Chen},\ and\ \citenamefont
  {Yang}}]{liu2018chiral}%
  \BibitemOpen
  \bibfield  {author} {\bibinfo {author} {\bibfnamefont {C.-C.}\ \bibnamefont
  {Liu}}, \bibinfo {author} {\bibfnamefont {L.-D.}\ \bibnamefont {Zhang}},
  \bibinfo {author} {\bibfnamefont {W.-Q.}\ \bibnamefont {Chen}}, \ and\
  \bibinfo {author} {\bibfnamefont {F.}~\bibnamefont {Yang}},\ }\href {\doibase
  10.1103/PhysRevLett.121.217001} {\bibfield  {journal} {\bibinfo  {journal}
  {Phys. Rev. Lett.}\ }\textbf {\bibinfo {volume} {121}},\ \bibinfo {pages}
  {217001} (\bibinfo {year} {2018})}\BibitemShut {NoStop}%
\bibitem [{\citenamefont {Gonzalez-Arraga}\ \emph {et~al.}(2017)\citenamefont
  {Gonzalez-Arraga}, \citenamefont {Lado}, \citenamefont {Guinea},\ and\
  \citenamefont {San-Jose}}]{LadoPablo}%
  \BibitemOpen
  \bibfield  {author} {\bibinfo {author} {\bibfnamefont {L.~A.}\ \bibnamefont
  {Gonzalez-Arraga}}, \bibinfo {author} {\bibfnamefont {J.~L.}\ \bibnamefont
  {Lado}}, \bibinfo {author} {\bibfnamefont {F.}~\bibnamefont {Guinea}}, \ and\
  \bibinfo {author} {\bibfnamefont {P.}~\bibnamefont {San-Jose}},\ }\href
  {\doibase 10.1103/PhysRevLett.119.107201} {\bibfield  {journal} {\bibinfo
  {journal} {Phys. Rev. Lett.}\ }\textbf {\bibinfo {volume} {119}},\ \bibinfo
  {pages} {107201} (\bibinfo {year} {2017})}\BibitemShut {NoStop}%
\bibitem [{\citenamefont {Ochi}\ \emph {et~al.}(2018)\citenamefont {Ochi},
  \citenamefont {Koshino},\ and\ \citenamefont {Kuroki}}]{ochi2018possible}%
  \BibitemOpen
  \bibfield  {author} {\bibinfo {author} {\bibfnamefont {M.}~\bibnamefont
  {Ochi}}, \bibinfo {author} {\bibfnamefont {M.}~\bibnamefont {Koshino}}, \
  and\ \bibinfo {author} {\bibfnamefont {K.}~\bibnamefont {Kuroki}},\ }\href
  {\doibase 10.1103/PhysRevB.98.081102} {\bibfield  {journal} {\bibinfo
  {journal} {Phys. Rev. B}\ }\textbf {\bibinfo {volume} {98}},\ \bibinfo
  {pages} {081102} (\bibinfo {year} {2018})}\BibitemShut {NoStop}%
\bibitem [{\citenamefont {Dodaro}\ \emph {et~al.}(2018)\citenamefont {Dodaro},
  \citenamefont {Kivelson}, \citenamefont {Schattner}, \citenamefont {Sun},\
  and\ \citenamefont {Wang}}]{dodaro2018phases}%
  \BibitemOpen
  \bibfield  {author} {\bibinfo {author} {\bibfnamefont {J.~F.}\ \bibnamefont
  {Dodaro}}, \bibinfo {author} {\bibfnamefont {S.~A.}\ \bibnamefont
  {Kivelson}}, \bibinfo {author} {\bibfnamefont {Y.}~\bibnamefont {Schattner}},
  \bibinfo {author} {\bibfnamefont {X.~Q.}\ \bibnamefont {Sun}}, \ and\
  \bibinfo {author} {\bibfnamefont {C.}~\bibnamefont {Wang}},\ }\href {\doibase
  10.1103/PhysRevB.98.075154} {\bibfield  {journal} {\bibinfo  {journal} {Phys.
  Rev. B}\ }\textbf {\bibinfo {volume} {98}},\ \bibinfo {pages} {075154}
  (\bibinfo {year} {2018})}\BibitemShut {NoStop}%
\bibitem [{\citenamefont {Guo}\ \emph {et~al.}(2018)\citenamefont {Guo},
  \citenamefont {Zhu}, \citenamefont {Feng},\ and\ \citenamefont
  {Scalettar}}]{guo2018pairing}%
  \BibitemOpen
  \bibfield  {author} {\bibinfo {author} {\bibfnamefont {H.}~\bibnamefont
  {Guo}}, \bibinfo {author} {\bibfnamefont {X.}~\bibnamefont {Zhu}}, \bibinfo
  {author} {\bibfnamefont {S.}~\bibnamefont {Feng}}, \ and\ \bibinfo {author}
  {\bibfnamefont {R.~T.}\ \bibnamefont {Scalettar}},\ }\href {\doibase
  10.1103/PhysRevB.97.235453} {\bibfield  {journal} {\bibinfo  {journal} {Phys.
  Rev. B}\ }\textbf {\bibinfo {volume} {97}},\ \bibinfo {pages} {235453}
  (\bibinfo {year} {2018})}\BibitemShut {NoStop}%
\bibitem [{\citenamefont {Isobe}\ \emph {et~al.}(2018)\citenamefont {Isobe},
  \citenamefont {Yuan},\ and\ \citenamefont {Fu}}]{isobe2018unconventional}%
  \BibitemOpen
  \bibfield  {author} {\bibinfo {author} {\bibfnamefont {H.}~\bibnamefont
  {Isobe}}, \bibinfo {author} {\bibfnamefont {N.~F.~Q.}\ \bibnamefont {Yuan}},
  \ and\ \bibinfo {author} {\bibfnamefont {L.}~\bibnamefont {Fu}},\ }\href
  {\doibase 10.1103/PhysRevX.8.041041} {\bibfield  {journal} {\bibinfo
  {journal} {Phys. Rev. X}\ }\textbf {\bibinfo {volume} {8}},\ \bibinfo {pages}
  {041041} (\bibinfo {year} {2018})}\BibitemShut {NoStop}%
\bibitem [{\citenamefont {Venderbos}\ and\ \citenamefont
  {Fernandes}(2018)}]{venderbos2018correlations}%
  \BibitemOpen
  \bibfield  {author} {\bibinfo {author} {\bibfnamefont {J.~W.~F.}\
  \bibnamefont {Venderbos}}\ and\ \bibinfo {author} {\bibfnamefont {R.~M.}\
  \bibnamefont {Fernandes}},\ }\href {\doibase 10.1103/PhysRevB.98.245103}
  {\bibfield  {journal} {\bibinfo  {journal} {Phys. Rev. B}\ }\textbf {\bibinfo
  {volume} {98}},\ \bibinfo {pages} {245103} (\bibinfo {year}
  {2018})}\BibitemShut {NoStop}%
\bibitem [{\citenamefont {Guinea}\ and\ \citenamefont
  {Walet}(2018)}]{guinea2018electrostatic}%
  \BibitemOpen
  \bibfield  {author} {\bibinfo {author} {\bibfnamefont {F.}~\bibnamefont
  {Guinea}}\ and\ \bibinfo {author} {\bibfnamefont {N.~R.}\ \bibnamefont
  {Walet}},\ }\href {https://www.pnas.org/content/115/52/13174} {\bibfield
  {journal} {\bibinfo  {journal} {Proceedings of the National Academy of
  Sciences}\ }\textbf {\bibinfo {volume} {115}},\ \bibinfo {pages} {13174}
  (\bibinfo {year} {2018})}\BibitemShut {NoStop}%
\bibitem [{\citenamefont {Liu}\ \emph {et~al.}(2019{\natexlab{a}})\citenamefont
  {Liu}, \citenamefont {Liu},\ and\ \citenamefont {Dai}}]{liu2018pseudo}%
  \BibitemOpen
  \bibfield  {author} {\bibinfo {author} {\bibfnamefont {J.}~\bibnamefont
  {Liu}}, \bibinfo {author} {\bibfnamefont {J.}~\bibnamefont {Liu}}, \ and\
  \bibinfo {author} {\bibfnamefont {X.}~\bibnamefont {Dai}},\ }\href {\doibase
  10.1103/PhysRevB.99.155415} {\bibfield  {journal} {\bibinfo  {journal} {Phys.
  Rev. B}\ }\textbf {\bibinfo {volume} {99}},\ \bibinfo {pages} {155415}
  (\bibinfo {year} {2019}{\natexlab{a}})}\BibitemShut {NoStop}%
\bibitem [{\citenamefont {Liu}\ \emph {et~al.}(2019{\natexlab{b}})\citenamefont
  {Liu}, \citenamefont {Ma}, \citenamefont {Gao},\ and\ \citenamefont
  {Dai}}]{LiuValley2019}%
  \BibitemOpen
  \bibfield  {author} {\bibinfo {author} {\bibfnamefont {J.}~\bibnamefont
  {Liu}}, \bibinfo {author} {\bibfnamefont {Z.}~\bibnamefont {Ma}}, \bibinfo
  {author} {\bibfnamefont {J.}~\bibnamefont {Gao}}, \ and\ \bibinfo {author}
  {\bibfnamefont {X.}~\bibnamefont {Dai}},\ }\href {\doibase
  10.1103/PhysRevX.9.031021} {\bibfield  {journal} {\bibinfo  {journal} {Phys.
  Rev. X}\ }\textbf {\bibinfo {volume} {9}},\ \bibinfo {pages} {031021}
  (\bibinfo {year} {2019}{\natexlab{b}})}\BibitemShut {NoStop}%
\bibitem [{\citenamefont {Cea}\ \emph {et~al.}(2019)\citenamefont {Cea},
  \citenamefont {Walet},\ and\ \citenamefont {Guinea}}]{Cea2019}%
  \BibitemOpen
  \bibfield  {author} {\bibinfo {author} {\bibfnamefont {T.}~\bibnamefont
  {Cea}}, \bibinfo {author} {\bibfnamefont {N.~R.}\ \bibnamefont {Walet}}, \
  and\ \bibinfo {author} {\bibfnamefont {F.}~\bibnamefont {Guinea}},\ }\href
  {\doibase 10.1103/PhysRevB.100.205113} {\bibfield  {journal} {\bibinfo
  {journal} {Phys. Rev. B}\ }\textbf {\bibinfo {volume} {100}},\ \bibinfo
  {pages} {205113} (\bibinfo {year} {2019})}\BibitemShut {NoStop}%
\bibitem [{\citenamefont {Tang}\ \emph {et~al.}(2019)\citenamefont {Tang},
  \citenamefont {Yang}, \citenamefont {Wang}, \citenamefont {Zhang},\ and\
  \citenamefont {Wang}}]{tang2019spin}%
  \BibitemOpen
  \bibfield  {author} {\bibinfo {author} {\bibfnamefont {Q.-K.}\ \bibnamefont
  {Tang}}, \bibinfo {author} {\bibfnamefont {L.}~\bibnamefont {Yang}}, \bibinfo
  {author} {\bibfnamefont {D.}~\bibnamefont {Wang}}, \bibinfo {author}
  {\bibfnamefont {F.-C.}\ \bibnamefont {Zhang}}, \ and\ \bibinfo {author}
  {\bibfnamefont {Q.-H.}\ \bibnamefont {Wang}},\ }\href {\doibase
  10.1103/PhysRevB.99.094521} {\bibfield  {journal} {\bibinfo  {journal} {Phys.
  Rev. B}\ }\textbf {\bibinfo {volume} {99}},\ \bibinfo {pages} {094521}
  (\bibinfo {year} {2019})}\BibitemShut {NoStop}%
\bibitem [{\citenamefont {Gonz\'alez}\ and\ \citenamefont
  {Stauber}(2019)}]{gonzalez2019kohn}%
  \BibitemOpen
  \bibfield  {author} {\bibinfo {author} {\bibfnamefont {J.}~\bibnamefont
  {Gonz\'alez}}\ and\ \bibinfo {author} {\bibfnamefont {T.}~\bibnamefont
  {Stauber}},\ }\href {\doibase 10.1103/PhysRevLett.122.026801} {\bibfield
  {journal} {\bibinfo  {journal} {Phys. Rev. Lett.}\ }\textbf {\bibinfo
  {volume} {122}},\ \bibinfo {pages} {026801} (\bibinfo {year}
  {2019})}\BibitemShut {NoStop}%
\bibitem [{\citenamefont {Kang}\ and\ \citenamefont
  {Vafek}(2019)}]{kang2019strong}%
  \BibitemOpen
  \bibfield  {author} {\bibinfo {author} {\bibfnamefont {J.}~\bibnamefont
  {Kang}}\ and\ \bibinfo {author} {\bibfnamefont {O.}~\bibnamefont {Vafek}},\
  }\href {\doibase 10.1103/PhysRevLett.122.246401} {\bibfield  {journal}
  {\bibinfo  {journal} {Phys. Rev. Lett.}\ }\textbf {\bibinfo {volume} {122}},\
  \bibinfo {pages} {246401} (\bibinfo {year} {2019})}\BibitemShut {NoStop}%
\bibitem [{\citenamefont {Seo}\ \emph {et~al.}(2019)\citenamefont {Seo},
  \citenamefont {Kotov},\ and\ \citenamefont {Uchoa}}]{seo2019ferromagnetic}%
  \BibitemOpen
  \bibfield  {author} {\bibinfo {author} {\bibfnamefont {K.}~\bibnamefont
  {Seo}}, \bibinfo {author} {\bibfnamefont {V.~N.}\ \bibnamefont {Kotov}}, \
  and\ \bibinfo {author} {\bibfnamefont {B.}~\bibnamefont {Uchoa}},\ }\href
  {\doibase 10.1103/PhysRevLett.122.246402} {\bibfield  {journal} {\bibinfo
  {journal} {Phys. Rev. Lett.}\ }\textbf {\bibinfo {volume} {122}},\ \bibinfo
  {pages} {246402} (\bibinfo {year} {2019})}\BibitemShut {NoStop}%
\bibitem [{\citenamefont {Zhang}\ \emph {et~al.}(2019)\citenamefont {Zhang},
  \citenamefont {Mao}, \citenamefont {Cao}, \citenamefont {Jarillo-Herrero},\
  and\ \citenamefont {Senthil}}]{zhang2019nearly}%
  \BibitemOpen
  \bibfield  {author} {\bibinfo {author} {\bibfnamefont {Y.-H.}\ \bibnamefont
  {Zhang}}, \bibinfo {author} {\bibfnamefont {D.}~\bibnamefont {Mao}}, \bibinfo
  {author} {\bibfnamefont {Y.}~\bibnamefont {Cao}}, \bibinfo {author}
  {\bibfnamefont {P.}~\bibnamefont {Jarillo-Herrero}}, \ and\ \bibinfo {author}
  {\bibfnamefont {T.}~\bibnamefont {Senthil}},\ }\href {\doibase
  10.1103/PhysRevB.99.075127} {\bibfield  {journal} {\bibinfo  {journal} {Phys.
  Rev. B}\ }\textbf {\bibinfo {volume} {99}},\ \bibinfo {pages} {075127}
  (\bibinfo {year} {2019})}\BibitemShut {NoStop}%
\bibitem [{\citenamefont {Lee}\ \emph {et~al.}(2019)\citenamefont {Lee},
  \citenamefont {Khalaf}, \citenamefont {Liu}, \citenamefont {Liu},
  \citenamefont {Hao}, \citenamefont {Kim},\ and\ \citenamefont
  {Vishwanath}}]{lee2019theory}%
  \BibitemOpen
  \bibfield  {author} {\bibinfo {author} {\bibfnamefont {J.~Y.}\ \bibnamefont
  {Lee}}, \bibinfo {author} {\bibfnamefont {E.}~\bibnamefont {Khalaf}},
  \bibinfo {author} {\bibfnamefont {S.}~\bibnamefont {Liu}}, \bibinfo {author}
  {\bibfnamefont {X.}~\bibnamefont {Liu}}, \bibinfo {author} {\bibfnamefont
  {Z.}~\bibnamefont {Hao}}, \bibinfo {author} {\bibfnamefont {P.}~\bibnamefont
  {Kim}}, \ and\ \bibinfo {author} {\bibfnamefont {A.}~\bibnamefont
  {Vishwanath}},\ }\href {https://www.nature.com/articles/s41467-019-12981-1}
  {\bibfield  {journal} {\bibinfo  {journal} {Nature communications}\ }\textbf
  {\bibinfo {volume} {10}},\ \bibinfo {pages} {1} (\bibinfo {year}
  {2019})}\BibitemShut {NoStop}%
\bibitem [{\citenamefont {Wu}\ and\ \citenamefont
  {Das~Sarma}(2020)}]{wucollective}%
  \BibitemOpen
  \bibfield  {author} {\bibinfo {author} {\bibfnamefont {F.}~\bibnamefont
  {Wu}}\ and\ \bibinfo {author} {\bibfnamefont {S.}~\bibnamefont {Das~Sarma}},\
  }\href {\doibase 10.1103/PhysRevLett.124.046403} {\bibfield  {journal}
  {\bibinfo  {journal} {Phys. Rev. Lett.}\ }\textbf {\bibinfo {volume} {124}},\
  \bibinfo {pages} {046403} (\bibinfo {year} {2020})}\BibitemShut {NoStop}%
\bibitem [{\citenamefont {Wu}\ \emph {et~al.}(2019)\citenamefont {Wu},
  \citenamefont {Keselman}, \citenamefont {Jian}, \citenamefont {Pawlak},\ and\
  \citenamefont {Xu}}]{wu2019ferromagnetism}%
  \BibitemOpen
  \bibfield  {author} {\bibinfo {author} {\bibfnamefont {X.-C.}\ \bibnamefont
  {Wu}}, \bibinfo {author} {\bibfnamefont {A.}~\bibnamefont {Keselman}},
  \bibinfo {author} {\bibfnamefont {C.-M.}\ \bibnamefont {Jian}}, \bibinfo
  {author} {\bibfnamefont {K.~A.}\ \bibnamefont {Pawlak}}, \ and\ \bibinfo
  {author} {\bibfnamefont {C.}~\bibnamefont {Xu}},\ }\href {\doibase
  10.1103/PhysRevB.100.024421} {\bibfield  {journal} {\bibinfo  {journal}
  {Phys. Rev. B}\ }\textbf {\bibinfo {volume} {100}},\ \bibinfo {pages}
  {024421} (\bibinfo {year} {2019})}\BibitemShut {NoStop}%
\bibitem [{\citenamefont {Bultinck}\ \emph
  {et~al.}(2020{\natexlab{a}})\citenamefont {Bultinck}, \citenamefont
  {Chatterjee},\ and\ \citenamefont {Zaletel}}]{bultinck2019anomalous}%
  \BibitemOpen
  \bibfield  {author} {\bibinfo {author} {\bibfnamefont {N.}~\bibnamefont
  {Bultinck}}, \bibinfo {author} {\bibfnamefont {S.}~\bibnamefont
  {Chatterjee}}, \ and\ \bibinfo {author} {\bibfnamefont {M.~P.}\ \bibnamefont
  {Zaletel}},\ }\href {\doibase 10.1103/PhysRevLett.124.166601} {\bibfield
  {journal} {\bibinfo  {journal} {Phys. Rev. Lett.}\ }\textbf {\bibinfo
  {volume} {124}},\ \bibinfo {pages} {166601} (\bibinfo {year}
  {2020}{\natexlab{a}})}\BibitemShut {NoStop}%
\bibitem [{\citenamefont {Liu}\ \emph {et~al.}(2019{\natexlab{c}})\citenamefont
  {Liu}, \citenamefont {Khalaf}, \citenamefont {Lee},\ and\ \citenamefont
  {Vishwanath}}]{liu2019nematic}%
  \BibitemOpen
  \bibfield  {author} {\bibinfo {author} {\bibfnamefont {S.}~\bibnamefont
  {Liu}}, \bibinfo {author} {\bibfnamefont {E.}~\bibnamefont {Khalaf}},
  \bibinfo {author} {\bibfnamefont {J.~Y.}\ \bibnamefont {Lee}}, \ and\
  \bibinfo {author} {\bibfnamefont {A.}~\bibnamefont {Vishwanath}},\ }\href
  {https://arxiv.org/abs/1905.07409} {\bibfield  {journal} {\bibinfo  {journal}
  {arXiv preprint arXiv:1905.07409}\ } (\bibinfo {year}
  {2019}{\natexlab{c}})}\BibitemShut {NoStop}%
\bibitem [{\citenamefont {Alavirad}\ and\ \citenamefont
  {Sau}(2019)}]{alavirad2019ferromagnetism}%
  \BibitemOpen
  \bibfield  {author} {\bibinfo {author} {\bibfnamefont {Y.}~\bibnamefont
  {Alavirad}}\ and\ \bibinfo {author} {\bibfnamefont {J.~D.}\ \bibnamefont
  {Sau}},\ }\href {https://arxiv.org/abs/1907.13633} {\bibfield  {journal}
  {\bibinfo  {journal} {arXiv preprint arXiv:1907.13633}\ } (\bibinfo {year}
  {2019})}\BibitemShut {NoStop}%
\bibitem [{\citenamefont {Chatterjee}\ \emph {et~al.}(2020)\citenamefont
  {Chatterjee}, \citenamefont {Bultinck},\ and\ \citenamefont
  {Zaletel}}]{chatterjee2019symmetry}%
  \BibitemOpen
  \bibfield  {author} {\bibinfo {author} {\bibfnamefont {S.}~\bibnamefont
  {Chatterjee}}, \bibinfo {author} {\bibfnamefont {N.}~\bibnamefont
  {Bultinck}}, \ and\ \bibinfo {author} {\bibfnamefont {M.~P.}\ \bibnamefont
  {Zaletel}},\ }\href {\doibase 10.1103/PhysRevB.101.165141} {\bibfield
  {journal} {\bibinfo  {journal} {Phys. Rev. B}\ }\textbf {\bibinfo {volume}
  {101}},\ \bibinfo {pages} {165141} (\bibinfo {year} {2020})}\BibitemShut
  {NoStop}%
\bibitem [{\citenamefont {Chichinadze}\ \emph {et~al.}(2020)\citenamefont
  {Chichinadze}, \citenamefont {Classen},\ and\ \citenamefont
  {Chubukov}}]{chichinadze2019nematic}%
  \BibitemOpen
  \bibfield  {author} {\bibinfo {author} {\bibfnamefont {D.~V.}\ \bibnamefont
  {Chichinadze}}, \bibinfo {author} {\bibfnamefont {L.}~\bibnamefont
  {Classen}}, \ and\ \bibinfo {author} {\bibfnamefont {A.~V.}\ \bibnamefont
  {Chubukov}},\ }\href {\doibase 10.1103/PhysRevB.101.224513} {\bibfield
  {journal} {\bibinfo  {journal} {Phys. Rev. B}\ }\textbf {\bibinfo {volume}
  {101}},\ \bibinfo {pages} {224513} (\bibinfo {year} {2020})}\BibitemShut
  {NoStop}%
\bibitem [{\citenamefont {Bultinck}\ \emph
  {et~al.}(2020{\natexlab{b}})\citenamefont {Bultinck}, \citenamefont {Khalaf},
  \citenamefont {Liu}, \citenamefont {Chatterjee}, \citenamefont {Vishwanath},\
  and\ \citenamefont {Zaletel}}]{bultinck2019ground}%
  \BibitemOpen
  \bibfield  {author} {\bibinfo {author} {\bibfnamefont {N.}~\bibnamefont
  {Bultinck}}, \bibinfo {author} {\bibfnamefont {E.}~\bibnamefont {Khalaf}},
  \bibinfo {author} {\bibfnamefont {S.}~\bibnamefont {Liu}}, \bibinfo {author}
  {\bibfnamefont {S.}~\bibnamefont {Chatterjee}}, \bibinfo {author}
  {\bibfnamefont {A.}~\bibnamefont {Vishwanath}}, \ and\ \bibinfo {author}
  {\bibfnamefont {M.~P.}\ \bibnamefont {Zaletel}},\ }\href {\doibase
  10.1103/PhysRevX.10.031034} {\bibfield  {journal} {\bibinfo  {journal} {Phys.
  Rev. X}\ }\textbf {\bibinfo {volume} {10}},\ \bibinfo {pages} {031034}
  (\bibinfo {year} {2020}{\natexlab{b}})}\BibitemShut {NoStop}%
\bibitem [{\citenamefont {{Liu}}\ and\ \citenamefont
  {{Dai}}(2019)}]{liu2019correlated}%
  \BibitemOpen
  \bibfield  {author} {\bibinfo {author} {\bibfnamefont {J.}~\bibnamefont
  {{Liu}}}\ and\ \bibinfo {author} {\bibfnamefont {X.}~\bibnamefont {{Dai}}},\
  }\href@noop {} {\bibfield  {journal} {\bibinfo  {journal} {arXiv e-prints}\
  ,\ \bibinfo {eid} {arXiv:1911.03760}} (\bibinfo {year} {2019})},\ \Eprint
  {http://arxiv.org/abs/1911.03760} {arXiv:1911.03760 [cond-mat.str-el]}
  \BibitemShut {NoStop}%
\bibitem [{\citenamefont {Fernandes}\ and\ \citenamefont
  {Venderbos}(2020)}]{fernandes2019nematicity}%
  \BibitemOpen
  \bibfield  {author} {\bibinfo {author} {\bibfnamefont {R.~M.}\ \bibnamefont
  {Fernandes}}\ and\ \bibinfo {author} {\bibfnamefont {J.~W.}\ \bibnamefont
  {Venderbos}},\ }\href {https://advances.sciencemag.org/content/6/32/eaba8834}
  {\bibfield  {journal} {\bibinfo  {journal} {Science Advances}\ }\textbf
  {\bibinfo {volume} {6}},\ \bibinfo {pages} {eaba8834} (\bibinfo {year}
  {2020})}\BibitemShut {NoStop}%
\bibitem [{\citenamefont {Zhang}\ \emph {et~al.}(2020)\citenamefont {Zhang},
  \citenamefont {Jiang}, \citenamefont {Wang},\ and\ \citenamefont
  {Zhang}}]{zhang2020correlated}%
  \BibitemOpen
  \bibfield  {author} {\bibinfo {author} {\bibfnamefont {Y.}~\bibnamefont
  {Zhang}}, \bibinfo {author} {\bibfnamefont {K.}~\bibnamefont {Jiang}},
  \bibinfo {author} {\bibfnamefont {Z.}~\bibnamefont {Wang}}, \ and\ \bibinfo
  {author} {\bibfnamefont {F.}~\bibnamefont {Zhang}},\ }\href {\doibase
  10.1103/PhysRevB.102.035136} {\bibfield  {journal} {\bibinfo  {journal}
  {Phys. Rev. B}\ }\textbf {\bibinfo {volume} {102}},\ \bibinfo {pages}
  {035136} (\bibinfo {year} {2020})}\BibitemShut {NoStop}%
\bibitem [{\citenamefont {Repellin}\ \emph {et~al.}(2020)\citenamefont
  {Repellin}, \citenamefont {Dong}, \citenamefont {Zhang},\ and\ \citenamefont
  {Senthil}}]{repellin2019ferromagnetism}%
  \BibitemOpen
  \bibfield  {author} {\bibinfo {author} {\bibfnamefont {C.}~\bibnamefont
  {Repellin}}, \bibinfo {author} {\bibfnamefont {Z.}~\bibnamefont {Dong}},
  \bibinfo {author} {\bibfnamefont {Y.-H.}\ \bibnamefont {Zhang}}, \ and\
  \bibinfo {author} {\bibfnamefont {T.}~\bibnamefont {Senthil}},\ }\href
  {\doibase 10.1103/PhysRevLett.124.187601} {\bibfield  {journal} {\bibinfo
  {journal} {Phys. Rev. Lett.}\ }\textbf {\bibinfo {volume} {124}},\ \bibinfo
  {pages} {187601} (\bibinfo {year} {2020})}\BibitemShut {NoStop}%
\bibitem [{\citenamefont {Liu}\ and\ \citenamefont
  {Dai}(2020)}]{LiuAnomalous2020}%
  \BibitemOpen
  \bibfield  {author} {\bibinfo {author} {\bibfnamefont {J.}~\bibnamefont
  {Liu}}\ and\ \bibinfo {author} {\bibfnamefont {X.}~\bibnamefont {Dai}},\
  }\href {\doibase 10.1038/s41524-020-0299-4} {\bibfield  {journal} {\bibinfo
  {journal} {npj Computational Materials}\ }\textbf {\bibinfo {volume} {6}},\
  \bibinfo {pages} {57} (\bibinfo {year} {2020})}\BibitemShut {NoStop}%
\bibitem [{\citenamefont {Kang}\ and\ \citenamefont
  {Vafek}(2020)}]{kang2020nonabelian}%
  \BibitemOpen
  \bibfield  {author} {\bibinfo {author} {\bibfnamefont {J.}~\bibnamefont
  {Kang}}\ and\ \bibinfo {author} {\bibfnamefont {O.}~\bibnamefont {Vafek}},\
  }\href {\doibase 10.1103/PhysRevB.102.035161} {\bibfield  {journal} {\bibinfo
   {journal} {Phys. Rev. B}\ }\textbf {\bibinfo {volume} {102}},\ \bibinfo
  {pages} {035161} (\bibinfo {year} {2020})}\BibitemShut {NoStop}%
\bibitem [{\citenamefont {Huang}\ \emph {et~al.}(2020)\citenamefont {Huang},
  \citenamefont {Huang},\ and\ \citenamefont {Lee}}]{huang2020slaverotor}%
  \BibitemOpen
  \bibfield  {author} {\bibinfo {author} {\bibfnamefont {S.-M.}\ \bibnamefont
  {Huang}}, \bibinfo {author} {\bibfnamefont {Y.-P.}\ \bibnamefont {Huang}}, \
  and\ \bibinfo {author} {\bibfnamefont {T.-K.}\ \bibnamefont {Lee}},\ }\href
  {\doibase 10.1103/PhysRevB.101.235140} {\bibfield  {journal} {\bibinfo
  {journal} {Phys. Rev. B}\ }\textbf {\bibinfo {volume} {101}},\ \bibinfo
  {pages} {235140} (\bibinfo {year} {2020})}\BibitemShut {NoStop}%
\bibitem [{\citenamefont {Lu}\ \emph {et~al.}(2020)\citenamefont {Lu},
  \citenamefont {Zhang}, \citenamefont {Zhang}, \citenamefont {Zhang},
  \citenamefont {Liu}, \citenamefont {Gu}, \citenamefont {Chen},\ and\
  \citenamefont {Yang}}]{lu2020chiral}%
  \BibitemOpen
  \bibfield  {author} {\bibinfo {author} {\bibfnamefont {C.}~\bibnamefont
  {Lu}}, \bibinfo {author} {\bibfnamefont {Y.}~\bibnamefont {Zhang}}, \bibinfo
  {author} {\bibfnamefont {Y.}~\bibnamefont {Zhang}}, \bibinfo {author}
  {\bibfnamefont {M.}~\bibnamefont {Zhang}}, \bibinfo {author} {\bibfnamefont
  {C.-C.}\ \bibnamefont {Liu}}, \bibinfo {author} {\bibfnamefont {Z.-C.}\
  \bibnamefont {Gu}}, \bibinfo {author} {\bibfnamefont {W.-Q.}\ \bibnamefont
  {Chen}}, \ and\ \bibinfo {author} {\bibfnamefont {F.}~\bibnamefont {Yang}},\
  }\href@noop {} {\  (\bibinfo {year} {2020})},\ \Eprint
  {http://arxiv.org/abs/2003.09513} {arXiv:2003.09513 [cond-mat.str-el]}
  \BibitemShut {NoStop}%
\bibitem [{\citenamefont {Li}\ \emph {et~al.}(2020)\citenamefont {Li},
  \citenamefont {Zhang}, \citenamefont {Ren}, \citenamefont {Liu},
  \citenamefont {Dai},\ and\ \citenamefont {He}}]{li2019experimental}%
  \BibitemOpen
  \bibfield  {author} {\bibinfo {author} {\bibfnamefont {S.-Y.}\ \bibnamefont
  {Li}}, \bibinfo {author} {\bibfnamefont {Y.}~\bibnamefont {Zhang}}, \bibinfo
  {author} {\bibfnamefont {Y.-N.}\ \bibnamefont {Ren}}, \bibinfo {author}
  {\bibfnamefont {J.}~\bibnamefont {Liu}}, \bibinfo {author} {\bibfnamefont
  {X.}~\bibnamefont {Dai}}, \ and\ \bibinfo {author} {\bibfnamefont
  {L.}~\bibnamefont {He}},\ }\href {\doibase 10.1103/PhysRevB.102.121406}
  {\bibfield  {journal} {\bibinfo  {journal} {Phys. Rev. B}\ }\textbf {\bibinfo
  {volume} {102}},\ \bibinfo {pages} {121406} (\bibinfo {year}
  {2020})}\BibitemShut {NoStop}%
\bibitem [{\citenamefont {{Wang}}\ \emph
  {et~al.}(2020{\natexlab{a}})\citenamefont {{Wang}}, \citenamefont {{Kang}},\
  and\ \citenamefont {{Fernandes}}}]{YuxuanWang2020}%
  \BibitemOpen
  \bibfield  {author} {\bibinfo {author} {\bibfnamefont {Y.}~\bibnamefont
  {{Wang}}}, \bibinfo {author} {\bibfnamefont {J.}~\bibnamefont {{Kang}}}, \
  and\ \bibinfo {author} {\bibfnamefont {R.~M.}\ \bibnamefont {{Fernandes}}},\
  }\href@noop {} {\bibfield  {journal} {\bibinfo  {journal} {arXiv e-prints}\
  ,\ \bibinfo {eid} {arXiv:2009.01237}} (\bibinfo {year}
  {2020}{\natexlab{a}})},\ \Eprint {http://arxiv.org/abs/2009.01237}
  {arXiv:2009.01237 [cond-mat.supr-con]} \BibitemShut {NoStop}%
\bibitem [{\citenamefont {{Wang}}\ \emph
  {et~al.}(2020{\natexlab{b}})\citenamefont {{Wang}}, \citenamefont
  {{Bultinck}},\ and\ \citenamefont {{Zaletel}}}]{TianleWang2020}%
  \BibitemOpen
  \bibfield  {author} {\bibinfo {author} {\bibfnamefont {T.}~\bibnamefont
  {{Wang}}}, \bibinfo {author} {\bibfnamefont {N.}~\bibnamefont {{Bultinck}}},
  \ and\ \bibinfo {author} {\bibfnamefont {M.~P.}\ \bibnamefont {{Zaletel}}},\
  }\href@noop {} {\bibfield  {journal} {\bibinfo  {journal} {arXiv e-prints}\
  ,\ \bibinfo {eid} {arXiv:2008.06528}} (\bibinfo {year}
  {2020}{\natexlab{b}})},\ \Eprint {http://arxiv.org/abs/2008.06528}
  {arXiv:2008.06528 [cond-mat.str-el]} \BibitemShut {NoStop}%
\bibitem [{\citenamefont {{Christos}}\ \emph {et~al.}(2020)\citenamefont
  {{Christos}}, \citenamefont {{Sachdev}},\ and\ \citenamefont
  {{Scheurer}}}]{Christos2020}%
  \BibitemOpen
  \bibfield  {author} {\bibinfo {author} {\bibfnamefont {M.}~\bibnamefont
  {{Christos}}}, \bibinfo {author} {\bibfnamefont {S.}~\bibnamefont
  {{Sachdev}}}, \ and\ \bibinfo {author} {\bibfnamefont {M.}~\bibnamefont
  {{Scheurer}}},\ }\href@noop {} {\bibfield  {journal} {\bibinfo  {journal}
  {arXiv e-prints}\ ,\ \bibinfo {eid} {arXiv:2007.00007}} (\bibinfo {year}
  {2020})},\ \Eprint {http://arxiv.org/abs/2007.00007} {arXiv:2007.00007
  [cond-mat.str-el]} \BibitemShut {NoStop}%
\bibitem [{\citenamefont {{Kozii}}\ \emph {et~al.}(2020)\citenamefont
  {{Kozii}}, \citenamefont {{Zaletel}},\ and\ \citenamefont
  {{Bultinck}}}]{VKozii2020}%
  \BibitemOpen
  \bibfield  {author} {\bibinfo {author} {\bibfnamefont {V.}~\bibnamefont
  {{Kozii}}}, \bibinfo {author} {\bibfnamefont {M.~P.}\ \bibnamefont
  {{Zaletel}}}, \ and\ \bibinfo {author} {\bibfnamefont {N.}~\bibnamefont
  {{Bultinck}}},\ }\href@noop {} {\bibfield  {journal} {\bibinfo  {journal}
  {arXiv e-prints}\ ,\ \bibinfo {eid} {arXiv:2005.12961}} (\bibinfo {year}
  {2020})},\ \Eprint {http://arxiv.org/abs/2005.12961} {arXiv:2005.12961
  [cond-mat.str-el]} \BibitemShut {NoStop}%
\bibitem [{\citenamefont {He}\ \emph {et~al.}(2020)\citenamefont {He},
  \citenamefont {Goldhaber-Gordon},\ and\ \citenamefont {Law}}]{WYHe2020}%
  \BibitemOpen
  \bibfield  {author} {\bibinfo {author} {\bibfnamefont {W.-Y.}\ \bibnamefont
  {He}}, \bibinfo {author} {\bibfnamefont {D.}~\bibnamefont
  {Goldhaber-Gordon}}, \ and\ \bibinfo {author} {\bibfnamefont {K.~T.}\
  \bibnamefont {Law}},\ }\href {\doibase 10.1038/s41467-020-15473-9} {\bibfield
   {journal} {\bibinfo  {journal} {Nature Communications}\ }\textbf {\bibinfo
  {volume} {11}},\ \bibinfo {pages} {1650} (\bibinfo {year}
  {2020})}\BibitemShut {NoStop}%
\bibitem [{\citenamefont {Xu}\ \emph {et~al.}(2018)\citenamefont {Xu},
  \citenamefont {Law},\ and\ \citenamefont {Lee}}]{xu2018kekule}%
  \BibitemOpen
  \bibfield  {author} {\bibinfo {author} {\bibfnamefont {X.~Y.}\ \bibnamefont
  {Xu}}, \bibinfo {author} {\bibfnamefont {K.~T.}\ \bibnamefont {Law}}, \ and\
  \bibinfo {author} {\bibfnamefont {P.~A.}\ \bibnamefont {Lee}},\ }\href
  {\doibase 10.1103/PhysRevB.98.121406} {\bibfield  {journal} {\bibinfo
  {journal} {Phys. Rev. B}\ }\textbf {\bibinfo {volume} {98}},\ \bibinfo
  {pages} {121406} (\bibinfo {year} {2018})}\BibitemShut {NoStop}%
\bibitem [{\citenamefont {Da~Liao}\ \emph {et~al.}(2019)\citenamefont
  {Da~Liao}, \citenamefont {Meng},\ and\ \citenamefont {Xu}}]{YDLiao2019}%
  \BibitemOpen
  \bibfield  {author} {\bibinfo {author} {\bibfnamefont {Y.}~\bibnamefont
  {Da~Liao}}, \bibinfo {author} {\bibfnamefont {Z.~Y.}\ \bibnamefont {Meng}}, \
  and\ \bibinfo {author} {\bibfnamefont {X.~Y.}\ \bibnamefont {Xu}},\ }\href
  {\doibase 10.1103/PhysRevLett.123.157601} {\bibfield  {journal} {\bibinfo
  {journal} {Phys. Rev. Lett.}\ }\textbf {\bibinfo {volume} {123}},\ \bibinfo
  {pages} {157601} (\bibinfo {year} {2019})}\BibitemShut {NoStop}%
\bibitem [{\citenamefont {{Da Liao}}\ \emph {et~al.}(2020)\citenamefont {{Da
  Liao}}, \citenamefont {{Xu}}, \citenamefont {{Meng}},\ and\ \citenamefont
  {{Kang}}}]{YDLiao2020review}%
  \BibitemOpen
  \bibfield  {author} {\bibinfo {author} {\bibfnamefont {Y.}~\bibnamefont {{Da
  Liao}}}, \bibinfo {author} {\bibfnamefont {X.~Y.}\ \bibnamefont {{Xu}}},
  \bibinfo {author} {\bibfnamefont {Z.~Y.}\ \bibnamefont {{Meng}}}, \ and\
  \bibinfo {author} {\bibfnamefont {J.}~\bibnamefont {{Kang}}},\ }\href@noop {}
  {\bibfield  {journal} {\bibinfo  {journal} {arXiv e-prints}\ ,\ \bibinfo
  {eid} {arXiv:2009.10076}} (\bibinfo {year} {2020})},\ \Eprint
  {http://arxiv.org/abs/2009.10076} {arXiv:2009.10076 [cond-mat.str-el]}
  \BibitemShut {NoStop}%
\bibitem [{\citenamefont {Soejima}\ \emph {et~al.}(2020)\citenamefont
  {Soejima}, \citenamefont {Parker}, \citenamefont {Bultinck}, \citenamefont
  {Hauschild},\ and\ \citenamefont {Zaletel}}]{soejima2020efficient}%
  \BibitemOpen
  \bibfield  {author} {\bibinfo {author} {\bibfnamefont {T.}~\bibnamefont
  {Soejima}}, \bibinfo {author} {\bibfnamefont {D.~E.}\ \bibnamefont {Parker}},
  \bibinfo {author} {\bibfnamefont {N.}~\bibnamefont {Bultinck}}, \bibinfo
  {author} {\bibfnamefont {J.}~\bibnamefont {Hauschild}}, \ and\ \bibinfo
  {author} {\bibfnamefont {M.~P.}\ \bibnamefont {Zaletel}},\ }\href {\doibase
  10.1103/PhysRevB.102.205111} {\bibfield  {journal} {\bibinfo  {journal}
  {Phys. Rev. B}\ }\textbf {\bibinfo {volume} {102}},\ \bibinfo {pages}
  {205111} (\bibinfo {year} {2020})}\BibitemShut {NoStop}%
\bibitem [{\citenamefont {{Xie}}\ \emph {et~al.}(2020)\citenamefont {{Xie}},
  \citenamefont {{Cowsik}}, \citenamefont {{Song}}, \citenamefont {{Lian}},
  \citenamefont {{Bernevig}},\ and\ \citenamefont {{Regnault}}}]{XieFang2020}%
  \BibitemOpen
  \bibfield  {author} {\bibinfo {author} {\bibfnamefont {F.}~\bibnamefont
  {{Xie}}}, \bibinfo {author} {\bibfnamefont {A.}~\bibnamefont {{Cowsik}}},
  \bibinfo {author} {\bibfnamefont {Z.-D.}\ \bibnamefont {{Song}}}, \bibinfo
  {author} {\bibfnamefont {B.}~\bibnamefont {{Lian}}}, \bibinfo {author}
  {\bibfnamefont {B.~A.}\ \bibnamefont {{Bernevig}}}, \ and\ \bibinfo {author}
  {\bibfnamefont {N.}~\bibnamefont {{Regnault}}},\ }\href@noop {} {\bibfield
  {journal} {\bibinfo  {journal} {arXiv e-prints}\ ,\ \bibinfo {eid}
  {arXiv:2010.00588}} (\bibinfo {year} {2020})},\ \Eprint
  {http://arxiv.org/abs/2010.00588} {arXiv:2010.00588 [cond-mat.str-el]}
  \BibitemShut {NoStop}%
\bibitem [{\citenamefont {{Vafek}}\ and\ \citenamefont
  {{Kang}}(2020)}]{VafekKang2020}%
  \BibitemOpen
  \bibfield  {author} {\bibinfo {author} {\bibfnamefont {O.}~\bibnamefont
  {{Vafek}}}\ and\ \bibinfo {author} {\bibfnamefont {J.}~\bibnamefont
  {{Kang}}},\ }\href@noop {} {\bibfield  {journal} {\bibinfo  {journal} {arXiv
  e-prints}\ ,\ \bibinfo {eid} {arXiv:2009.09413}} (\bibinfo {year} {2020})},\
  \Eprint {http://arxiv.org/abs/2009.09413} {arXiv:2009.09413
  [cond-mat.str-el]} \BibitemShut {NoStop}%
\bibitem [{\citenamefont {Sharpe}\ \emph {et~al.}(2019)\citenamefont {Sharpe},
  \citenamefont {Fox}, \citenamefont {Barnard}, \citenamefont {Finney},
  \citenamefont {Watanabe}, \citenamefont {Taniguchi}, \citenamefont
  {Kastner},\ and\ \citenamefont {Goldhaber-Gordon}}]{sharpe2019emergent}%
  \BibitemOpen
  \bibfield  {author} {\bibinfo {author} {\bibfnamefont {A.~L.}\ \bibnamefont
  {Sharpe}}, \bibinfo {author} {\bibfnamefont {E.~J.}\ \bibnamefont {Fox}},
  \bibinfo {author} {\bibfnamefont {A.~W.}\ \bibnamefont {Barnard}}, \bibinfo
  {author} {\bibfnamefont {J.}~\bibnamefont {Finney}}, \bibinfo {author}
  {\bibfnamefont {K.}~\bibnamefont {Watanabe}}, \bibinfo {author}
  {\bibfnamefont {T.}~\bibnamefont {Taniguchi}}, \bibinfo {author}
  {\bibfnamefont {M.}~\bibnamefont {Kastner}}, \ and\ \bibinfo {author}
  {\bibfnamefont {D.}~\bibnamefont {Goldhaber-Gordon}},\ }\href
  {https://science.sciencemag.org/content/365/6453/605} {\bibfield  {journal}
  {\bibinfo  {journal} {Science}\ }\textbf {\bibinfo {volume} {365}},\ \bibinfo
  {pages} {605} (\bibinfo {year} {2019})}\BibitemShut {NoStop}%
\bibitem [{\citenamefont {Serlin}\ \emph {et~al.}(2020)\citenamefont {Serlin},
  \citenamefont {Tschirhart}, \citenamefont {Polshyn}, \citenamefont {Zhang},
  \citenamefont {Zhu}, \citenamefont {Watanabe}, \citenamefont {Taniguchi},
  \citenamefont {Balents},\ and\ \citenamefont {Young}}]{serlin2020intrinsic}%
  \BibitemOpen
  \bibfield  {author} {\bibinfo {author} {\bibfnamefont {M.}~\bibnamefont
  {Serlin}}, \bibinfo {author} {\bibfnamefont {C.}~\bibnamefont {Tschirhart}},
  \bibinfo {author} {\bibfnamefont {H.}~\bibnamefont {Polshyn}}, \bibinfo
  {author} {\bibfnamefont {Y.}~\bibnamefont {Zhang}}, \bibinfo {author}
  {\bibfnamefont {J.}~\bibnamefont {Zhu}}, \bibinfo {author} {\bibfnamefont
  {K.}~\bibnamefont {Watanabe}}, \bibinfo {author} {\bibfnamefont
  {T.}~\bibnamefont {Taniguchi}}, \bibinfo {author} {\bibfnamefont
  {L.}~\bibnamefont {Balents}}, \ and\ \bibinfo {author} {\bibfnamefont
  {A.}~\bibnamefont {Young}},\ }\href
  {https://science.sciencemag.org/content/367/6480/900} {\bibfield  {journal}
  {\bibinfo  {journal} {Science}\ }\textbf {\bibinfo {volume} {367}},\ \bibinfo
  {pages} {900} (\bibinfo {year} {2020})}\BibitemShut {NoStop}%
\bibitem [{\citenamefont {Xie}\ and\ \citenamefont
  {MacDonald}(2020)}]{MacDonald2020}%
  \BibitemOpen
  \bibfield  {author} {\bibinfo {author} {\bibfnamefont {M.}~\bibnamefont
  {Xie}}\ and\ \bibinfo {author} {\bibfnamefont {A.~H.}\ \bibnamefont
  {MacDonald}},\ }\href {\doibase 10.1103/PhysRevLett.124.097601} {\bibfield
  {journal} {\bibinfo  {journal} {Phys. Rev. Lett.}\ }\textbf {\bibinfo
  {volume} {124}},\ \bibinfo {pages} {097601} (\bibinfo {year}
  {2020})}\BibitemShut {NoStop}%
\bibitem [{\citenamefont {Cea}\ and\ \citenamefont
  {Guinea}(2020)}]{Guinea2020}%
  \BibitemOpen
  \bibfield  {author} {\bibinfo {author} {\bibfnamefont {T.}~\bibnamefont
  {Cea}}\ and\ \bibinfo {author} {\bibfnamefont {F.}~\bibnamefont {Guinea}},\
  }\href {\doibase 10.1103/PhysRevB.102.045107} {\bibfield  {journal} {\bibinfo
   {journal} {Phys. Rev. B}\ }\textbf {\bibinfo {volume} {102}},\ \bibinfo
  {pages} {045107} (\bibinfo {year} {2020})}\BibitemShut {NoStop}%
\bibitem [{\citenamefont {Lang}\ \emph {et~al.}(2013)\citenamefont {Lang},
  \citenamefont {Meng}, \citenamefont {Muramatsu}, \citenamefont {Wessel},\
  and\ \citenamefont {Assaad}}]{Lang2013}%
  \BibitemOpen
  \bibfield  {author} {\bibinfo {author} {\bibfnamefont {T.~C.}\ \bibnamefont
  {Lang}}, \bibinfo {author} {\bibfnamefont {Z.~Y.}\ \bibnamefont {Meng}},
  \bibinfo {author} {\bibfnamefont {A.}~\bibnamefont {Muramatsu}}, \bibinfo
  {author} {\bibfnamefont {S.}~\bibnamefont {Wessel}}, \ and\ \bibinfo {author}
  {\bibfnamefont {F.~F.}\ \bibnamefont {Assaad}},\ }\href {\doibase
  10.1103/PhysRevLett.111.066401} {\bibfield  {journal} {\bibinfo  {journal}
  {Phys. Rev. Lett.}\ }\textbf {\bibinfo {volume} {111}},\ \bibinfo {pages}
  {066401} (\bibinfo {year} {2013})}\BibitemShut {NoStop}%
\bibitem [{\citenamefont {Zhou}\ \emph {et~al.}(2016)\citenamefont {Zhou},
  \citenamefont {Wang}, \citenamefont {Meng}, \citenamefont {Wang},\ and\
  \citenamefont {Wu}}]{ZCZhou2016}%
  \BibitemOpen
  \bibfield  {author} {\bibinfo {author} {\bibfnamefont {Z.}~\bibnamefont
  {Zhou}}, \bibinfo {author} {\bibfnamefont {D.}~\bibnamefont {Wang}}, \bibinfo
  {author} {\bibfnamefont {Z.~Y.}\ \bibnamefont {Meng}}, \bibinfo {author}
  {\bibfnamefont {Y.}~\bibnamefont {Wang}}, \ and\ \bibinfo {author}
  {\bibfnamefont {C.}~\bibnamefont {Wu}},\ }\href {\doibase
  10.1103/PhysRevB.93.245157} {\bibfield  {journal} {\bibinfo  {journal} {Phys.
  Rev. B}\ }\textbf {\bibinfo {volume} {93}},\ \bibinfo {pages} {245157}
  (\bibinfo {year} {2016})}\BibitemShut {NoStop}%
\bibitem [{\citenamefont {Moon}\ and\ \citenamefont
  {Koshino}(2012)}]{Koshino2012}%
  \BibitemOpen
  \bibfield  {author} {\bibinfo {author} {\bibfnamefont {P.}~\bibnamefont
  {Moon}}\ and\ \bibinfo {author} {\bibfnamefont {M.}~\bibnamefont {Koshino}},\
  }\href {\doibase 10.1103/PhysRevB.85.195458} {\bibfield  {journal} {\bibinfo
  {journal} {Phys. Rev. B}\ }\textbf {\bibinfo {volume} {85}},\ \bibinfo
  {pages} {195458} (\bibinfo {year} {2012})}\BibitemShut {NoStop}%
\bibitem [{\citenamefont {Meng}\ \emph {et~al.}(2010)\citenamefont {Meng},
  \citenamefont {Lang}, \citenamefont {Wessel}, \citenamefont {Assaad},\ and\
  \citenamefont {Muramatsu}}]{meng2010quantum}%
  \BibitemOpen
  \bibfield  {author} {\bibinfo {author} {\bibfnamefont {Z.~Y.}\ \bibnamefont
  {Meng}}, \bibinfo {author} {\bibfnamefont {T.~C.}\ \bibnamefont {Lang}},
  \bibinfo {author} {\bibfnamefont {S.}~\bibnamefont {Wessel}}, \bibinfo
  {author} {\bibfnamefont {F.~F.}\ \bibnamefont {Assaad}}, \ and\ \bibinfo
  {author} {\bibfnamefont {A.}~\bibnamefont {Muramatsu}},\ }\href {\doibase
  10.1038/nature08942} {\bibfield  {journal} {\bibinfo  {journal} {Nature}\
  }\textbf {\bibinfo {volume} {464}},\ \bibinfo {pages} {847} (\bibinfo {year}
  {2010})}\BibitemShut {NoStop}%
\bibitem [{\citenamefont {Xu}\ \emph {et~al.}(2017)\citenamefont {Xu},
  \citenamefont {Beach}, \citenamefont {Sun}, \citenamefont {Assaad},\ and\
  \citenamefont {Meng}}]{xu2017topo}%
  \BibitemOpen
  \bibfield  {author} {\bibinfo {author} {\bibfnamefont {X.~Y.}\ \bibnamefont
  {Xu}}, \bibinfo {author} {\bibfnamefont {K.~S.~D.}\ \bibnamefont {Beach}},
  \bibinfo {author} {\bibfnamefont {K.}~\bibnamefont {Sun}}, \bibinfo {author}
  {\bibfnamefont {F.~F.}\ \bibnamefont {Assaad}}, \ and\ \bibinfo {author}
  {\bibfnamefont {Z.~Y.}\ \bibnamefont {Meng}},\ }\href {\doibase
  10.1103/PhysRevB.95.085110} {\bibfield  {journal} {\bibinfo  {journal} {Phys.
  Rev. B}\ }\textbf {\bibinfo {volume} {95}},\ \bibinfo {pages} {085110}
  (\bibinfo {year} {2017})}\BibitemShut {NoStop}%
\bibitem [{\citenamefont {He}\ \emph {et~al.}(2018)\citenamefont {He},
  \citenamefont {Xu}, \citenamefont {Sun}, \citenamefont {Assaad},
  \citenamefont {Meng},\ and\ \citenamefont {Lu}}]{YuanYaoHe2018}%
  \BibitemOpen
  \bibfield  {author} {\bibinfo {author} {\bibfnamefont {Y.-Y.}\ \bibnamefont
  {He}}, \bibinfo {author} {\bibfnamefont {X.~Y.}\ \bibnamefont {Xu}}, \bibinfo
  {author} {\bibfnamefont {K.}~\bibnamefont {Sun}}, \bibinfo {author}
  {\bibfnamefont {F.~F.}\ \bibnamefont {Assaad}}, \bibinfo {author}
  {\bibfnamefont {Z.~Y.}\ \bibnamefont {Meng}}, \ and\ \bibinfo {author}
  {\bibfnamefont {Z.-Y.}\ \bibnamefont {Lu}},\ }\href {\doibase
  10.1103/PhysRevB.97.081110} {\bibfield  {journal} {\bibinfo  {journal} {Phys.
  Rev. B}\ }\textbf {\bibinfo {volume} {97}},\ \bibinfo {pages} {081110}
  (\bibinfo {year} {2018})}\BibitemShut {NoStop}%
\bibitem [{\citenamefont {Liu}\ \emph {et~al.}(2020{\natexlab{b}})\citenamefont
  {Liu}, \citenamefont {Wang}, \citenamefont {Sun},\ and\ \citenamefont
  {Meng}}]{YZLiu2020}%
  \BibitemOpen
  \bibfield  {author} {\bibinfo {author} {\bibfnamefont {Y.}~\bibnamefont
  {Liu}}, \bibinfo {author} {\bibfnamefont {W.}~\bibnamefont {Wang}}, \bibinfo
  {author} {\bibfnamefont {K.}~\bibnamefont {Sun}}, \ and\ \bibinfo {author}
  {\bibfnamefont {Z.~Y.}\ \bibnamefont {Meng}},\ }\href {\doibase
  10.1103/PhysRevB.101.064308} {\bibfield  {journal} {\bibinfo  {journal}
  {Phys. Rev. B}\ }\textbf {\bibinfo {volume} {101}},\ \bibinfo {pages}
  {064308} (\bibinfo {year} {2020}{\natexlab{b}})}\BibitemShut {NoStop}%
\bibitem [{\citenamefont {Haldane}(1988)}]{Haldane88}%
  \BibitemOpen
  \bibfield  {author} {\bibinfo {author} {\bibfnamefont {F.~D.~M.}\
  \bibnamefont {Haldane}},\ }\href {\doibase 10.1103/PhysRevLett.61.2015}
  {\bibfield  {journal} {\bibinfo  {journal} {Phys. Rev. Lett.}\ }\textbf
  {\bibinfo {volume} {61}},\ \bibinfo {pages} {2015} (\bibinfo {year}
  {1988})}\BibitemShut {NoStop}%
\bibitem [{\citenamefont {Hohenadler}\ \emph {et~al.}(2012)\citenamefont
  {Hohenadler}, \citenamefont {Meng}, \citenamefont {Lang}, \citenamefont
  {Wessel}, \citenamefont {Muramatsu},\ and\ \citenamefont
  {Assaad}}]{Hohenadler2012}%
  \BibitemOpen
  \bibfield  {author} {\bibinfo {author} {\bibfnamefont {M.}~\bibnamefont
  {Hohenadler}}, \bibinfo {author} {\bibfnamefont {Z.~Y.}\ \bibnamefont
  {Meng}}, \bibinfo {author} {\bibfnamefont {T.~C.}\ \bibnamefont {Lang}},
  \bibinfo {author} {\bibfnamefont {S.}~\bibnamefont {Wessel}}, \bibinfo
  {author} {\bibfnamefont {A.}~\bibnamefont {Muramatsu}}, \ and\ \bibinfo
  {author} {\bibfnamefont {F.~F.}\ \bibnamefont {Assaad}},\ }\href {\doibase
  10.1103/PhysRevB.85.115132} {\bibfield  {journal} {\bibinfo  {journal} {Phys.
  Rev. B}\ }\textbf {\bibinfo {volume} {85}},\ \bibinfo {pages} {115132}
  (\bibinfo {year} {2012})}\BibitemShut {NoStop}%
\bibitem [{\citenamefont {He}\ \emph {et~al.}(2016)\citenamefont {He},
  \citenamefont {Wu}, \citenamefont {You}, \citenamefont {Xu}, \citenamefont
  {Meng},\ and\ \citenamefont {Lu}}]{YYHe2016}%
  \BibitemOpen
  \bibfield  {author} {\bibinfo {author} {\bibfnamefont {Y.-Y.}\ \bibnamefont
  {He}}, \bibinfo {author} {\bibfnamefont {H.-Q.}\ \bibnamefont {Wu}}, \bibinfo
  {author} {\bibfnamefont {Y.-Z.}\ \bibnamefont {You}}, \bibinfo {author}
  {\bibfnamefont {C.}~\bibnamefont {Xu}}, \bibinfo {author} {\bibfnamefont
  {Z.~Y.}\ \bibnamefont {Meng}}, \ and\ \bibinfo {author} {\bibfnamefont
  {Z.-Y.}\ \bibnamefont {Lu}},\ }\href {\doibase 10.1103/PhysRevB.93.115150}
  {\bibfield  {journal} {\bibinfo  {journal} {Phys. Rev. B}\ }\textbf {\bibinfo
  {volume} {93}},\ \bibinfo {pages} {115150} (\bibinfo {year}
  {2016})}\BibitemShut {NoStop}%
\bibitem [{\citenamefont {Po}\ \emph {et~al.}(2018{\natexlab{b}})\citenamefont
  {Po}, \citenamefont {Watanabe},\ and\ \citenamefont
  {Vishwanath}}]{fragile_topology}%
  \BibitemOpen
  \bibfield  {author} {\bibinfo {author} {\bibfnamefont {H.~C.}\ \bibnamefont
  {Po}}, \bibinfo {author} {\bibfnamefont {H.}~\bibnamefont {Watanabe}}, \ and\
  \bibinfo {author} {\bibfnamefont {A.}~\bibnamefont {Vishwanath}},\ }\href
  {\doibase 10.1103/PhysRevLett.121.126402} {\bibfield  {journal} {\bibinfo
  {journal} {Phys. Rev. Lett.}\ }\textbf {\bibinfo {volume} {121}},\ \bibinfo
  {pages} {126402} (\bibinfo {year} {2018}{\natexlab{b}})}\BibitemShut
  {NoStop}%
\bibitem [{\citenamefont {Assaad}\ and\ \citenamefont
  {Evertz}(2008)}]{Assaad2008}%
  \BibitemOpen
  \bibfield  {author} {\bibinfo {author} {\bibfnamefont {F.}~\bibnamefont
  {Assaad}}\ and\ \bibinfo {author} {\bibfnamefont {H.}~\bibnamefont
  {Evertz}},\ }\enquote {\bibinfo {title} {World-line and determinantal quantum
  monte carlo methods for spins, phonons and electrons},}\ in\ \href {\doibase
  10.1007/978-3-540-74686-7_10} {\emph {\bibinfo {booktitle} {Computational
  Many-Particle Physics}}},\ \bibinfo {editor} {edited by\ \bibinfo {editor}
  {\bibfnamefont {H.}~\bibnamefont {Fehske}}, \bibinfo {editor} {\bibfnamefont
  {R.}~\bibnamefont {Schneider}}, \ and\ \bibinfo {editor} {\bibfnamefont
  {A.}~\bibnamefont {Wei{\ss}e}}}\ (\bibinfo  {publisher} {Springer Berlin
  Heidelberg},\ \bibinfo {address} {Berlin, Heidelberg},\ \bibinfo {year}
  {2008})\ pp.\ \bibinfo {pages} {277--356}\BibitemShut {NoStop}%
\end{thebibliography}%
\end{document}